\documentclass[longauth]{aa}

\usepackage{xcolor}
\usepackage{graphicx}
\usepackage{multirow}
\usepackage{txfonts}
\usepackage{hyperref}
\hypersetup{breaklinks=true,colorlinks=true,urlcolor=blue, linkcolor=blue,  citecolor=blue}
\usepackage{makecell}

\definecolor{Gray0}{gray}{0.95}
\definecolor{Gray1}{gray}{0.85}
\definecolor{Gray2}{gray}{0.75}
\definecolor{Gray3}{gray}{0.65}
\definecolor{LightCyan}{rgb}{0.88,1,1}
\definecolor{color1}{rgb}{0.74,0.9,0.9}
\definecolor{color2}{rgb}{0.58,0.9,0.9}
\definecolor{color3}{rgb}{0.36,0.9,0.9}

\begin{document}

\title{MATISSE, the VLTI mid-infrared imaging spectro-interferometer}


\author{ 
B. Lopez \inst{\ref{inst_O}}  \and S. Lagarde \inst{\ref{inst_O}} \and R.G. Petrov \inst{\ref{inst_O}} \and W. Jaffe \inst{\ref{inst_L}} \and P. Antonelli \inst{\ref{inst_O}}\and F. Allouche \inst{\ref{inst_O}} \and P. Berio \inst{\ref{inst_O}}  \and A. Matter \inst{\ref{inst_O}}  \and A. Meilland \inst{\ref{inst_O}} \and F. Millour \inst{\ref{inst_O}}  \and  S. Robbe-Dubois \inst{\ref{inst_O}}  \and Th. Henning \inst{\ref{inst_H}} \and G. Weigelt \inst{\ref{inst_B}} \and A. Glindemann \inst{\ref{inst_Garch}} \and T. Agocs \inst{\ref{inst_A}} \and Ch. Bailet \inst{\ref{inst_O}} \and U. Beckmann  \inst{\ref{inst_B}} \and F. Bettonvil  \inst{\ref{inst_A}} \and R. van~Boekel\inst{\ref{inst_H}} \and P. Bourget  \inst{\ref{inst_E}} \and Y. Bresson \inst{\ref{inst_O}} \and P. Bristow\inst{\ref{inst_Garch}} \and P. Cruzal\`ebes \inst{\ref{inst_O}} 
\and E. Eldswijk \inst{\ref{inst_A}} \and Y. Fanteï Caujolle \inst{\ref{inst_O}}   \and
J.C. Gonz\'alez Herrera\inst{\ref{inst_Garch}}\and U. Graser  \inst{\ref{inst_H}} \and
P. Guajardo\inst{\ref{inst_E}} \and M. Heininger \inst{5} \and K.-H. Hofmann \inst{\ref{inst_B}}  \and G. Kroes \inst{\ref{inst_A}} \and W. Laun \inst{\ref{inst_H}}  \and M. Lehmitz \inst{\ref{inst_H}} \and
C. Leinert\inst{\ref{inst_H}}
\and K. Meisenheimer \inst{\ref{inst_H}}  \and S. Morel \inst{\ref{inst_O}}  \and U. Neumann \inst{\ref{inst_H}} \and C. Paladini \inst{\ref{inst_E}}
\and I. Percheron
\inst{\ref{inst_Garch}} \and M. Riquelme \inst{\ref{inst_E}}\and M. Schoeller \inst{\ref{inst_Garch}} \and Ph. Stee \inst{\ref{inst_O}} \and
L. Venema\inst{\ref{inst_A}} \and
J. Woillez\inst{\ref{inst_Garch}} \and G. Zins \inst{\ref{inst_Garch},\ref{inst_E}}     \and 
P. \'Abrah\'am\inst{\ref{inst_K}}\and
S. Abadie\inst{\ref{inst_E}}
\and
R. Abuter\inst{\ref{inst_Garch}}\and
M. Accardo\inst{\ref{inst_Garch}}\and
T. Adler\inst{\ref{inst_H}}\and
J. Alonso\inst{\ref{inst_E}}\and
J.-C. Augereau\inst{\ref{inst_I}}\and
A. B\"ohm\inst{\ref{inst_H}}\and
G. Bazin\inst{\ref{inst_Garch}}\and
J. Beltran\inst{\ref{inst_E}} \and
A. Bensberg \inst{\ref{inst_Ki}}
\and
W. Boland\inst{\ref{inst_L}}\and
R. Brast\inst{\ref{inst_Garch}}\and
L. Burtscher\inst{\ref{inst_L}}\and
R. Castillo\inst{\ref{inst_E}}\and
A. Chelli\inst{\ref{inst_O}}\and
C. Cid\inst{\ref{inst_E}}\and
J.-M. Clausse\inst{\ref{inst_O}}\and
C. Connot\inst{\ref{inst_B}}\and
R.D. Conzelmann\inst{\ref{inst_Garch}}\and
W.-C. Danchi\inst{\ref{inst_Nas}}\and
M. Delbo\inst{\ref{inst_O}}
\and  J. Drevon \inst{\ref{inst_O}}
\and
C. Dominik\inst{\ref{inst_P}}\and
A. van Duin\inst{\ref{inst_A}}\and
M. Ebert\inst{\ref{inst_H}}\and
F. Eisenhauer\inst{\ref{inst_MPE}}\and
S. Flament\inst{\ref{inst_O}}\and
R. Frahm\inst{\ref{inst_Garch}}\and
V. G\'amez Rosas\inst{\ref{inst_L}}\and
A. Gabasch\inst{\ref{inst_Garch}}\and
A. Gallenne\inst{\ref{inst_Co},\ref{inst_Wa},\ref{inst_Sa}}\and
E. Garces\inst{\ref{inst_E}}\and
P. Girard\inst{\ref{inst_O}}\and
A. Glazenborg\inst{\ref{inst_A}}\and
F.Y.J. Gont\'e\inst{\ref{inst_E}}\and
F. Guitton\inst{\ref{inst_O}}\and
M. de Haan\inst{\ref{inst_A}}\and
H. Hanenburg\inst{\ref{inst_A}}\and
X. Haubois\inst{\ref{inst_E}} \and  V. Hocdé \inst{\ref{inst_O}}
\and
M. Hogerheijde\inst{\ref{inst_L},\ref{inst_P}}\and
R. ter Horst\inst{\ref{inst_A}}\and
J. Hron\inst{\ref{inst_V}}\and
C.A. Hummel \inst{\ref{inst_Garch}} \and
N. Hubin\inst{\ref{inst_Garch}}\and
R. Huerta\inst{\ref{inst_E}}
\and
J. Idserda\inst{\ref{inst_A}}\and
J. W. Isbell\inst{\ref{inst_H}}\and
D. Ives\inst{\ref{inst_Garch}}\and
G. Jakob\inst{\ref{inst_Garch}}\and
A. Jask\'o\inst{\ref{inst_A},\ref{inst_K}}\and
L. Jochum\inst{\ref{inst_E}}\and  L. Klarmann \inst{\ref{inst_H}}
\and
R. Klein\inst{\ref{inst_H}}\and
J. Kragt\inst{\ref{inst_A}}\and
S. Kuindersma\inst{\ref{inst_A}} \and  E. Kokoulina \inst{\ref{inst_O}} \and
L. Labadie\inst{\ref{inst_C}}\and
S. Lacour\inst{\ref{inst_LESIA}}\and
J. Leftley \inst{\ref{inst_O}} \and
R. Le Poole\inst{\ref{inst_L}}\and
J.-L. Lizon\inst{\ref{inst_Garch}}\and
M. Lopez\inst{\ref{inst_E}}\and
F. Lykou\inst{\ref{inst_K}}\and
A. M\'erand\inst{\ref{inst_Garch}}\and
A. Marcotto \inst{\ref{inst_O}} \and
N. Mauclert \inst{\ref{inst_O}} \and
T. Maurer\inst{\ref{inst_H}}\and
L.H. Mehrgan\inst{\ref{inst_Garch}}\and
J. Meisner\inst{\ref{inst_L}}\and
K. Meixner\inst{\ref{inst_H}}\and
 M. Mellein\inst{\ref{inst_H}}\and
J.L. Menut\inst{\ref{inst_GEOAZUR}}\and
L. Mohr\inst{\ref{inst_H}}\and
L. Mosoni\inst{\ref{inst_Zs},\ref{inst_K}}\and
R. Navarro\inst{\ref{inst_A}}\and
E. Nu{\ss}baum\inst{\ref{inst_B}}\and
L. Pallanca\inst{\ref{inst_E}}\and
E. Pantin\inst{\ref{inst_Pa}}\and
L. Pasquini\inst{\ref{inst_Garch}}\and
T. Phan Duc\inst{\ref{inst_Garch}}\and
J.-U. Pott\inst{\ref{inst_H}}\and
E. Pozna\inst{\ref{inst_Garch}}\and
A. Richichi\inst{\ref{inst_Th}}\and
A. Ridinger\inst{\ref{inst_H}}\and
F. Rigal\inst{\ref{inst_A}}\and
Th. Rivinius\inst{\ref{inst_E}}\and
R. Roelfsema\inst{\ref{inst_A}}\and
R.-R. Rohloff\inst{\ref{inst_H}}\and
S. Rousseau\inst{\ref{inst_O}}\and
D. Salabert\inst{\ref{inst_O}}\and
D. Schertl\inst{\ref{inst_B}}\and
N. Schuhler\inst{\ref{inst_E}}\and
M. Schuil\inst{\ref{inst_A}}\and
K. Shabun\inst{\ref{inst_Garch}}\and
A. Soulain\inst{\ref{inst_Sy}}\and
C. Stephan\inst{\ref{inst_E}} \and
P. Toledo\inst{\ref{inst_E}} \and
K. Tristram\inst{\ref{inst_E}} \and
N. Tromp\inst{\ref{inst_A}}\and
F. Vakili \inst{\ref{inst_O}} \and
J. Varga\inst{\ref{inst_L},\ref{inst_K}}\and
J. Vinther\inst{\ref{inst_Garch}}\and
L.B.F.M. Waters\inst{\ref{inst_Ra},\ref{inst_U}}\and
M. Wittkowski\inst{\ref{inst_Garch}}
\and
S. Wolf\inst{\ref{inst_Ki}}
\and
F. Wrhel\inst{\ref{inst_H}}
\and  G. Yoffe \inst{\ref{inst_H}}
}

\date{Received October --, 2020; accepted --, 2020}

\institute{
Laboratoire Lagrange, Universit\'e C\^ote d'Azur, Observatoire de la C\^ote d'Azur, CNRS, Boulevard de l'Observatoire, CS 34229, 06304 Nice Cedex 4, France\label{inst_O} \and Leiden Observatory, Leiden University, Niels Bohrweg 2, NL-2333 CA Leiden, the Netherlands\label{inst_L} \and
Max-Planck-Institut f\"ur Radioastronomie, Auf dem H\"ugel 69, D-53121 Bonn, Germany\label{inst_B} \and
NOVA Optical IR Instrumentation Group at ASTRON (Netherlands)\label{inst_A} \and
Max Planck Institute for Astronomy, K\"onigstuhl 17, D-69117 Heidelberg, Germany\label{inst_H} \and  European Southern Observatory Headquarters, Karl-Schwarzschild-Stra\ss e 2, 85748 Garching bei M\"unchen, Germany \label{inst_Garch}
\and
European Southern Observatory, Alonso de Cordova 3107, Vitacura, Santiago, Chile\label{inst_E}
\and
Konkoly Observatory, Research Centre for Astronomy and Earth Sciences, E\"otv\"os Lor\'and Research Network (ELKH), Konkoly-Thege Mikl\'os \'ut 15-17, H-1121 Budapest, Hungary,\label{inst_K} 
\and
Univ. Grenoble Alpes, CNRS, IPAG, 38000, Grenoble, France\label{inst_I} 
\and
NASA Goddard Space Flight Center
Astrophysics Science Division, Code 660
Greenbelt, MD 20771 \label{inst_Nas} 
\and
Anton Pannekoek Institute for Astronomy, University of Amsterdam, Science Park 904, 1090 GE Amsterdam, The Netherlands\label{inst_P} \and
MPE Max Planck Institute for Extraterrestrial Physics, Giessenbachstrasse, 85741 Garching, Germany\label{inst_MPE} \and
Departamento de Astronom\'ia, Universidad de Concepci\'on, Casilla 160-C, Concepci\'on, Chile\label{inst_Co} \and
Nicolaus Copernicus Astronomical Centre, Polish Academy of Sciences, Bartycka 18, 00-716 Warszawa, Poland\label{inst_Wa} \and
Unidad Mixta Internacional Franco-Chilena de Astronom\'ia (CNRS UMI 3386), Departamento de Astronom\'ia, Universidad de Chile, Camino El Observatorio 1515, Las Condes, Santiago, Chile\label{inst_Sa}
\and
Department of Astrophysics, University of Vienna, Türkenschanzstrasse 17, A-1180 Vienna, Austria\label{inst_V}
\and
I. Physikalisches Institut, Universit\"at zu K\"oln, Z\"ulpicher Str. 77, 50937, K\"oln, Germany\label{inst_C} \and
LESIA - Laboratoire d'études spatiales et d'instrumentation en astrophysique, Observatoire de Paris, 5 place Jules Janssen, 92190 Meudon,France\label{inst_LESIA}\and
GEOAZUR, Campus CNRS, 1 Sophia Antipolis, 06560 Valbonne\label{inst_GEOAZUR} \and
Zselic Park of Stars, 064/2 hrsz., 7477 Zselickisfalud, Hungary\label{inst_Zs} \and
AIM, CEA, CNRS, Universit\'e Paris-Saclay, Universit\'e Paris Diderot, Sorbonne Paris Cit\'e, F-91191 Gif-sur-Yvette, France\label{inst_Pa} 
\and 
National Astronomical Research Institute of Thailand 191 Siriphanich Bldg., Huay Kaew Rd., Suthep, Muang Chiang Mai 50200 Thailand\label{inst_Th} 
\and
Sydney Institute for Astronomy, School of Physics, A28, The University of Sydney, NSW 2006, Australia\label{inst_Sy}
\and
Institute for Mathematics, Astrophysics and Particle Physics, Radboud University, P.O. Box 9010, MC 62 NL-6500 GL Nijmegen, the Netherlands\label{inst_Ra} \and
SRON Netherlands Institute for Space Research, Sorbonnelaan 2, NL-3584 CA Utrecht, the Netherlands\label{inst_U} \and
Institute of Theoretical Physics and Astrophysics
University of Kiel, 24118 Kiel, Germany
\label{inst_Ki}
}

\abstract
{Optical interferometry is at a key development stage. The Very Large Telescope Interferometer (VLTI) has established a stable, robust infrastructure for long-baseline interferometry that is usable by general astronomical observers. The present second-generation instruments offer a wide wavelength coverage and improved performance. Their sensitivity and measurement accuracy lead to data and images of high reliability.}
{We have developed the Multi AperTure mid-Infrared SpectroScopic Experiment (MATISSE) to access, for the first time, high resolution imaging in a wide spectral domain. Many front-line topics are explored with this new equipment, including: stellar activity and mass loss; planet formation and evolution in the gas and dust disks around young stars; and environment interaction and accretion processes around super massive black holes in active galactic nuclei (AGN).}
{The instrument is a spectro-interferometric imager in the transmission windows called L, M, and N, from 2.8 to 13.0 microns, combining four optical beams from the VLTI's unit or auxiliary telescopes. Its concept, related observing procedure, data reduction, and calibration approach, is the product of 30 years of instrumental research and has benefitted from the expertise developed in the frame of the VLTI’s first generation instruments. The instrument utilises a multi-axial beam combination that delivers spectrally dispersed fringes. The signal 
provides the following quantities at several spectral resolutions: photometric flux, coherent fluxes, visibilities, closure phases, wavelength differential visibilities and phases, and aperture-synthesis imaging.}
 {This article provides an overview of the physical principle of the instrument and its functionalities. The motivation of the choice of the instrumental concept and the characteristics of the delivered signal are detailed with a description of the observing modes and of their performance limit. 
 {MATISSE offers four spectral resolutions in L\&M bands, namely 30, 500, 1000 and 3400, and 30 and 220 in the N band, and it provides an angular resolution down to 3 $mas$ for the shortest wavelengths. The MATISSE stand-alone sensitivity
limits are 60 $mJy$ in L and 300 $mJy$ in N. The paper gives details of the sensitivity limits for the different measurables and their related precision criteria, considering telescope configurations and spectral resolutions. 
 We also discuss the gain provided with the GRA4MAT fringe tracker.}
 An ensemble of data and reconstructed images illustrate the first acquired key observations.}
{The instrument has been in operation at Cerro Paranal, ESO, Chile, since 2018, and has been open for science use by the international community since April 2019. The first scientific results are being published now.} 
   \maketitle

 \section{Introduction}
The mid-infrared optical interferometry methods, coupling an array of telescopes in a wavelength domain sensitive to the environmental background emission (i.e. from 3 to 13 $\mu m$), were developed 30 years ago. It can still be seen as a young observing technique which benefitted greatly from significant research and technological progresses, and yet it has left room for improvement for observing procedures and data analysis. Whereas in the past, observations were limited to bright evolved stars \citep{Danchi1994}, the performance limits achieved over the last two decades have permitted observations of faint sources including extra-galactic targets \citep{jaffe2004, Swain2003}.
\\
Pioneering work was done in the field of heterodyne detection by {\citet{Johnson1974} and} \citet{Sutton1978} in the US, and by \citet{Assus1979} in Europe. The earliest interferometer successfully used for stellar interferometry was the ISI interferometer \citep{Danchi1988, Hale2000}, installed at the Mount Wilson Observatory and operating in a heterodyne detection scheme in the N band atmospheric transmission window. {The first closure phase measurements in the mid-infrared with a set of telescopes were obtained with this interferometer \citep{weiner2006}}. In parallel, successful attempts for homodyne or a direct detection scheme were also demonstrated in Europe by \citet{Mekarnia1990} on the SOIRDETE interferometer.
In the L band mid-infrared atmospheric window, it was a few years later that a fibre-type instrument called TISIS on the IOTA interferometer \citep{Menesson1999} carried out interferometric observations of Mira stars and supergiants \citep{Perrin2004}.
\\

The current modern era for mid-infrared interferometry, with international access to large observatory facilities, started in the 2000s. Between 2002 and 2012, two new mid-infrared interferometers entered into operation: the MID-Infrared  instrument at the VLTI \citep[MIDI,][]{Leinert2004} and the Keck Interferometer \citep[KI,][]{Ragland+2009}, at the Keck observatory, initiating observations of young stellar objects (YSOs) in the L band of the mid-infrared simultaneously to the K near-infrared band. The Keck Interferometer Nuller \citep[KIN,][]{colavita2010,KIN2012} entered into operation soon after. At that time, 
the  interferometric instruments were combining the light coming from two telescopes (1.8 m or 8 m diameter telescopes at the VLTI and 10 m telescopes at the Keck observatory, equipped with adaptive optics systems). MIDI 
became the most scientifically productive mid-infrared interferometric instrument, with 162 peer-reviewed publications as of January 2022.

 The Multi AperTure mid-Infrared SpectroScopic Experiment (MATISSE) is the mid-infrared spectrograph and imager of the VLTI. This second generation interferometric instrument, built on the expertise acquired on MIDI \citep{Leinert2003} and AMBER \citep{AMBER2007}, will significantly contribute to several fundamental research topics in astrophysics. It focusses, for instance, on the inner regions of disks around young stars where planets form and evolve, the surface structure and mass loss of stars at different evolutionary stages and frequently in binary interactions, and the environment of black holes in active galactic nuclei (AGN).
The inner region of protoplanetary disks fulfils the key conditions required for the formation of rocky planets such as the Earth. It contains the dust material that represents the elementary bricks to initiate and grow the planet core embryos.
MATISSE is designed to observe the small spatial scales in YSO environments 
and in all types of circumstellar environments in general, such as the ones that formed during the post-main sequence stellar stages where the mass loss processes in action are at the origin of the recycling of the material in our Galaxy. Stellar winds are taking place at the stellar photosphere level, and the tight interaction in binary systems frequently enhances the mass loss rate at
the periastron and shapes the resulting nebula.
MATISSE will also be of importance to probe the inner dust structure in  AGN. The first generation VLTI instrument MIDI 
has substantially challenged the simple 'dust torus' of the AGN unified model. That model is supposedly edge-on and hiding the central source in type 2 AGN and it is face-on and dominated by the inner dust sublimation radius in type 1 AGN.
Clumpy structures with important polar extensions are expected to trace the dusty wind driven by radiation pressure in the brightest AGN. 
The 2-telescope MIDI instrument was unable to make images and could only give angular sizes as a function of the position angle. It was limited to the N band where it is difficult to discriminate the effects of the temperature and composition of the emitting dust from the foreground absorbing medium. MATISSE solves these two problems by making images in the N band, in addition to in L and M bands, which are much more specific to dust emission. The inner dust structure in AGN is the tracer of the exchanges between the central engine, its surrounding broad line region (BLR), and the host galaxy, and it can reveal the mechanisms of co-evolution of the super massive black holes and galaxies.
Understanding the geometry of the nuclei well can allow independent distance estimates in combination with infrared (IR) reverberation mapping \citep['dust parallax',][]{Honig2014}.
The recent breakthrough made by GRAVITY on the spectro-astrometry of BLRs \citep{GravityColl2018} makes this MATISSE information of an even higher value.\\

For all these research topics, MATISSE characteristics offer unique worldwide capabilities. The first is opening the L and M bands (2.8–4.2 and 4.5–5.0 $\mu$m bands, respectively), which together with the N band offer access to very specific gas and dust material spectral signatures for long-baseline IR interferometry. The angular resolution in the L band is about 3 milliarcseconds (mas), and various spectral resolutions between R $\sim$ 30 and R $\sim$ 3500 are available. The second unique capability is the mid-infrared imaging — closure-phase aperture-synthesis imaging — performed with the four unit telescopes (UTs) or auxiliary telescopes (ATs) of the VLTI array.

With MATISSE, ESO and our consortium\footnote{MATISSE was designed, funded, and built in close collaboration with ESO by a consortium composed of institutes in France (J.-L. Lagrange Laboratory — INSU-CNRS — Côte d’Azur Observatory — the University of Nice Sophia-Antipolis), Germany (MPIA, MPIfR, and the University of Kiel), the Netherlands (NOVA and the University of Leiden), and Austria (the University of Vienna). The Konkoly Observatory and the University of Cologne have also provided support in manufacturing the instrument.} are contributing to the new generation of mid-infrared instrumentation becoming available to the astronomical community. The instrument has been available in open time since April 2019. Its integration at Paranal took place from November 2017 to February 2018. A two-year commissioning period followed and allowed for the instrument performance to be assessed in all modes with ATs and UTs, and within an upgraded VLTI infrastructure \citep{Woillez+2015} involving the implementation of the ATs' adaptive optics, NAOMI \citep{Naomi2019}, and of the fringe tracker mode, GRA4MAT (Woillez et al. in preparation). 

\begin{figure*}[htbp]
   \resizebox{\hsize}{!}
        {\includegraphics{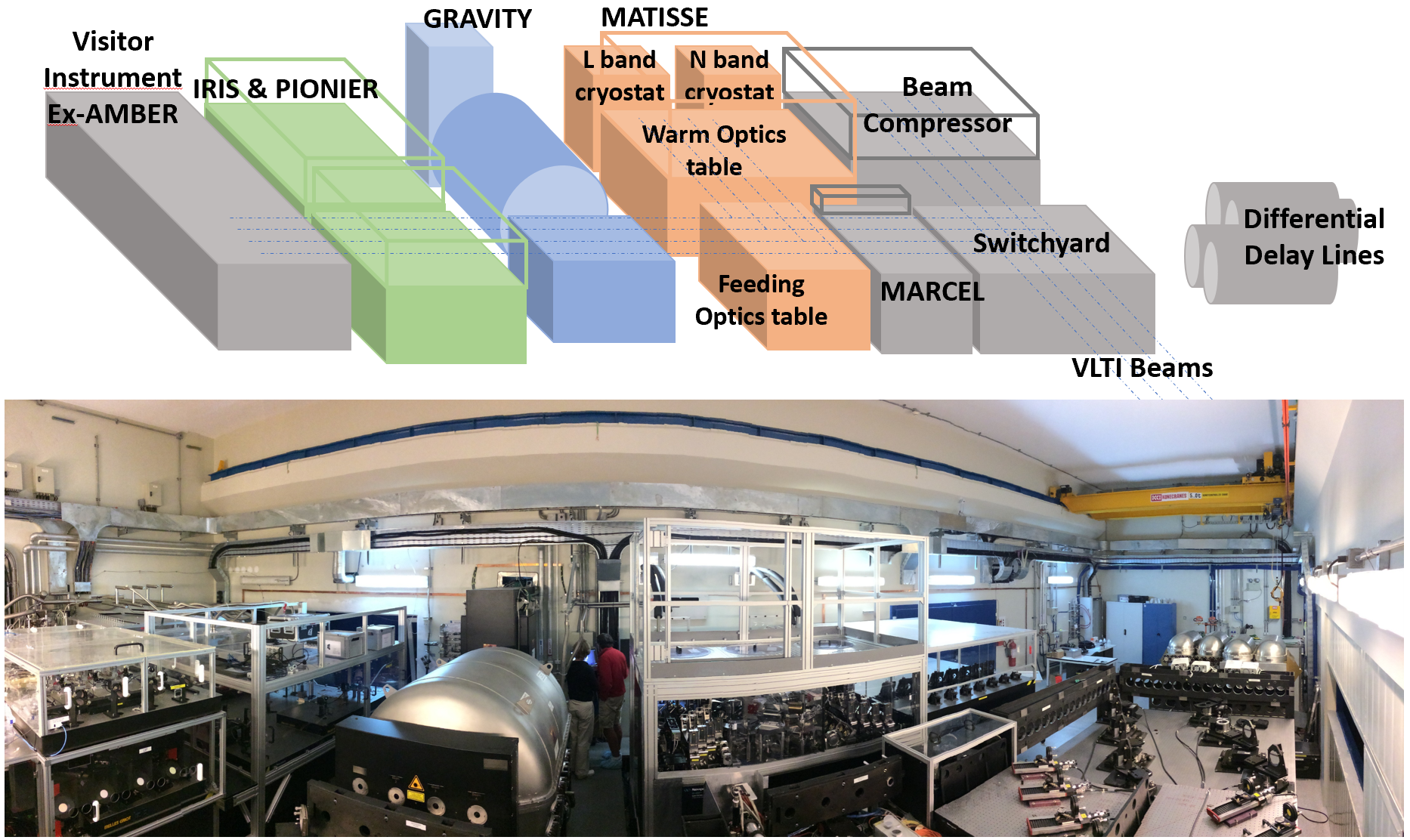}}
           
      \caption{MATISSE in the VLTI focal laboratory. Top: layout of the instruments and devices. In front of MATISSE, the feeding optics table reflects either
four UT or AT beams towards the instrument, after their transportation through the VLTI optical train. Bottom: view of MATISSE warm optics table called the WOP (and its cover) in the foreground.}
         \label{matisseinlab}
   \end{figure*}

\begin{figure}[htbp]
\centering
\includegraphics[width=0.485\textwidth]{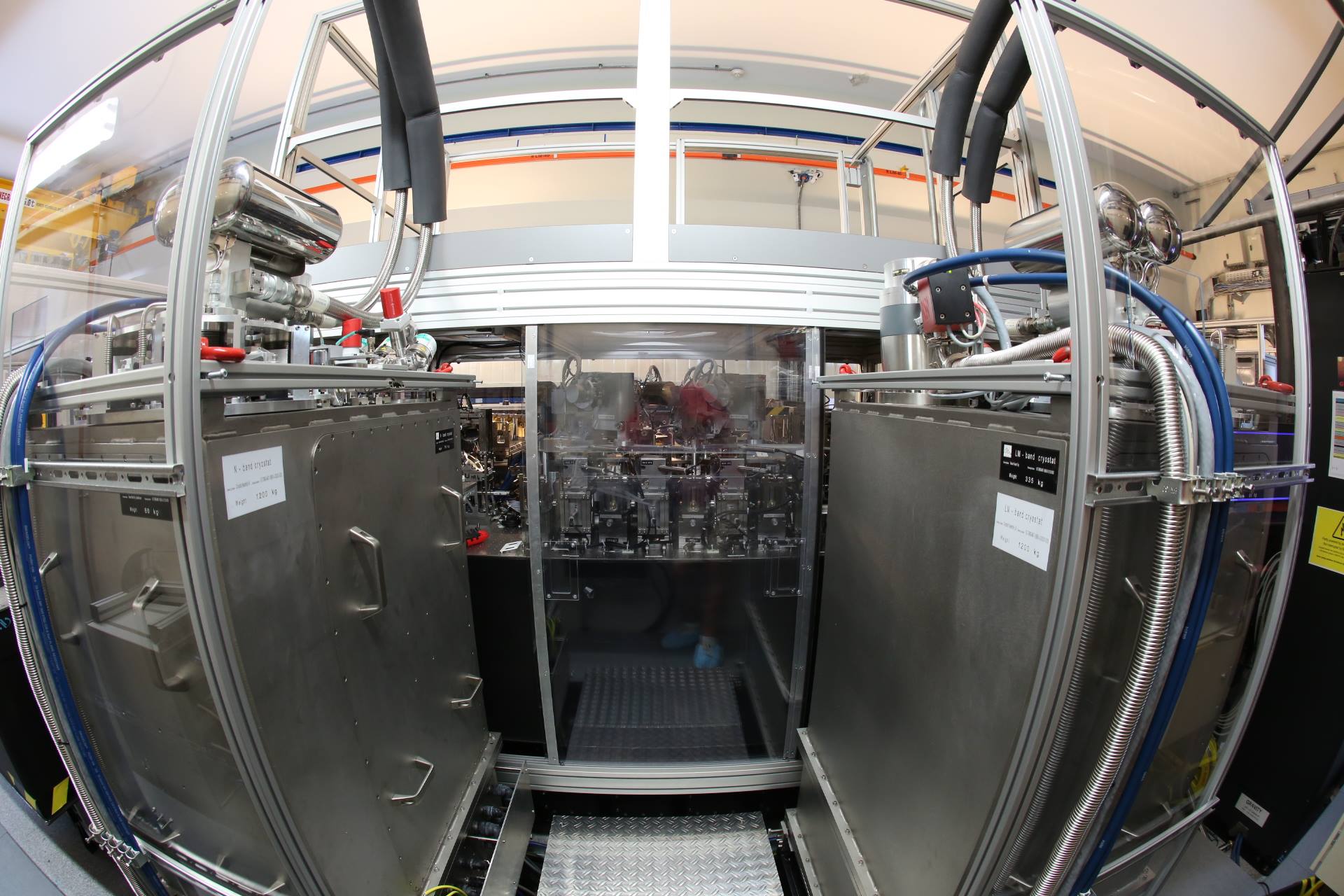}
    \caption{Fish-eye view of the backside of MATISSE in the VLTI laboratory. The two stainless-steel grey structures are the two cryostats, the left one housing the N band cold optical bench (COB) and detector, and the right one housing the L and M band COB and detector.}
        \label{matisseinlab2}
\end{figure}

The MATISSE instrument consists of a warm optical system (WOP, see Fig. \ref{matisseinlab}) and two cold optical benches (COBs) housed in two separate cryostats, one for the L-M band, and one for the N band (Fig.~\ref{matisseinlab2}). The WOP operates at an ambient temperature. It is fed by four beams from the VLTI, which are propagated via dichroic beam splitters to the two COBs.

MATISSE is an essential piece completing a worldwide ensemble of astronomical instruments. From spectral considerations and regarding the near-infrared domain (NIR), MATISSE extends the operating wavelength range of existing instruments such as PIONIER \citep{PIONIER2011} and GRAVITY \citep{GRAVITY2017}, which are both also sensitive to IR radiation but at a shorter wavelength.  On the other side of the wavelength domain, MATISSE complements the (sub)millimetre domain in which high angular resolution observations are made with the {radio interferometer called ALMA} \citep{ALMA2020}. {MATISSE gives access to long baseline lengths compared to the mid-infrared interferometric instrument of the Large Binocular Telescope \citep[LBTI,][]{Hinz2008} currently entering into operation \citep{Sallum2021}}.
MATISSE is thus a bridge in wavelengths of strong scientific importance as well as an exploratory path for high angular resolution in its spectral bands.
The scientific community is currently awaiting {the JWST observations} \citep{JWST2006} and the installation of METIS, the ELT mid-infrared instrument \citep{METIS2018}, both of which will operate in the same wavelength domain as MATISSE. In terms of sensitivity, these instruments will have advantages over MATISSE. JWST will benefit from the absence of the atmosphere and low background observing conditions due to its location in space, while METIS will benefit from the large collecting area of the ELT. However, from an imaging point of view, MATISSE has a resolving power 20$\times$ and 4-5$\times$ greater than JWST and METIS, respectively. 
Indeed, MATISSE can access the decisive mas scale, which:
\begin{itemize}
    \item translates to the au scale (and sub-au) at the typical distance of star-forming regions ($\sim 150$ pc). In particular, accessing the au scale is required to reach the water ice line in disks, which is thought to represent the outer boundary of the telluric planet forming region.
    \item 
    complements the resolving power of single-aperture instruments by probing the dust sublimation radii area.
    \item images with the best achievable resolution, together the stellar photospheres and their features, simultaneously with the dust formation zones often involved in the mass-loss phenomenon. 
\end{itemize}

MATISSE will allow one to trace not only different spatial regions of the astrophysical objects, but also their specific spectral signatures (Tab. \ref{tab:spectralsignature}), and it will thus provide insights into previously unexplored areas such as the investigation of the distribution of volatiles in addition to that of the dust featuring various solid components (silicates, carbon) and their mineralogy.

Section \ref{secPrinciple} is an overview of the main instrument characteristics. The observing procedure, the real-time coherencing method, and the data reduction software are described in Section \ref{secobsdrs}. Section \ref{secperf} summarises the main performance of MATISSE in terms of sensitivity limits in all observing modes. Some early observations done during the MATISSE first light and commissioning sessions are presented in Section \ref{secastro}. They illustrate the capabilities and performance of this instrument. Section \ref{secapersp} tackles the ongoing work, which is maximising the performance.

\section{MATISSE principle}
\label{secPrinciple}
\subsection{Instrument characteristics, spectral domains, and resolutions} 
\begin{figure}[htbp]
    \centering

    \includegraphics[width=0.45\textwidth]{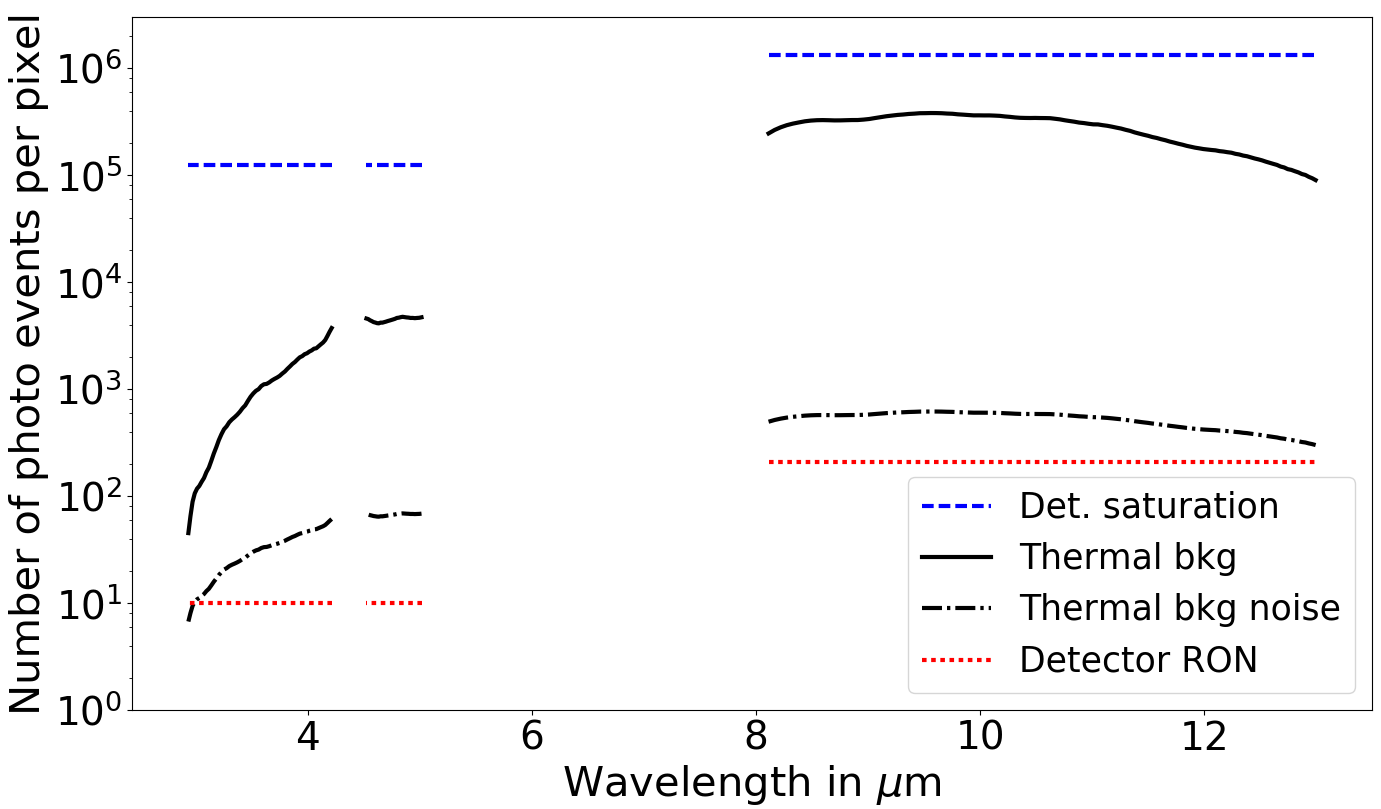}
    \caption{Comparison between the level of the thermal background (solid black line), the detector saturation limit (blue dashed line), the detector read out noise (red dotted line), and the thermal background noise (black dash-dotted line). 
    {These values are obtained in LOW spectral resolution. In the L and M bands, the DIT is 125ms and in the N band the DIT is 20ms.}
    In the L band, the DIT is limited by the coherence time. In the N band, the background level implies short DITs to avoid detector saturation.}
    \label{fig:skyandsat}
\end{figure}

MATISSE is a four-beam instrument operating in three atmospheric transmission windows, the L, M, and N bands of the mid-infrared. It produces dispersed fringes on two different detectors simultaneously, on the HAWAII-2RG from Teledyne Technologies for the L\&M bands and on AQUARIUS from the Raytheon Company for the N band. The instrument is optimised for the L and N bands. The L band is specified from 2.8 to 4.2~$\mu$m, and the N band from 8.0 to 13.0~$\mu$m. In addition, MATISSE also operates in the M band, from 4.5 to 5.0~$\mu$m. The L, M, and N bands can be observed simultaneously.
\\
MATISSE is composed of two separate twin instruments because of the specificity of the materials used for the optical transmissive elements, the limited spectral bandwidth sensitivity of IR detectors, and the fact that the signal sampling in time and space has to be optimised on a large spectral domain with different noise regimes.  
In the L band, the detector integration time is driven by the seeing coherence time or by the fringe tracking time (if an external fringe tracker is used). In the N band, it is driven by the high thermal background level, which imposes the individual detector integration time to be short enough to avoid saturation, as shown in Fig.~\ref{fig:skyandsat}. Indeed the maximum exposure time in the N band is less than 30 ms while in the L band, several seconds of integration time are allowed. 
\\

In the L band, the seeing-dependent Strehl ratio dominates the visibility calibration errors as in AMBER, while in the N band, as in MIDI, MATISSE is sensitive to the temporal fluctuations of the thermal background.
 An important aspect of the concept and of the observing methods relies on the implementations of solutions to minimise these background effects. 
\\

\renewcommand{\tabcolsep}{1pt}
\begin{table}[htbp]
\caption{Selected spectral signatures accessible with MATISSE}
\centering
\label{tab:spectralsignature}
\begin{tabular}{cc}
\hline\hline
Components & Wavelengths ($\mu$m)  \\
\hline\hline

H$_2$O (ice)&3.14 \\
H$_2$O (gas)&2.8--4.0 \\
H recombination lines&4.05 (Br$\alpha$),4.65 (Pf$\beta$)\\
Polycyclic Aromatic Hydrocarbons &3.3--3.4\\
Nano-diamonds & 3.43--3.53 \\
CO fundamental transition series & 4.6--4.78\\
CO (ice) & 4.6--4.7\\
\hline
Amorphous silicates& 9.8\\
Crystalline silicates (olivines, pyroxenes)& 9.7,10.6,11.3,11.6\\
PAHs&8.6,11.4,12.2,12.8\\
Fine structure lines (e.g. [NeII])&10.5,10.9,12.8\\
\hline\hline
\end{tabular}
\end{table}    
\renewcommand{\tabcolsep}{4pt}



Moreover, the offered spectral resolutions, aiming to access the various spectral signatures listed in Tab. \ref{tab:spectralsignature}, differ between the L and N bands. The resolutions offered are presented in Tab. \ref{tab:spectralresolutions} .

One limitation of the HAWAII-2RG detector (2Kx2K array of 18 $\mu$m sized pixels), used for L and M band observations, is its slow pixel readout clock (100 kHz) at low read out noise (RON < $15e^-$). A fast read mode, with a RON of about $75e^-$ per pixel and per read, exists. However it is used only to avoid detector saturation in Low spectral Resolution (LR) for targets brighter than 300 Jy in the L band with ATs. The slow read mode implies a full reading of the detector in about 1.3 seconds for the correlated double sampling read-out mode. This duration exceeds the coherence time of the atmosphere, which is of the order of 120 ms in the L band on good Paranal nights. Without a fringe tracker, this readout time only allows one to read the full pixel range from 2.8 to 4.2 $\mu$m covering the L and M spectral bands in LR. For higher spectral resolutions, which generate a spectrum illuminating the entire detector width, only a limited spectral bandwidth of 0.2 $\mu$m at medium resolution and 0.1~$\mu$m at high resolution can be read by the detector during the atmosphere coherence time. This limitation in the L and M band can be overcome only when operating with a fringe tracker.
In the N band, the pixel readout clock of the AQUARIUS detector (1Kx1K array of 30 $\mu$m sized pixels) at 2.4MHz is sufficient to read all the detector pixels in 20 ms and hence all the spectral channels before reaching the saturation limit due to the background.

\begin{table}
\caption{Spectral resolutions.} 
\centering
\label{tab:spectralresolutions}
\begin{tabular}{ccc}
\hline\hline
Spectral Mode & Spectral bandwidth ($\mu$m) &  Resolution \\
\hline\hline
LOW-N  & 8.0-13.0 & 31.5 \\
HIGH-N & 8.0-13.0 & 218  \\
\hline  
LOW-LM & 2.8-4.2; 4.5-5.0 & 31.5 \\
MED-LM & 2.8-4.2; 4.5-5.0 & 499 \\
HIGH-L & 2.8-4.2 & 979 \\
\hline
VHIGH-L & 3.95-4.2 & 3370 \\
VHIGH-M & 4.5-5.0 & 3370 \\
\hline\hline
\end{tabular}
\tablefoot{
Calculated with a pinhole of 1.5$\lambda$/D in the L and M band, and 2$\lambda$/D in the N band.}
\end{table}

\subsection{Beam combination scheme}
The choice of the combination scheme of MATISSE was based on a signal-to-noise ratio (S/N) analysis aiming at comparing two basic principles that 
are called the global combination and the pair-wise combination, and they are also based on considerations related to the design architecture. The related equations for the S/N analysis were  provided in \cite{lagarde2008}.

In the global combination, all the telescope beams are mixed together to produce a common interference pattern. For this reason, it is also called a multi-axial all-in-one combination, while the pair-wise combination mixes the beams pair by pair. The MIDI instrument is an example of both a pair-wise combination and an all-in-one combination since it has only two input beams. Starting from the MIDI experience, a thorough  analysis was {comparing} the advantages and drawbacks of the two combinations in the case of a four-beam instrument.

One of the advantages of the pair-wise combination would have been the possibility to extract the photometric information without using specific photometric channels \textbf{\citep[e.g.][]{Lacour2008}}. Indeed, if three or more input telescope beams ($n_{T}\geq 3$) are used, $n_T(n_T-1)/2$ pairs or equations are produced, then the individual unknown photometries can be deduced from a linear combination of the pair-wise interferometric channels.

For the multi-axial global scheme,  specific photometric channels are needed to measure the photometry, simultaneously aimed at viewing the beam input quality and normalising the visibilities from the coherent flux quantities.
To obtain the photometry of the source or to optimise the measurement of the coherent flux, two modes are considered for 
the global combination: the SiPhot and the HighSens modes.
In MATISSE, the flux ratio $\alpha$~=~2/3 corresponds to the SiPhot mode: two-thirds of the flux is sent into the interferometric channel, and one-third is sent into the individual photometric channels. With $\alpha$ = 1, the HighSens mode, all the flux is sent to the interferometric channel. 

In the mid-infrared domain, the error due to the thermal background ($\sqrt{n_{B}}$ where $n_{B}$ is the thermal background per telescope beam) dominates the other sources of errors in
most cases, such as the target photon noise ($\sqrt{n_{*}}$ where $n_{*}$ is the number of photons of the astronomical target per beam) or the detector RON. 
In the case of global combination, the theoretical coherent flux error in this background-limited regime is given
by~ $\sigma^2_{Cij}$~=~2$N_{T}n_{B}/\alpha$. The error on the photometry, which is obtained in SiPhot, is given by~ $\sigma^2_{ni}$~=~$N_{T}n_{B}/(1-\alpha)$, with $N_T$ being the number of telescopes and (1~-~$\alpha$) being the corresponding fraction of light in the photometric channels.
On the other hand, in the case of the pair-wise combination, the error on the coherent fluxes is~ $\sigma^2_{Cij}$~=~[$N_{T}$($N_T$-1)/$\alpha]n_B$, and the errors on the photometries are~ $\sigma^2_{ni}$~=~10$n_B$.\\

\begin{table}[htbp]
\caption{Quantification of the signal-to-noise ratio (S/N) in the case of observations with four telescopes in the thermal background regime.
}
\centering
\label{tab:concept_choice}
\begin{tabular}{cccc}
\hline\hline
    &           & \multicolumn{2}{c}{All-in-one}\\
S/N & Pair-wise & HighSens & SiPhot\\
\hline\hline
Photometry & $n_{*}/\sqrt{10 n_{B}}$&   -   & $n_{*}/\sqrt{12 n_{B}}$\\
\hline
Coherent Flux  & $n_{*}V/\sqrt{12 n_{B}}$ & $n_{*}V/\sqrt{8 n_{B}}$ & $n_{*}V/\sqrt{12 n_{B}}$\\
\hline\hline
\end{tabular}
\tablefoot{The global recombination scheme, in HighSens and SiPhot modes, is compared to the pair-wise recombination scheme for a chopping rate of 50\%. The $n_{*}$ and $n_{B}$ are the average numbers of photons per telescope from the source and the thermal background, respectively, and $V$ represents the visibility.
}
\end{table}

According to the content of Table \ref{tab:concept_choice}, in the background-limited noise regime, the error level on the coherent flux slightly favours the choice of an all-in-one scheme compared to a pair-wise scheme. 
In order to benefit from both advantages of the multi-axial global scheme for the coherent flux measurements and for the visibility measurements, MATISSE includes a slider allowing the instrument to switch from $\alpha$ = 1 to $\alpha$ = 2/3 with the use of mirrors and beam splitters, respectively.
Moreover, in addition to this advantage of maximised S/N for the coherent flux, the multi-axial global all-in-one scheme is more robust with regard to the construction of closure-phase measurements.  Triplets of beams, which are required for the closure-phase measurements, are travelling through the same optical combiner in the multi-axial all-in-one scheme. In the pair-wise separation it is not the case and for this reason, the quality of the closure phase measurement in the pair-wise separation depends on the instrument stability.
Finally, the multi-axial all-in-one scheme allowed for a more compact and simple design with less optical elements than in the case of a pair-wise scheme.

\subsection{Signal encoding}
\begin{figure}[htbp]
    \centering

    \includegraphics[width=0.45\textwidth]{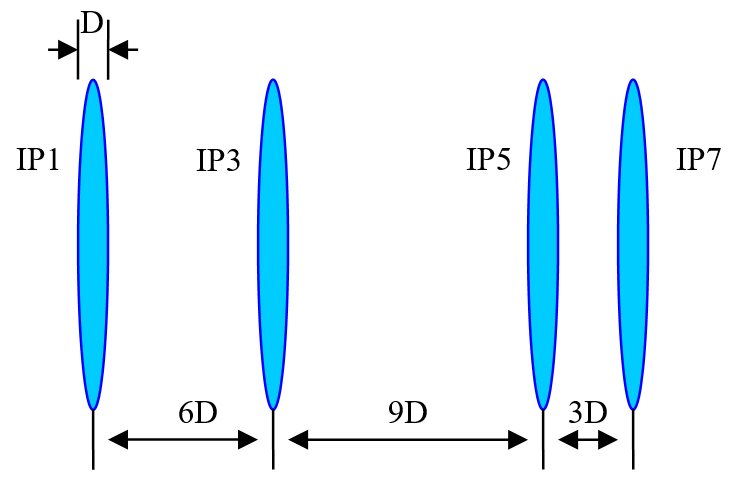}
    \caption{Pupil configuration arrangement after anamorphosis and before the camera optics.}
    \label{fig:beam_config}
\end{figure}

\begin{figure}[htbp]
    \centering
    \includegraphics[width=0.45\textwidth]{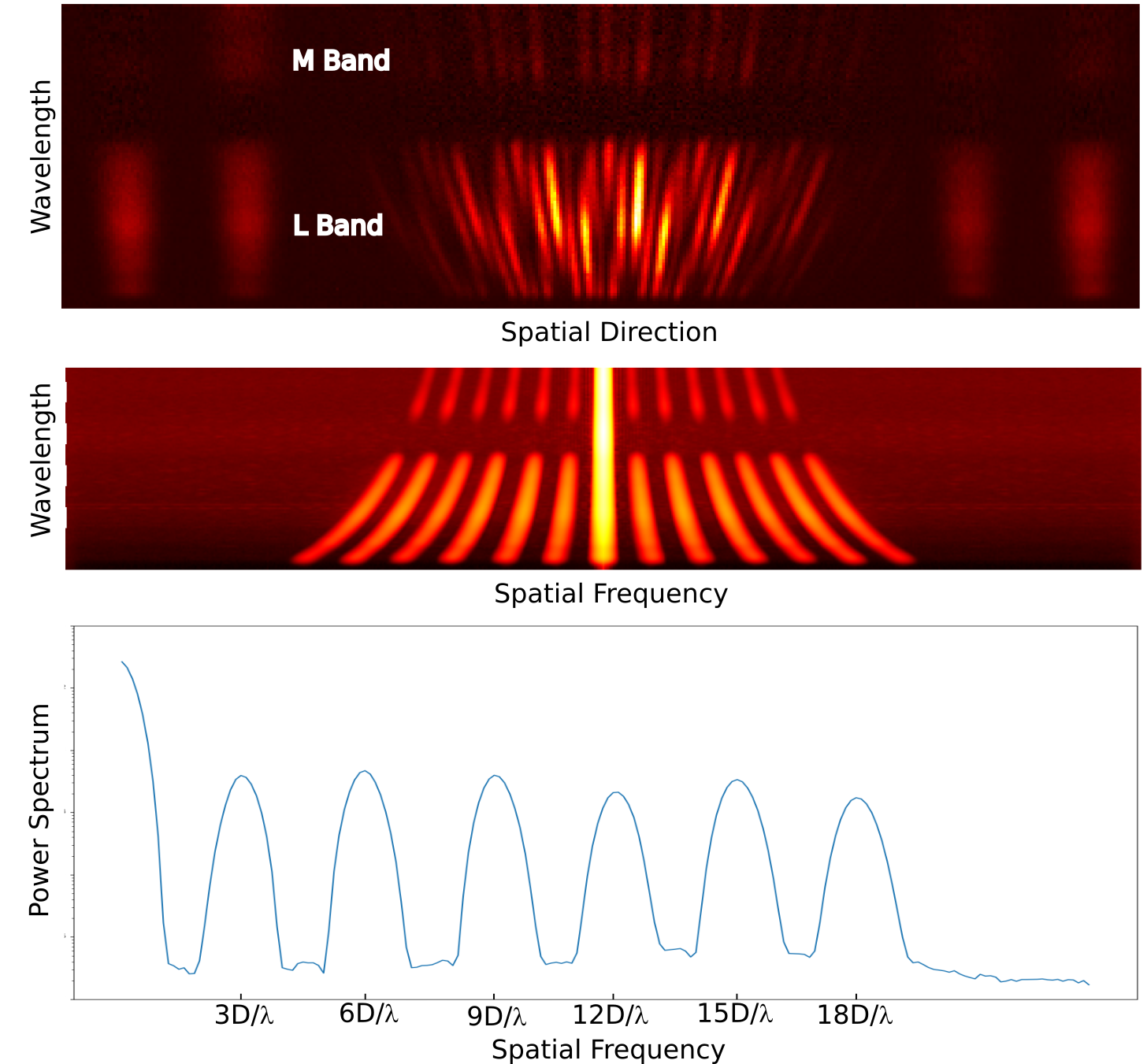}
    \caption{Signal encoding. Top: Detector image showing the interferogram and the four photometric signals on one of the two MATISSE detectors, i.e. the HAWAII-2RG detector in the L band and M band. The observation was made on the calibration star HD31529 with the four ATs in low-spectral resolution. Middle: Power spectrum of each spectral channel. Bottom: A cut through the power spectrum at $3.57~\mu m$ (in logscale).}
    \label{fig:signal_on_detector}
\end{figure}

 The beam combination is performed by a camera lens placed in front of each detector. A non-redundant beam configuration arrangement with separations of B$_{ij}$ between beams i and j, respectively, equals 3D, 9D, and 6D (where D is the beam diameter, see Fig. \ref{fig:beam_config}), and this allows the fringe peaks to be separated in Fourier space.
 The Fourier transform of the interferogram produces six fringe peaks centred at different frequencies of B$_{ij}$/$\lambda$ (3D/$\lambda$ 6D/$\lambda$, 9D/$\lambda$, 12D/$\lambda$, 15D/$\lambda$, and 18D/$\lambda$), as well as a low-frequency peak that contains the object photometry and the thermal background emission coming from the four telescopes (see Fig. \ref{fig:signal_on_detector} ). 

The analytical expressions of the 1D Fourier transform of the interferometric signal $I(u)$ and the photometric signals $P_{i}(u)$, per spectral channel, are given by
\begin{equation}
\label{eq:eqgen}
I(u)=M_B(u)\sum_{i=1}^4n^I_{Bi}+M(u)\sum_{i=1}^4n^I_{*i}+\sum_{i=1}^4\sum_{j=2,j>i}^4M(u-u_{ij})\sqrt{n^I_{*i}n^I_{*j}}V_{ij}~,
\end{equation}
\begin{equation}
P_i(u)=M_B(u)n^P_{Bi}+M(u)n_{*i}^P~,
\end{equation}
where $i$ and $j$ are the {telescope} beam indices and {$ij$ relates to the input physical baseline between the considered pair of telescopes. We note that} $n^I_{Bi}$ is the number of photons produced by the thermal background for beam i in the interferometric channel; $n^P_{Bi}$ is the number of photons produced by the thermal background in the photometric channel $i$; $n^I_{*i}$ and $n^I_{*j}$ are the numbers of photons produced by the observed object for each beam in the interferometric channel; $n^P_{*i}$ is the number of photons produced by the observed object in the photometric channel $i$; and $V_{ij}$ is the complex visibility. Furthermore,
$M_B(u)$ is the low-frequency peak associated with the thermal background emission; $M(u)$ is the low frequency peak of the interferometer transfer function; and $M(u-u_{ij})$ is the fringe peak of the interferometer transfer function at the spatial frequency $u_{ij}$ ($u_{ij}$ = $B_{ij}$/$\lambda$), {where $B_{ij}$ are the output instrument baselines here}.
As the illumination of the background is close to being uniform on the detector, the shapes of $M$ and $M_B$ are different. 
At the zero frequency, $M(0)$=$M_B(0)$=$1$. In Eq. \ref{eq:eqgen} the third term represents the coherent fluxes $C_{ij}$ of the six baselines. 

Regarding the signal encoding for the interferometric channel, the detector sampling in the spatial direction is the following: the point spread function (PSF) is sampled over 72 pixels per $\lambda/D$, and the narrowest and widest fringes are sampled with 4 and 24 pixels, respectively. That sampling applied to both the L and N bands at $\lambda$ equals 3.2 and 8.0~$\mu$m, respectively. 
In the spectral direction, the sampling is 3 pixels per $\lambda$/D. Since the PSF sampling requires 72 pixels in the spatial direction and 3 pixels in the spectral direction, an anamorphosis factor of 24 (illustrated in Fig. \ref{fig:beam_config}) is applied in the interferometric channel.

For the photometric channels in the spatial direction, the magnification of the photometric PSF is lower than the interferometric one (12 pixels per $\lambda/D$) in order to obtain the same flux level per pixel in the five images (the four photometries and the interferometry). In the spectral direction, the sampling is the same as in the interferometric channels (3 pixels per $\lambda/D$). The anamorphosis factor for the photometric beam is therefore limited to a factor 4. In consequence, the spatial size of the interferometric channel is larger than the photometric one in order to optimise the sampling of the six fringe patterns contained in the interferogram. 

In the SiPhot mode used for the L band, the observations of the interferometric signal and of the four individual photometric signals are carried out simultaneously, making it possible to perform an accurate visibility estimation. During such an observation, five images are formed on the detector: the multi-axial all-in-one interferometric signal surrounded by the four individual photometric beams as shown in  Fig.~\ref{fig:signal_on_detector}. In the SiPhot mode, the interferometric beam and the photometric beams receive  two-thirds and one-third of the incoming flux, respectively. The interferogram is dispersed in the spectral direction as prescribed by a pioneering interferometric set-up \citep{Labeyrie1975}.
Spectro-spatial-instrumental studies by \cite{Beckers1982}, \cite{PetrovPhD}, \cite{mourard1989}, and \cite{LagardePhD} have shown the interest of spectrally dispersed fringes for differential interferometry, providing spectro-astrometry measurements of the source as formalised in \citet{AMBER2007}.  

\subsection{Signal optimisation in the mid-infrared environment}
\label{sect:SignOpt}
\subsection*{{\bf Spatial filtering}}
To measure the coherent fluxes with a good accuracy, the design utilises spatial filters, including image and pupil stops inside the cryostats. The telescopes deliver beams with partially corrected wavefront errors. The Strehl ratio in the N band can vary between 0.8 and 1 with the seeing. This corresponds to coherent flux variations of 4\% without spatial filtering. In the L band, the Strehl can fluctuate between 0.5 and 0.9, corresponding to a coherent flux error of 12\% without filtering. To reach a coherent flux accuracy of 1\%, a spatial filter with a pinhole of 2$\lambda$/D in diameter is required in the N band, and of 1.5$\lambda$/D in the L band. 

\begin{figure*}[htbp]
    \centering
    \includegraphics[width=0.9\textwidth]{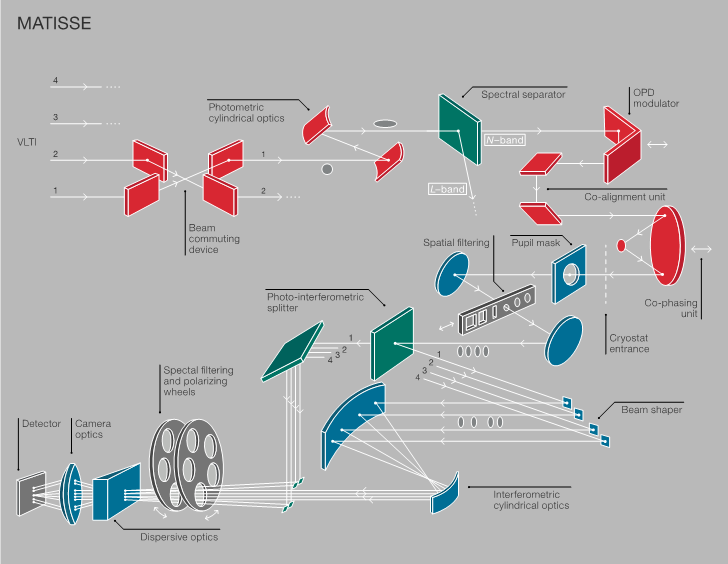}
    \caption{Schematic layout of the MATISSE instrument concept \citep[Credit:][]{2014Msngr}. The red parts represent optical elements located on the WOP at an ambient temperature. The blue parts represent optical elements of the cold optics bench located in the cryostats. Only one COB with its elements and detector is shown.}
    \label{fig:genlayout}
\end{figure*}

\subsection*{{\bf Background removal}}
\label{backgroundremoval}
The thermal background in the mid-infrared is variable and exceeds the target coherent flux by far
in most cases (see Fig. \ref{fig:skyandsat}). The thermal background contribution is present in the low frequency peak (see Eq. \ref{eq:eqgen}) and, due to the high level background, special care is taken to avoid cross-talk with the high frequency peaks containing the source coherent flux information.
Two methods are used
to reduce the cross-talk to a level lower than the background generated photon noise existing at all spatial frequencies in the Fourier plane: a) spatial modulation, as in the VLTI near-infrared spectrometer AMBER \citep{AMBER2007} and b) a combination with temporal modulation as in the VLTI mid-infrared spectrometer MIDI \citep{Leinert2003}. The spatial modulation is the essence  of the multi-axial concept. Without windowing, in Fourier space, the fringe peaks would be completely separated from each other and from the low-frequency peak. In practice, the theoretical spectral density of the interferogram is convolved by the Fourier transform of the finite window that delimits  the detector pixels to be read. Consequently, the low-frequency energy peak contaminates the fringe peaks. To decrease that contamination, that is to increase the rejection factor, a temporal modulation is performed. This modulation consists in modulating each fringe signal by applying an artificial optical path difference (OPD) with a periodic sequence during the coherence time. It is a generalisation of the natural 0-$\pi$ scheme of the MIDI co-axial combination. This function is provided by a set of piezo actuators. The modulation is made in ten steps of $\lambda$ /10 during the coherence time in which the background remains almost constant. The resulting ten signals are then numerically re-phased and summed. The lower frequency peak, which is not affected by the OPD modulation, and its contamination to the adjacent fringe peaks is then eliminated and only the high frequency peaks remain.
Moreover, to measure the visibility, we also need to extract the source photometry, which consists in separating the stellar flux from the sky background using chopping, that is a sequential observation at a frequency in the range 0.5-1 Hz between the target itself and the nearby region of the sky.

\subsection*{{\bf Phase optimisation}}
 MATISSE features a specific module called the beam commuting device (BCD), which has already been used in AMBER \citep{AMBER2007}, to calibrate closure and differential phases. Two BCDs are installed at the entrance of MATISSE for two pairs of beams. By commuting beams 1 and 2, and/or 3 and 4, we invert all the phase contributions before the instrument's BCD (i.e. including the contributions of the astrophysical source, the atmosphere, and the VLTI beam trains), while keeping fixed the phase contributions after the BCD device {(such as the optical distortions and the detector features)}. With a proper combination of phase signals obtained with the different BCD configurations, the instrument phase residuals located in the optical train between the BCD and the detector can be removed {for two differential phases and four closure phases}. 
 {For the closure-phases}, the BCDs allow a reduction of the phase residuals from several degrees to less than one degree {
(as discussed in Section \ref{sect:InstrCP})}.

\subsection{Instrument description}
\label{sect:InstrDecr}
Figure \ref{fig:genlayout} shows the general layout of the instrument with its main functions.
MATISSE is made of several parts: optical elements at ambient temperature on an optical table at the entrance of the instrument (WOP) and two separate cryostats, one for the L and M bands and another for the N band, including optics (COB) and detectors cooled at temperatures lower than 35 K. 

\begin{figure}[htbp]
    \centering

    \includegraphics[width=0.45\textwidth]{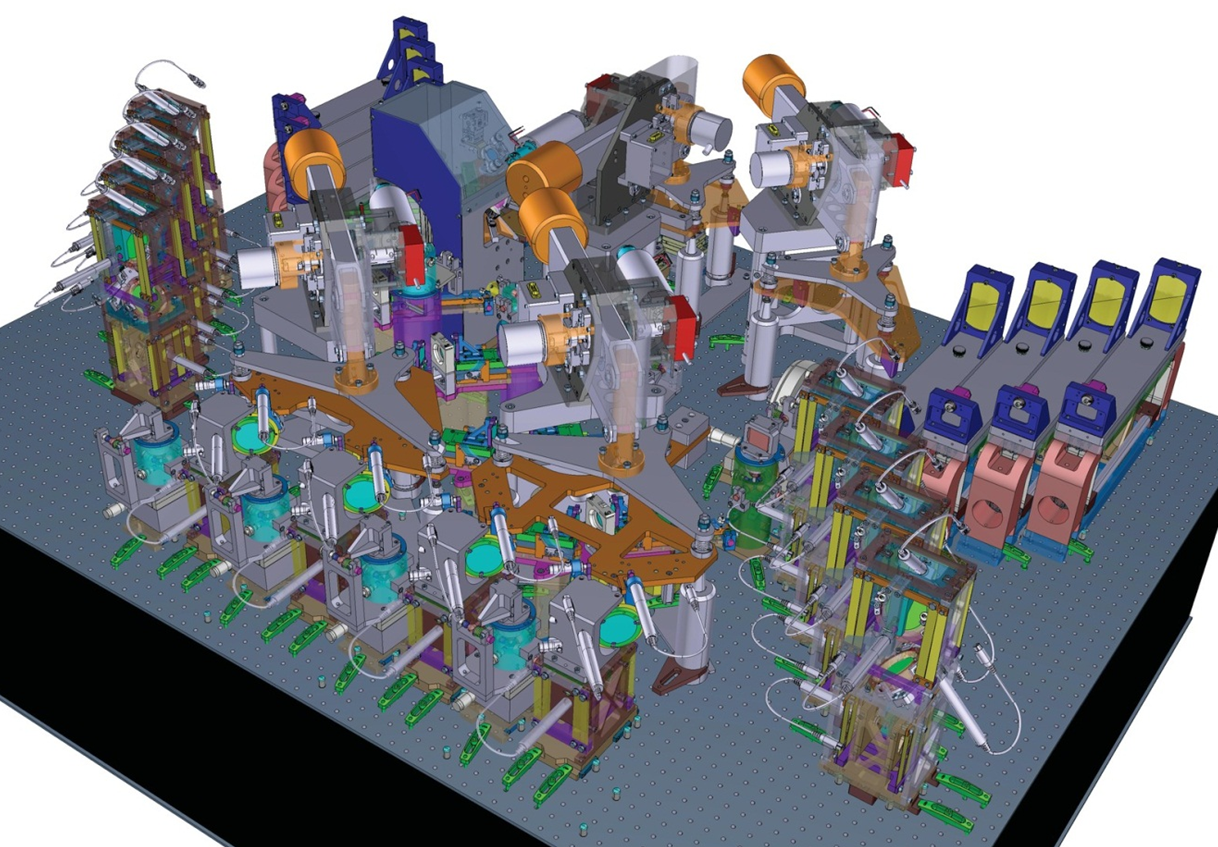}
    \caption{Overview of the warm optics modules on the optical table.}
    \label{fig:woptable}
\end{figure}

The WOP rests on a 2 m x 1.5 m optical table (see Fig. \ref{fig:woptable}). It receives four beams from the VLTI feeding optics coming from either UTs or ATs.  
These four beams enter first into the two independent BCDs 
(see Sect. \ref{sect:SignOpt}). One BCD is in `IN' mode when it is inserted in the optical path and it commutes its corresponding pair of beams. One BCD is in `OUT' mode when it is out of the optical path. The following four BCD configurations are thus possible: `OUT-OUT', `OUT-IN', `IN-OUT', and `IN-IN'. {The shift between different BCD positions is executed in a few seconds.}
Then the beams are individually anamorphosed with a ratio of 1:4 thanks to cylindrical optics. 
The beams are then spectrally separated with individual dichroics in order to form the L\&M band and the N band beams. 
Then the temporal modulation, which is different for each spectral band (see Sect. \ref{sect:SignOpt}), is made using piezo actuators.
Before entering into the cryostats, each beam passes through two modules: 
the first one includes the periscopes, which are used for the co-alignment (image and pupil) between the warm optics and the cold optics; and      the second one includes the delay lines, which deliver the pupil plane at the correct position into the cold optics and equalise the optical path differences between the beams and in particular the differential optical path between the L\&M band and the N band.   
In addition, the warm optics is made of internal optical sources (one visible for alignment and one IR for calibration purposes). These internal sources deliver four identical beams (identical to the VLTI ones), injected into the instrument using motorised mirrors. This sub-assembly is located on the warm optics table. 

\begin{figure}[htbp]
    \centering

    \includegraphics[width=0.45\textwidth]{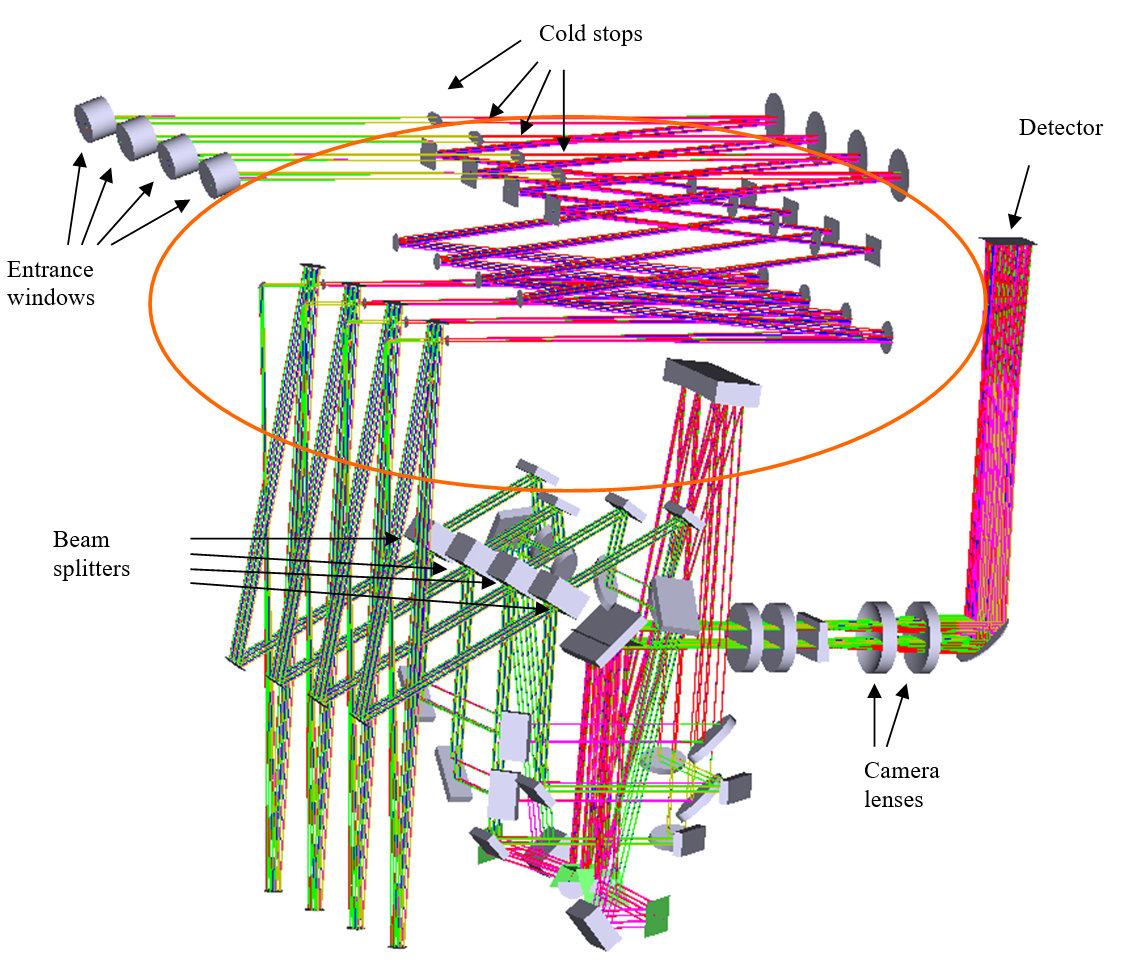}
    \caption{Layout of the interferometric light path inside a cryostat.}
    \label{fig:coboptics}
\end{figure}

The two cryostats L\&M and N bands are similar. Figure \ref{fig:coboptics} gives a view of the layout of the cold optics. 
The recombination of the four beams through the camera onto the detector is represented in this view. This recombination produces the so-called interferometric channel. 

Light enters the entrance windows of the cryostat from the upper left with an anamorphic factor of 4, passing the cold stops and the off-axis optics and spatial filtering module (slit or pinhole) until it reaches a motorised slider allowing the use of beamsplitters or mirrors. If beamsplitters are chosen, light is split into the interferometric channels and the photometric channels. The anamorphism of the interferometric channels is further increased with a factor 6, to a total of 24 by an additional anamorphic optics. Finally after passing the spectral filter and dispersion elements, light reaches the detector via the camera optics.  

\begin{figure}[htbp]
    \centering

    \includegraphics[width=0.3\textwidth]{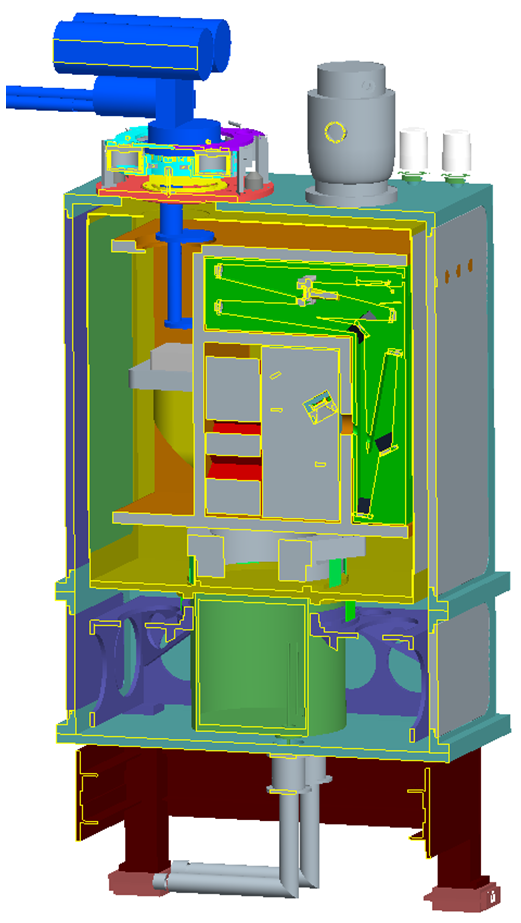}
    \caption{Cross section through one MATISSE cryostat.}
    \label{fig:cryostat}
\end{figure}

Figure \ref{fig:cryostat} shows the cryostat completely assembled with its components and the interior of the cryostat.
The operating temperature of the cold optics is 35 K. 
The operating temperature of the N band detector is 10 K. 
The operating temperature of the L\&M band detector is 40 K.
The vacuum ($<1.10^{-5}$ mbar) is realised by a turbomolecular pump fixed on the top of the cryostat. 
The operating temperatures are realised thanks to pulse tube coolers (with helium circulation) fixed on the top of the cryostat. The pulse tube coolers have two stages: 
the first stage cools the cold optics down and the second stage cools the detector down. 
At the bottom of the two cryostats, a liquid nitrogen vessel cools a radiation shield down at 80K surrounding the colder benches.

\section{Observing procedure and data reduction}
\label{secobsdrs}
\subsection{Observing modes and sequence}
MATISSE is designed to operate in both L\&M and N bands simultaneously. Whatever the band, it is possible to send all the photons into the interferometric channel (HighSens mode) or into the interferometric channel and the four photometric channels 
(SiPhot mode). The non-linear behaviour of the N band detector, when two separate zones or the edges of the registers are illuminated, makes a proper calibration of the Kappa matrix impossible (see Sect. \ref{secdrs}). That renders the SiPhot mode non-operational in the N band. Therefore, the N band photometry from each telescope is recorded in the interferometric channel, after the fringes are recorded, with one shutter open at a time. Thanks to the good Strehl stability in the N band, it is possible to calibrate the visibilities with photometries recorded at a different time. 
This is not the case in the L band in which fringes and photometries must be recorded simultaneously to ensure accurate visibility measurements. 
To summarise, the standard observing mode with MATISSE is the so-called Hybrid mode: SiPhot in the L band and HighSens in the N band. 

In the N band, the detector integration time (DIT) is set to 20~ms in low spectral resolution and to 75~ms in high spectral resolution due to the thermal background level. 
Regarding the L band, the DIT can go from 75-125 ms (the coherence time of the atmosphere) to 10 s (when a fringe tracker is used).
\begin{figure*}[htbp]
    \centering
    \includegraphics[width=0.9\textwidth]{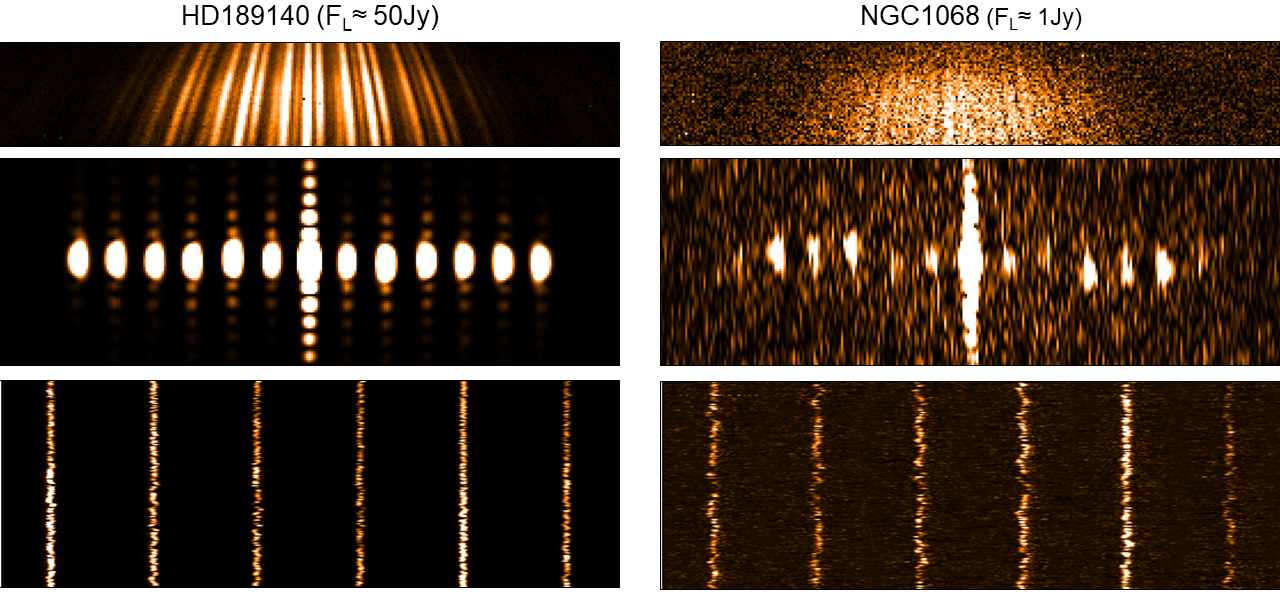}
    \caption{Images from MATISSE's NRTS processing for the calibrator HD\,189140 and the AGN NGC\,1068, both observed during MATISSE ATs commissioning on July 11, 2018. From top to bottom: Clean fringes frame the mean sky removed, with FFT2D showing the fringe peaks in x as a function of the optical path difference in y, and the waterfall of the six fringe-peak vectors as a function of time (in y) for a 1 min exposure.}
    \label{Fig:NRTS}
\end{figure*}

One observational block (OB) generates 14 exposures  for each band. In the N band, the sequence is the following: 
\begin{itemize}
    \item two sky exposures of 30 s recorded in the IN-IN and OUT-OUT BCD configurations,
    \item four interferometric exposures of 60s for each of the BCD configurations (IN-IN, OUT-IN, IN-OUT, and OUT-OUT) without chopping,
    \item eight photometric exposures of 60s with chopping.
\end{itemize}  

In the L band, the sequence is as follows: 
\begin{itemize}
    \item two sky exposures of 30s recorded in the IN-IN and OUT-OUT BCD configurations,
    \item four interferometric + photometric exposures of 60s for each of the BCD configurations (IN-IN, OUT-IN, IN-OUT, and OUT-OUT) without chopping,
    \item 2 $\times$ 4 interferometric + photometric exposures of 60 s for each of the BCD configurations (IN-IN, OUT-IN, IN-OUT, and OUT-OUT) with chopping.
\end{itemize}

We have in total 13 minutes of 'open shutter' in an OB, which is typically 27 minutes long, including the telescope preset process and the acquisition of images, pupils and fringes.



\subsection{Fringe detection and real-time coherencing}
\label{NRTS_detail}
During observations, fringes are searched, detected, and automatically kept within a fraction of the coherence length ($\lambda$R where $\lambda$ is the observing wavelength and R the spectral resolution) by the MATISSE Near-Real Time Software (NRTS). \\

The method used to measure the optical path difference between the beams is based on a modified version of the dispersed fringe tracking algorithm presented in \cite{1996ApOpt..35.3002K}. 
The steps of the peak detection algorithm are the following:
\begin{enumerate}
\item A mean sky is subtracted from each frame obtained on the target. It is computed either from the two sky exposures taken at the beginning of the observation sequence, or from sky-tagged frames during the exposure of chopped fringes.
\item A target frame without a fringe can also be subtracted. This frame is created by averaging frames with fringes on a full cycle of OPD modulation. This subtraction of a `foreground' image improves the fringe signal quality. 
\item Fast-Fourier transform along the spatial (x) dimension is performed. The resulting 1D Fourier Transform in x (FFTX) shows 13 peaks (one zero frequency and 6 $\times$ 2 symmetric fringe peaks). 
\item The six fringe peaks are extracted from the FFTX as well as seven noise slices taken between the fringe peaks.
\item The FFTX fringe and noise slices are interpolated on a grid with a regular wavenumber ($\sigma$=1/$\lambda$).
\item If OPD modulation is used, demodulation is performed on the fringe peaks by multiplying the vector by the quantity exp(-2i$\pi\Delta\sigma$), where $\Delta$ is the modulated OPD.
\item {A second FFT is performed on each re-sampled demodulated fringe and noise slice.}
\item Coherent integration (in the complex plane) is performed on N$_C$ frames.
\item Then, fringes and a noise peak can be integrated incoherently. This is performed by summing the square of the vectors on N$_I$ frames. 
\item Integrated fringe-peak vectors are finally cleaned by removing the noise vectors interpolated at the fringe vector frequencies.
\item The positions of the maximum, which correspond to the achromatic atmospheric OPD (or 'piston' or 'group delay') values to be corrected  \citep{1996ApOpt..35.3002K} are detected by fitting a Gaussian distribution and converted from pixels to OPDs. 

\item Finally, simulated annealing, used as a converging method towards the minimum $\chi^2$ between the calculated OPDs and the data, is used to compute the path per telescope from the per baseline OPD and S/N.
\end{enumerate}
We note that as coherent integration is very sensitive to OPD variations, N$_C$ should be, at maximum, twice the number of frames recorded during the atmospheric coherence time at the considered wavelength. As the DIT on the L\&M detector is already of the order of the coherence time in that band,  coherent integration should not be used on the L\&M arm. In the N band, however, the DIT is small compared to the atmospheric coherence time (of the order of a few hundreds of milliseconds) and N$_C$ is set to 40. Incoherent integration is only used for very faint objects close to the sensitivity limit.\\
Figure~\ref{Fig:NRTS} presents two examples of the NRTS processing taken from MATISSE ATs commissioning on July 11, 2018: the bright and unresolved star HD\,189140 (F$_L\sim$ 50 Jy and V$^2\geq$ 0.9) and the dimmer and partially resolved AGN NGC\,1068 (F$_L\sim$ 1Jy and V$^2\sim$ 0.1). In that example, only the L band observation is shown and the N$_C$ and N$_I$ parameters are both set to 1 so that the FFT and fringe peak waterfalls are not time-averaged.  


\subsection{Data reduction pipeline}
\label{secdrs}
The MATISSE data reduction software (DRS) is fully implemented in the ESO pipeline environment \citep{McKay2004}. It is interfaced with EsoRex and Reflex. It produces the calibration maps, reduces the raw data into OIFITS files, calibrates the data, and finally reconstructs an image from the reduced data. The processing is done independently for L\&M and for N.
The data reduction makes use of calibration maps to remove instrumental and electronic signatures. Regarding the detector's response, the following maps are used: the bias, the non-linearity, the flat-field, and the bad pixel maps. Added to that are:
\begin{itemize}
    \item the Shift map, which calibrates the dispersion law on both detectors for various spectral resolutions and computes the distortion of the photometric and interferometric channels,
    \item the Kappa matrix, which computes the anamorphic factor between the interferometric and photometric channel, the intensity ratio between interferometric and the photometric channel, and the spatial shift. 
\end{itemize}
From cleaned-up fringes and photometric frames, the pipeline estimates the coherent flux and the photometry using the multiple-stage modulation of MATISSE for the sky background removal (chopping, spatial modulation, and temporal modulation). Then, the raw visibilities and phases are estimated using two kinds of estimators: incoherent estimators (speckle-like), which can be computed without knowledge of the optical path difference, and coherent estimators that need an appropriate atmospheric OPD estimate. \\
The incoherent estimators are:
\begin{itemize}
    \item {\bf Squared coherent flux}
    \begin{equation}
        C^2_{ij}(\lambda)=\sum_u\left<|I(u,\lambda,t)|^2-\beta\right>_t
    \end{equation}
    where $C^2_{ij}(\lambda)$ is the squared coherent flux at wavelength $\lambda$ for the {input physical baseline} $ij$ corresponding to the telescope {beams} $i$ and $j$. Also, $\left<...\right>_t$ defines the time average; $I(u,\lambda,t)$ is the Fourier Transform of the interferogram at $\lambda$; $u$ is {the output spatial frequency integrated between $(B_{ij}-D)/\lambda$ and $(B_{ij}+D)/\lambda$ where $B_{ij}$ and $D$ are the output baseline and pupil diameter}, respectively; and $\beta$ is the bias produced when computing the power spectral density of the fringes, and estimated on $I$ between and after the fringe peaks.
    \\
    \item {\bf Squared visibility}
    \begin{equation}
        V^2_{ij}(\lambda)=\frac{C^2_{ij}(\lambda)}{\sum_x\left<P_{ij}(x,\lambda,t)\right>_t}
    \label{eq:V2}
    \end{equation}
     where $V^2_{ij}$ is the squared visibility for the {input physical baseline} $ij$,  $P_{ij}(x,\lambda,t)=P_i(x,\lambda,t)P_j(x,\lambda,t)$ with $P_i(x,\lambda,t)$ being the estimates of the photometric fluxes for the telescope $i$ corrected from the Kappa matrix.
     \\
     \item {\bf Closure phase}
     \begin{equation}
         \psi_{ijk}(\lambda)=Arg\left[\sum_{u,v} <I(u,\lambda,t)I(v,\lambda,t)I^*(u+v,\lambda,t)>_t-\gamma\right]
     \end{equation}
     where $\psi_{ijk}$ is the closure phase for the telescope triplet $ijk$; and $u$, $v$, and $u+v$ are the {output spatial frequency $B_{ij}$, $B_{jk}$, and $B_{ik}$, integrated, for instance for $B_{ij}$, between $(B_{ij}-D)/\lambda$ and $(B_{ij}+D)/\lambda$, where $B_{ij}$ is the output baseline and $D$ the output pupil diameter}. Furthermore, 
     $\gamma$ is the photon bias present in the bispectrum. 
\end{itemize}
For the coherent processing, the first step is to estimate the residual atmospheric OPD (or `piston') in each frame, which is left after the fringes' coherencing is performed during the observations. The residual atmospheric phase contribution in the measured interferometric data can be modelled as follows:
\begin{equation}
    \phi_{atm}(\lambda,t)\approx\phi_0(t)+2\pi\frac{\delta(t)}{\lambda}
,\end{equation}
where $\phi_0$ is an achromatic phase term due to the first order dispersion effects from the atmosphere, and $\delta(t)$ is the residual atmospheric piston. This piston is commonly referred as the `group delay'. Indeed, in the frame of such a phase model, and given the definition of the group delay G, we have the following:
\begin{equation}
G=\frac{\partial\phi_{atm}(t,\lambda)}{\partial k}=\delta(t).
\label{eqG}
\end{equation} 
The $\phi_0$ and $\delta(t)$ are estimated through the global estimation algorithm proposed by \cite{schutz2016}.  Then the frames are integrated coherently after having corrected the residual atmospheric phase providing the following estimators: 
\begin{itemize}
    \item {\bf Modulus of the coherent flux}
    \begin{equation}
        \left|C_{ij}(\lambda)\right|=\Re\left[\sum_u<I(u,\lambda,t) e^{-i\phi_{atm}(\lambda,t)}>_t\right]
    \end{equation}
    \item {\bf Differential Phase}
    \begin{equation}
        \varphi_{ij}(\lambda)=Arg\left[\sum_u<I(u,\lambda,t) e^{-i\phi_{atm}(\lambda,t)}>_t\right]
    \end{equation}
    \item {\bf Visibility}
    \begin{equation}
        V_{ij}(\lambda)=\frac{\left|C_{ij}(\lambda)\right|}{\sqrt{\sum_x \left<P_{ij}(x,\lambda,t)\right>_t}}
    \end{equation}
    \item {\bf Differential Visibility}
    \begin{equation}
        V_{Dij}(\lambda)=\frac{\sum_u |<CS_{ij}(u,\lambda,t)>_t|}{\sqrt{\sum_x \left<P_{ij}(x,\lambda,t)\right>_t\sum_x \left<\sum_{\lambda_{ref}}P_{ij}(x,\lambda_{ref},t)\right>_t}}
    \end{equation}
    where $CS_{ij}(u,\lambda,t)$ is the cross spectrum between one spectral channel (at $\lambda$) and a reference channel (with $\lambda_{ref} \neq \lambda$):
    \begin{equation}
        CS_{ij}(u,\lambda,t)=I(u,\lambda,t)\sum_{\lambda_{ref}}{}I^*(u,\lambda_{ref},t) e^{-i\phi^\prime_{atm}}
    \end{equation}
    and $\phi^\prime_{atm}=\phi_{atm}(\lambda,t)-\phi_{atm}(\lambda_{ref},t).$
\end{itemize}

The raw interferometric observables are then calibrated using the standard method based on the observation of calibrator stars. For the phase estimates (differential phase and closure phase), the BCD allows one to apply an additional calibration in order to remove any instrumental signatures. This is done by applying a specific linear combination of the phases in the different configurations of the BCD. 
For the visibility, it is important to calibrate the transfer function before combining the different BCD configurations since the transfer function depends on them. For the phase, it is not necessary to remove the instrumental phase from the calibrators since the BCD calibration is done for that. The last step of the pipeline is the image reconstruction. MATISSE uses the IRBis algorithm \citep{IRBis}, which was part of the instrument deliveries. 


\section{MATISSE performance}
\label{secperf}

{The limiting magnitudes offered by ESO to the community of users, as shown in Table \ref{tab:fairandgood}, are set by 
the following precision criteria, which are defined per spectral channel for one minute of observation:} 

\begin{itemize}
    \item Visibility: $\sigma_{V}=0.1$ 
    \item Closure phase: $\sigma_{\psi}=5^\circ$ 
    \item Differential phase: $\sigma_{\varphi}=4^\circ$ 
\end{itemize}
The next subsections describe how these MATISSE limits were estimated and then their values are provided. Along the VLTI chain, additional sensitivity limits are set in the R band by the adaptive optics, and in the K band by focal laboratory guiding. 
The limit of NAOMI on ATs is R=12.5 mag, where the adaptive optics has good performance, and R=15 mag for the loop closed with degraded performance. The limit with MACAO on UTs is V$\sim$15. In parallel, the limiting magnitude is up to 14 (10.7, respectively) in any of the sensing IRIS bands (J, H, and K band) on the UTs (on the ATs, respectively) for image tracking in the focal laboratory. 
 
\subsection{Contributions to MATISSE measurement errors}
The different contributions to the errors, which affect the measurements, are:

\begin{itemize}
    \item The fundamental noise errors that affect the precision of all measurements.
    \item The broadband seeing errors that affect the absolute visibility and closure phase independently from the source brightness.
    \item The broadband photometric errors likely caused by imperfect thermal background subtraction, which produce a source flux dependent error in the absolute visibilities.
\end{itemize}


\subsubsection{Fundamental noise errors}
\begin{figure}
    \centering
      {\includegraphics[width=0.45\textwidth]{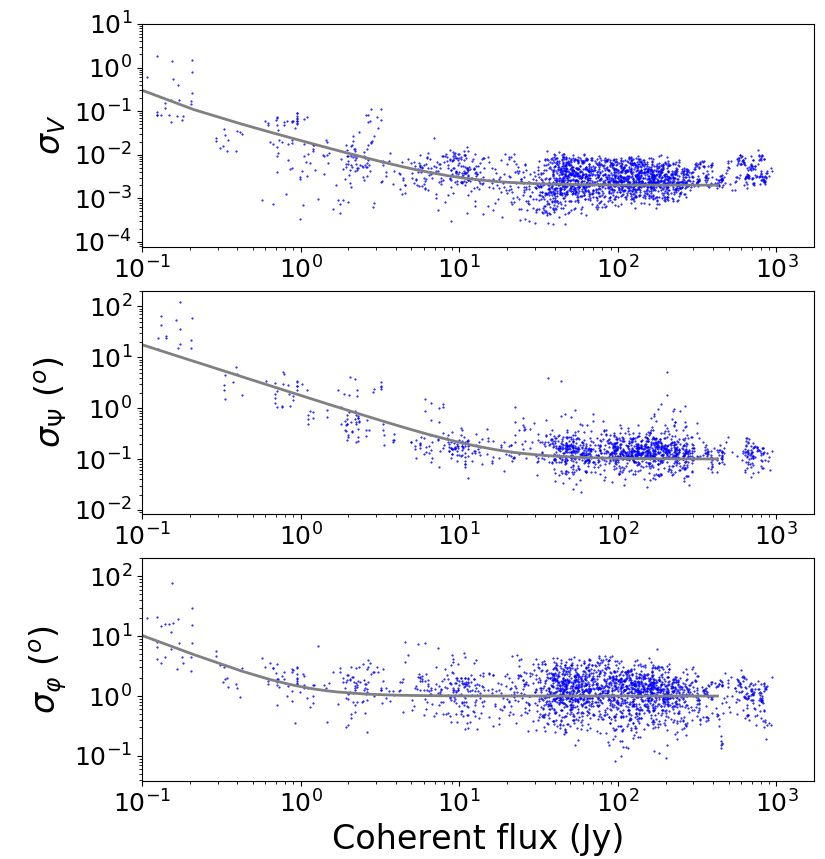}}
      \caption{Illustration of the fundamental noises' contribution for L band low spectral resolution observations with the ATs. The blue dots represent the measurement precisions per spectral channel and per 1mn exposure as a function of the source coherent flux for the absolute visibility (top), the closure phase (middle), and the differential phase (bottom). The solid grey line shows the tuned error prediction that goes through the median of the measures. 
      }
    \label{fig:blutdote_LLOWAT}
\end{figure}

The contribution of the fundamental noises in the measurement precisions per spectral channel and per exposure 
were evaluated on sky on a large number of calibrators with different magnitudes\footnote{These interferometric calibrators were selected from the MDFC catalogue \citep{2019MNRAS.490.3158C}.}. 
To disentangle the contribution of the fundamental noise in the data from the broadband seeing errors and the broadband photometric errors, which are both analysed in the next two subsections, we fitted, for every 1 min exposure and every observable (absolute visibility, closure phase, and differential phase), the measurement, $m(\lambda)$, with a low order polynomial function $f(\lambda)$. 
Then we computed the standard deviation, over the wavelength, of the residual $m(\lambda)-f(\lambda)$, denoted as $\sigma_{M}$. Here, $\sigma_{M}$ thus represents the estimated standard deviation of the error generated by the fundamental noises. Each blue dot in Fig. \ref{fig:blutdote_LLOWAT} corresponds to one $\sigma_{M}$ value estimated on a given calibrator and is plotted against the calibrator coherent flux. At high coherent fluxes, $\sigma_{M}$ does not seem to decrease as a function of the source flux anymore and it reaches a plateau for most of the observables.

Then, to evaluate the fundamental measurement precision properly for a given source coherent flux, from the cloud of blue dots shown in Fig. \ref{fig:blutdote_LLOWAT}, it was necessary to use a noise model that was tuned to the measured precisions to extract an average trend. This noise model is described in Appendix~\ref{sec:noise_model}. As a result of that noise model, $\sigma_{F}(\lambda)$ is the predicted standard deviation of the error generated by the fundamental noises for each observable.



Figure \ref{fig:blutdote_LLOWAT} shows the predicted variance $\sigma_{F}$ of the fundamental noise error, which was scaled to match the measured precision at best and thus extract an average trend. 
That average trend then constituted the basis to evaluate the measurement precision for a given source coherent flux. A flux independent variance was added to $\sigma_{F}^2$ to reproduce the observed plateau at high flux. This provides a lower limit on the measurement precision, given in Tab. \ref{tab:highflux}. The level of that plateau is independent of the spectral resolution or the telescope type, although the plateau appears at different source flux levels for ATs and UTs. 
\begin{table}[htbp]
\caption{Minimum MATISSE errors achievable at high flux. For the differential phase, these values are given for a broad reference channel. For a narrow continuum-line-continuum measurement in L, the differential phase can be as precise as $0.2^\circ$ }
\centering
\label{tab:highflux}
\begin{tabular}{cccc}
\hline\hline
Band  & L & M & N \\
\hline\hline
Visibility & $3 \times 10^{-3}$ & $5 \times 10^{-3}$ & $3 \times 10^{-3}$ \\
\hline
Closure phase & $0.1^\circ$ & $0.2^\circ$ & $0.1^\circ$ \\
\hline
Differential Phase & $1^\circ$ &$1^\circ$&$0.2^\circ$ \\
\hline\hline
\end{tabular}
\end{table} 

\subsubsection{Broadband seeing errors {on the visibility}}
\label{sect:InstrVis}
\begin{figure}[htbp]
\centering
            {\includegraphics[width=0.45\textwidth]{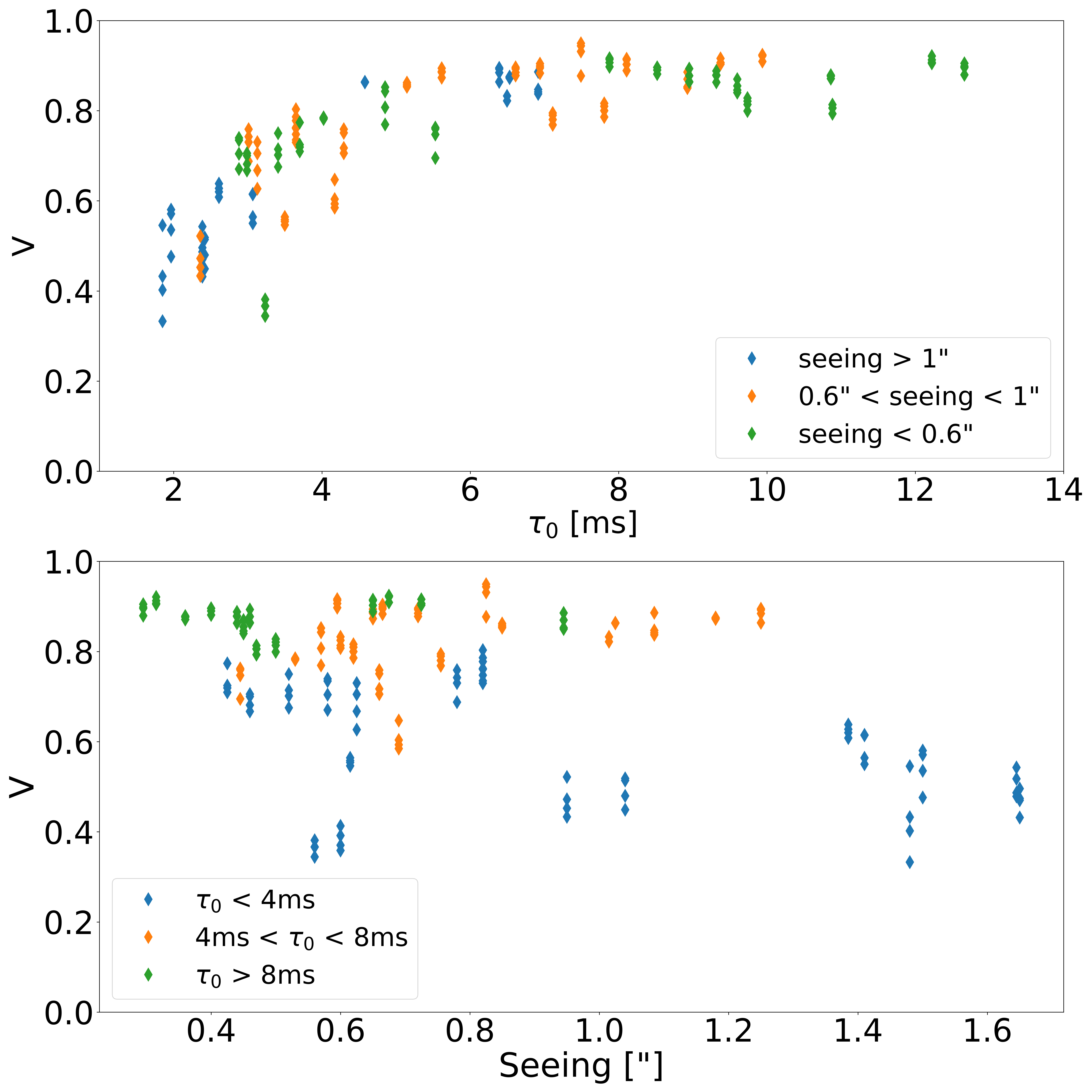}}
      \caption{Variation of the L band instrument + atmosphere visibility, averaged between 3.1 and 3.8~$\mu$m, as a function of seeing in arcsecond (bottom) and of the atmospheric coherence time measured in millisecond in the visible (top). The colour code indicates the seeing ranges and the coherence time. The observations were carried out with a DIT of 111~ms on calibrators with L band fluxes between 1 and 100~Jy.}
         \label{fig:vvsst0tau}
\end{figure}

Changes in seeing conditions and time-varying instrumental features such as OPD changes resulting from transient telescope vibrations can affect the instrument + atmosphere response during an observing night and degrade the fringes' contrast. 
It is the time variation of this transfer function that induces calibration errors. 
Since those affect all spectral channels  during the observation of either the science target or its calibration stars, these errors, called broadband seeing errors, are broadband calibration errors.
As MATISSE is a single mode instrument that carefully maintains its internal optical quality, it was expected that these time sensitive broadband calibration errors would be dominated by changes in the atmospheric piston jitter and would mainly depend on the atmospheric coherence time. This was confirmed during the commissioning and is illustrated in Fig. \ref{fig:vvsst0tau} in the L band. The displayed instrumental visibilities were estimated from calibrators and corrected for their diameter. 
We note that the response is much more sensitive to $\tau_{0}$ than to seeing, and it becomes less sensitive to atmospheric changes for $\tau_{0}$~>~5~ms, measured in the visible by ESO seeing monitors. Overall, that shows that the L band instrumental visibility (transfer function) can decrease by a factor 2 when moving from `fair' conditions ($\tau_{0}$~>~5~ms) to bad conditions ($\tau_{0}$~<~3~ms).

\begin{figure}[htbp]
\centering
             {\includegraphics[width=0.45\textwidth]{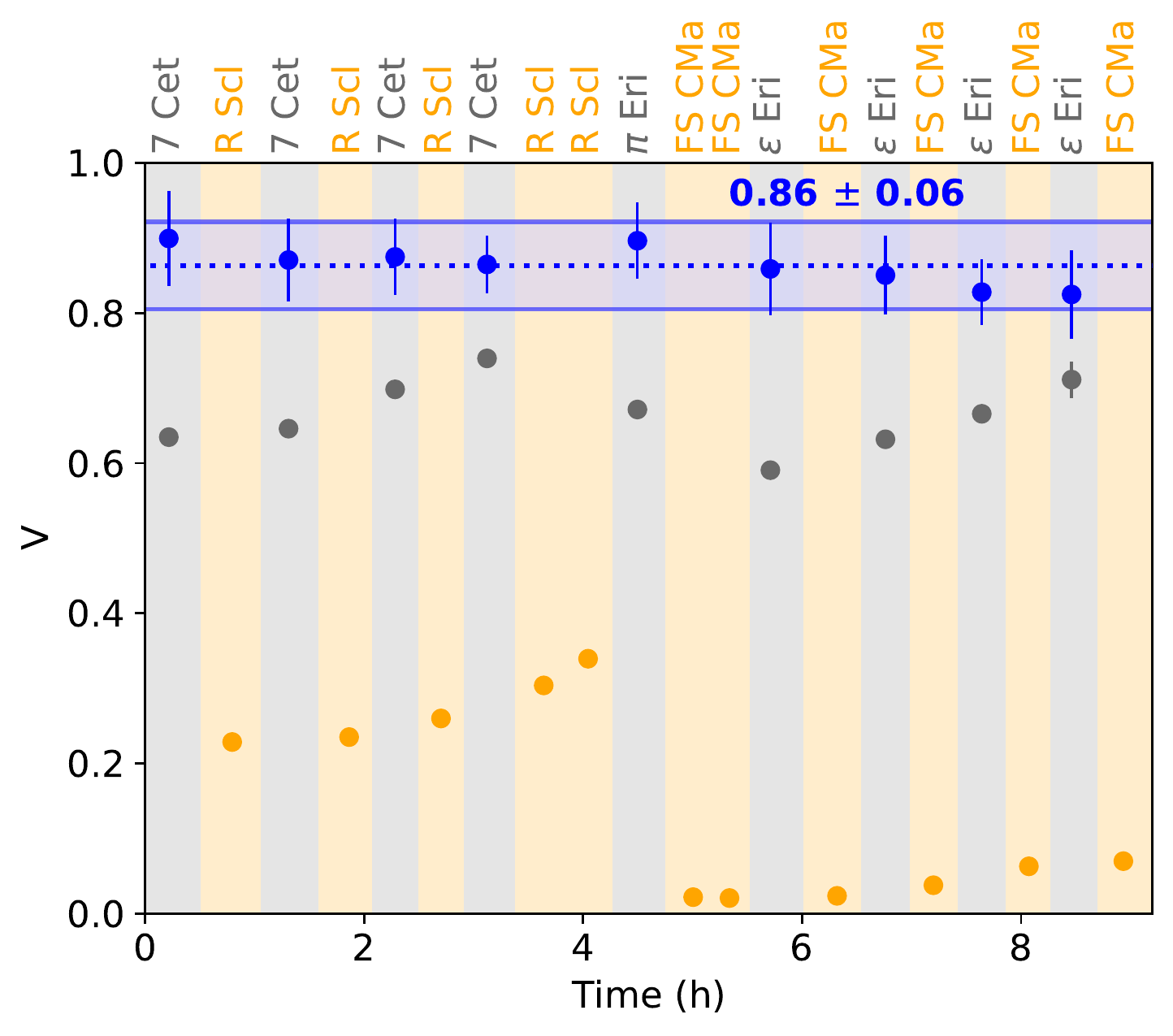}}
      \caption{ 
      Mean L-band visibility (averaged between 3.1 and 3.8~$\mu$m) for the K0-G2 baseline plotted as a function of time for a full night on December 9, 2018. The science objects are shown in orange, and the calibrators are in grey. The estimated transfer function in visibility (i.e. corrected from the partial resolution of the calibrators) is shown in blue. The variation of the transfer function represented by the modulation of the blue dots is a slow temporal variation.}
         \label{fig:TFvsTime}
\end{figure}

Figure \ref{fig:TFvsTime} shows the time variation of the transfer function (visibility and closure phase averaged over the spectral band of interest) on bright calibrators. The RMS of the variations (corrected from a low order polynomial fit) over one night gives an estimation of the error. The results are presented in Tab. \ref{tab:broadbanderrors}.

\begin{table*}
\centering
\caption{Broadband seeing errors for different seeing conditions. }
\label{tab:broadbanderrors}
\begin{tabular}{cccccccc}
\hline\hline
\multicolumn{2}{c}{Seeing conditions}  & \multicolumn{2}{c}{L (DIT=111ms)} & \multicolumn{2}{c}{M (DIT=111ms)} & \multicolumn{2}{c}{N (DIT*N$_{MOD}$=240ms)} \\
\hline
 $\tau_{0}$ (ms) & Seeing(") &   \makecell{Visibility \\ error} & \makecell{Closure phase \\ error}& \makecell{Visibility \\ error} & \makecell{Closure phase \\error}& \makecell{Visibility \\ error} & \makecell{Closure phase \\ error}\\
\hline\hline
3.2$\pm$0.5 &   0.96$\pm$0.1&   0.08 &  0.30$^\circ$&   0.05&   0.25$^\circ$&   0.045&  1.75$^\circ$
 \\
\hline
6.8$\pm$0.5     &0.74$\pm$0.1&  0.02&   0.26$^\circ$&   0.020&  0.24$^\circ$&   0.02&   0.49$^\circ$ \\
\hline
7.5$\pm$0.5     &0.56$\pm$0.1&  0.02&   0.16$^\circ$&   0.015&  0.15$^\circ$&   0.015  &0.29$^\circ$ \\
\hline\hline
\end{tabular}
\tablefoot{
These low closure-phase errors were obtained with the BCD correction.}
\end{table*}
These results are given for the standard frame times: 111 ms for the L and M bands, and 20 ms in the N band. In fact, in the N band, a coherent integration of ten elementary frames (a modulation cycle) yields an equivalent frame time close to the N band coherence time. {These visibility errors correspond to an external calibration and should improve with the square root of the number of science-calibrator cycles.}

\subsubsection{\textbf{Closure-phase calibration errors}}
\label{sect:InstrCP}

\begin{figure*}[htbp]
    \centering
    \includegraphics[width=0.9\textwidth]{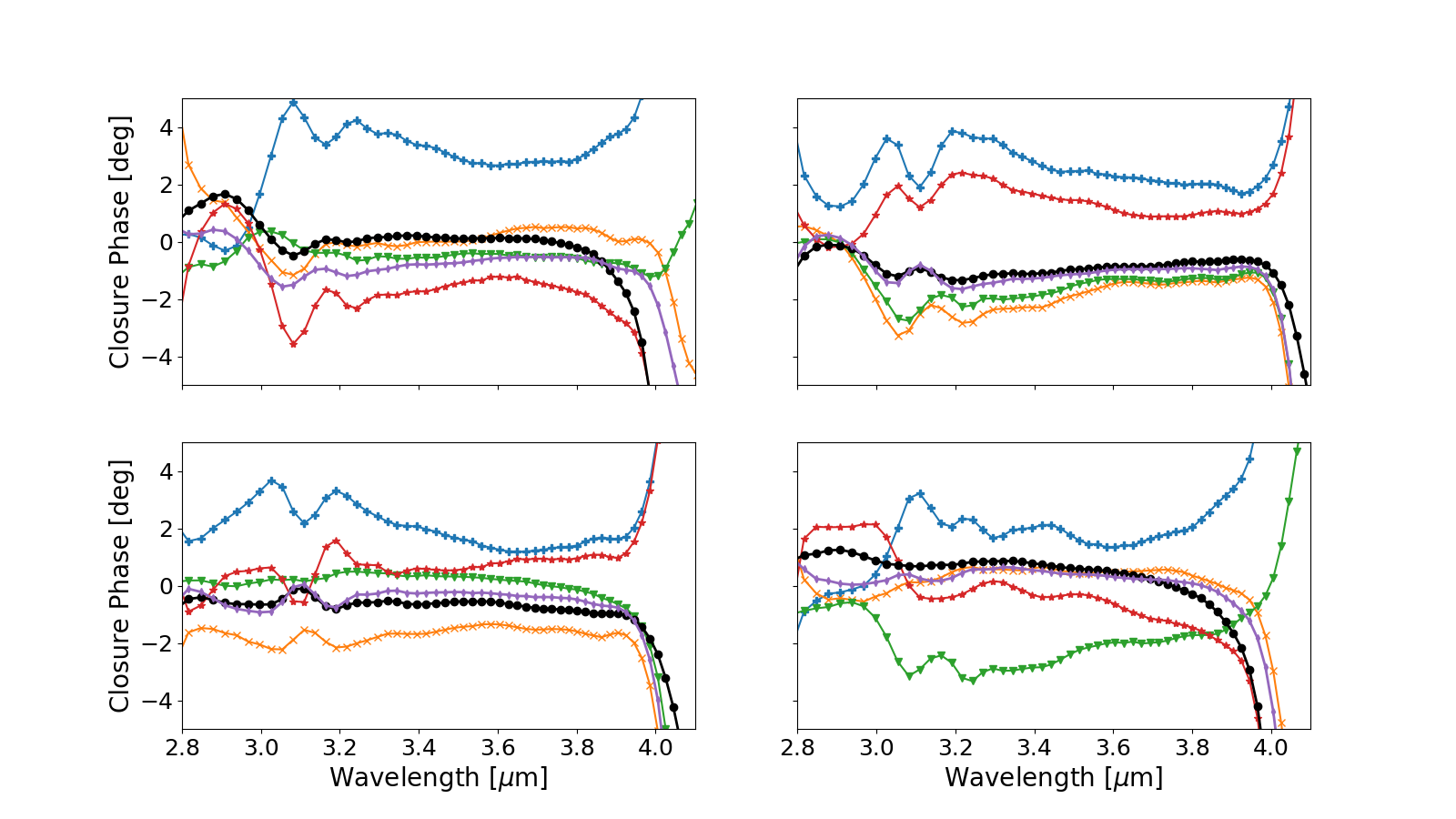}
    \caption{{Four closure phases obtained on a bright star, IRAS 10153-5540, as a function of the wavelength. The blue(\textbf{+}), red($\star$), green(\textbf{$\triangledown$}), and orange(\textbf{$\times$}) lines represent the CP for the different BCD position calibrated by VV396 Cen. The black($\bullet$) line represents the CP corrected by the BCDs only. The purple($\diamond$) line represents the CP of IRAS 10153-5540 calibrated by VV396 Cen and corrected by the BCDs.
   }}
    \label{Fig:calibcloturebcd}
\end{figure*}

{As presented before, the combination of the four BCD positions removes the instrumental contributions between the BCDs and the detector \citep{Millour2008}.} 
{To illustrate that, we present the case of the following two bright calibrators: IRAS 10153-5540 (73 Jy) and VV396 Cen (81 Jy). Fig. \ref{Fig:calibcloturebcd} shows the closure phases of IRAS 10153-5540 calibrated by the calibrator VV396 Cen  (in blue, orange, green, and red). If there was no systematic residual phase error, the closure phases should be zero and with a negligible fundamental noise contribution. The strong deviation observed after 3.9 $\mu m$\ is due to the chromatic OPD in the VLTI tunnels. Those closure phases are compared with the closure phases resulting from the calibration by the BCDs only (black) and from a calibration first by the calibrator and then by the BCDs (purple). The BCD correction reduces the measurement errors from a few degrees to a fraction of a degree; furthermore, the combination of calibrating with a calibrator and the BCDs provides no additional improvement. This result confirms that for closure-phase measurements, the BCD calibration is sufficient and no calibrator observation is necessary. Table \ref{tab:broadbanderrors} shows the closure-phase calibration errors obtained from a BCD calibration cycle.}

\subsubsection{Broadband photometric errors}
\begin{table}[htbp]

\caption{Broadband photometric error (in Jansky) }
\centering
\label{tab:broadbandphotometry}
\begin{tabular}{cccc}
\hline\hline
 Band  & L & M & N \\
\hline\hline
$\epsilon_p$ AT (Jy) &0.11 &0.19&2.3 \\
\hline
$\epsilon_p$ UT (Jy) & 0.008&0.016&0.08 \\
\hline\hline
\end{tabular}
\end{table}   
The computation of the absolute visibility requires an estimation of the source photometry, which requires telescope chopping to measure and subtract the thermal background contribution. However, due to a combination of time fluctuations and chopping imperfections that produce seeing contamination from the thermal background, a broadband photometric error $\epsilon_p$ can be introduced in the process. Such a broadband photometric error (given in Tab. \ref{tab:broadbandphotometry}) produces an additional source flux dependent error on the absolute visibility error budget. Its inclusion is described in Section \ref{sec:noise_model}. 
The estimation of the broadband photometric error due to the background fluctuations was based on a statistical analysis of the sky frames obtained from telescope chopping for the full L and M bands and for 1 $\mu$m width sub-bands in the N band.

\subsection{Global performances}
\begin{figure}
\centering
         
             {\includegraphics[width=0.45\textwidth]{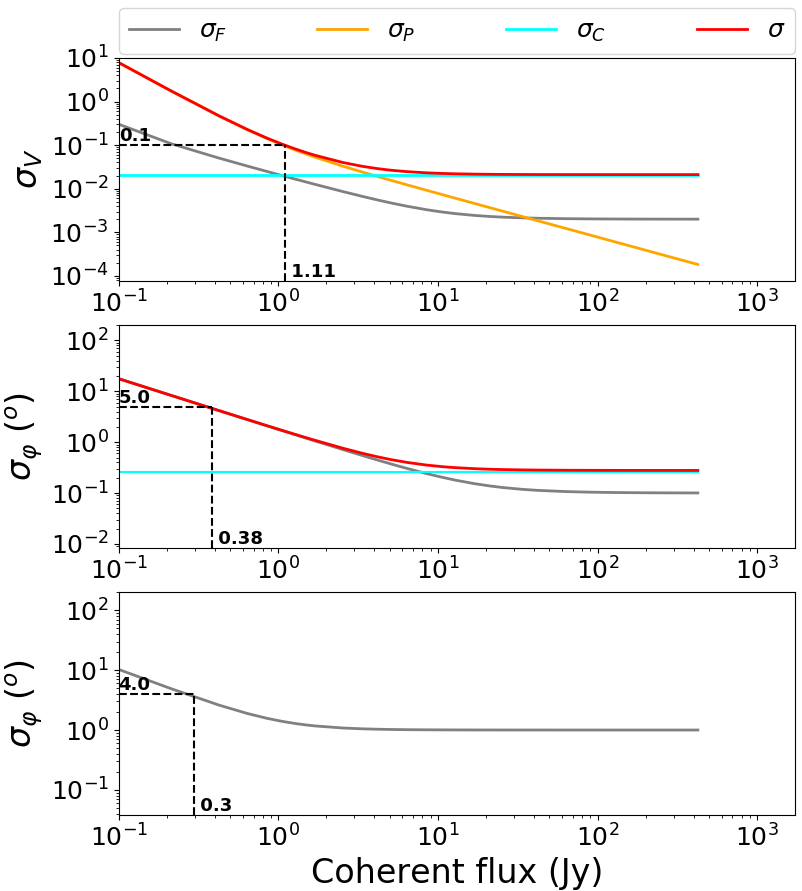}}
      \caption{Illustration of the determination of the global measurement errors in low spectral resolution in the L band with the ATs. Top: Visibility fundamental noise standard deviation ($\sigma_F$ in grey), flux dependent error due to the broadband photometric error ($\sigma_P$ in yellow), broadband calibration error ($\sigma_C$ in cyan), and their combination ($\sigma$ in red). Middle: Same for the closure phase, without $\sigma_P$. Bottom:  Differential phase fundamental noise standard deviation ($\sigma_F$ in grey).}
         \label{fig:perfoL}
\end{figure}

Fig. \ref{fig:perfoL} illustrates the combination of the different contributions to the global MATISSE measurement errors. 
For the differential phase, the sensitivity limit is directly given by the precision on the fundamental noise. For the closure phase, we added the variance of the fundamental noise per channel and the transfer function error deduced from the time transfer function of the measure as described in Sect. \ref{sect:InstrVis}. Then the plot of these combined variances was used to find the coherent flux that yields a closure phase accuracy per spectral channel better than 5$^\circ$ in 50\% of the cases. For the absolute visibility, we added the variances of the fundamental noise errors, the transfer function error, and the broadband photometric error. The plot of the combined variance was used to find the coherent flux that yields a visibility accuracy per spectral channel better than 0.1 in 50\% of the cases.\\

\begin{figure*}
\centering
         
             {\includegraphics[width=\textwidth]{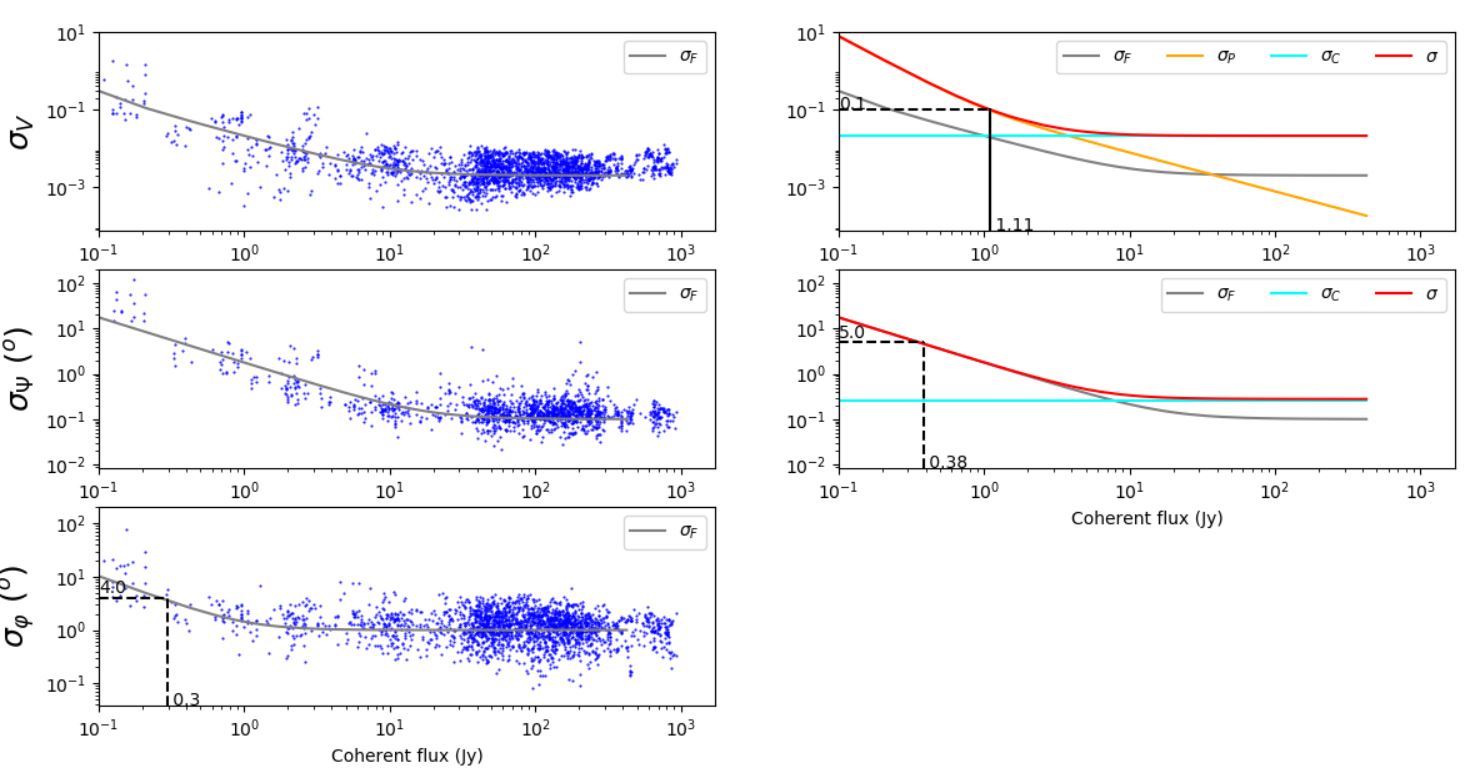}}
      \caption{Illustration of the general estimation of MATISSE precision and accuracy. Here we show the low-resolution setup in the L band with ATs. In the left column we plot the measured fundamental noise per spectral channel (blue dots) as a function of the source coherent flux. The grey line shows the value of the tuned noise prediction that goes through the median of the measures. Visibility on top in visibility units, closure phase in the middle and differential phase on bottom, in degrees. The coherent flux is in Jy. The top right panel shows the visibility fundamental noise standard deviation (grey), the flux dependent broad band photometric error (yellow) and the broad band calibration error due to seeing changes that is independent from the flux (cyan) as well as their combination (red) assuming these are independent variables.}
         \label{fig:generalETCAT}
\end{figure*}
\begin{figure*}[htbp]
\centering
{\includegraphics[width=\textwidth]{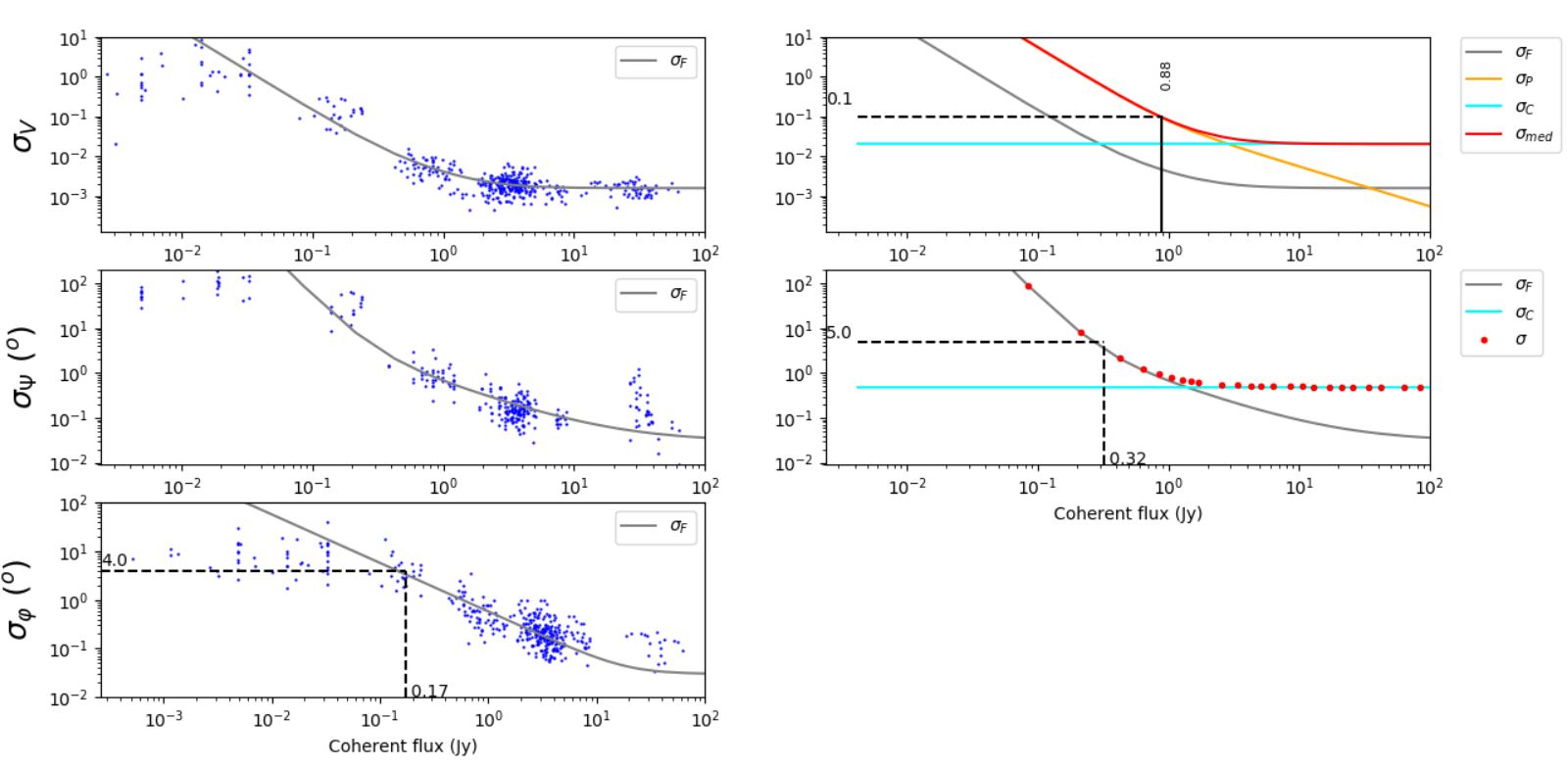}}
      \caption{Illustration of the general estimation of MATISSE precision and accuracy in the low-resolution setup in the N (8-9 $\mu$m) band with UTs. The plots and axes are the same as in Fig. \ref{fig:generalETCAT}.}
         \label{fig:generalETCUT}
\end{figure*}
\begin{table}
\caption{Spectral bands used for the performance qualification. 
}
\centering
\label{tab:Band_Perfo}
\begin{tabular}{ll}
\hline\hline
Spectral Mode & Wavelengths in $\mu m$ \\
\hline\hline
LOW-L   & $\lambda_{0}=3.5, \Delta\lambda= 0.6$ \\
MED-L   & $\lambda_{0}\in[3.2,4.1], \Delta\lambda=0.2$ \\
HIGH-L  & $\lambda_{0}\in[3.9,4.2], \Delta\lambda=0.1$ \\
\hline
LOW-M   & $\lambda_{0}=4.75, \Delta\lambda= 0.4$ \\
MED-M   & $\lambda_{0}\in[4.55,4.95], \Delta\lambda=0.2$ \\
\hline
LOW-N   & $\lambda_{0}=8.5, \Delta\lambda= 0.8$ \\
HIGH-N  & $\lambda_{0}=8.5, \Delta\lambda= 0.8$ \\
\hline\hline
\end{tabular}
\end{table}
 The MATISSE stand-alone performances measured in the spectral bands given in Tab. \ref{tab:Band_Perfo} are summarised in Tab. \ref{tab:fairandgood} for seeing conditions~<~0.9 arcsecond and $\tau_0>5$~ms. 
For poorer conditions ($\tau_0<3$~ms), all limits in Table \ref{tab:fairandgood} increase by typically 2 Jy.

\begin{table*}
\caption{MATISSE stand-alone (i.e. without fringe tracker) sensitivity limits in Jansky in L, M, and N bands.}
\centering
\label{tab:fairandgood}
\begin{tabular}{ccccccccccc}
\hline\hline
\multirow{2}{*}{Telescopes}     & \multirow{2}{*}{Resolution}   &\multicolumn{3}{c}{$V$}        &\multicolumn{3}{c}{$\psi$ }&\multicolumn{3}{c}{$\varphi$}
\\
& &     L&      M       & N     &L&     M&      N&      L&      M&      N \\
\hline\hline
\multirow{3}{*}{ATs}&   LOW     &1.1 &  2.1&    16.8&   0.4     & 1.9   &9.4&   0.3&    1.1&    5.0 \\
\cline{2-11}
&MED&   3.8 & 16.0 & - &        3.3     & 15.6 & - & 2.4 & 11.0 & - \\
\cline{2-11}
&       HIGH&   20.1 &  - &     30.3&   14.7& - &       29.9&   10.8    & - &     25.3  \\
\hline
\multirow{3}{*}{UTs}&   LOW&    0.3&    0.4&    0.9&    0.07&   0.2&    0.3&    0.06&   0.15& 0.3  \\
\cline{2-11}
&MED&   1.1&    1.1& - &        0.8&    0.9&    -&      0.6&    0.7& - \\
\cline{2-11}
&HIGH&  2.4&    -&      1.6 &   1.7& - & 1.5 & 1.2&     -&      1.1 \\
\hline\hline
\end{tabular}
\tablefoot{
The sensitivity limits are defined for the following precision criteria: $\sigma_{V}=0.1$,  $\sigma_{\psi}=5^\circ$, and $\sigma_{\varphi}=4^\circ$, to be achieved per spectral channel after one minute of observation. Any source brighter than these limits will provide a better precision. The symbol '-' means that the combination resolution-versus-band does not exist. The N band bias error contribution is dominant in the differential phase. Magnitude 0 in L (at 3.5$\mu m$), M (at 4.75$\mu m$), and N (at 8.5$\mu m$) bands corresponds to 290$Jy$, 165$Jy$, and 50$Jy$, respectively.}
\end{table*}

\begin{figure}
\centering{\includegraphics[width=0.45\textwidth]{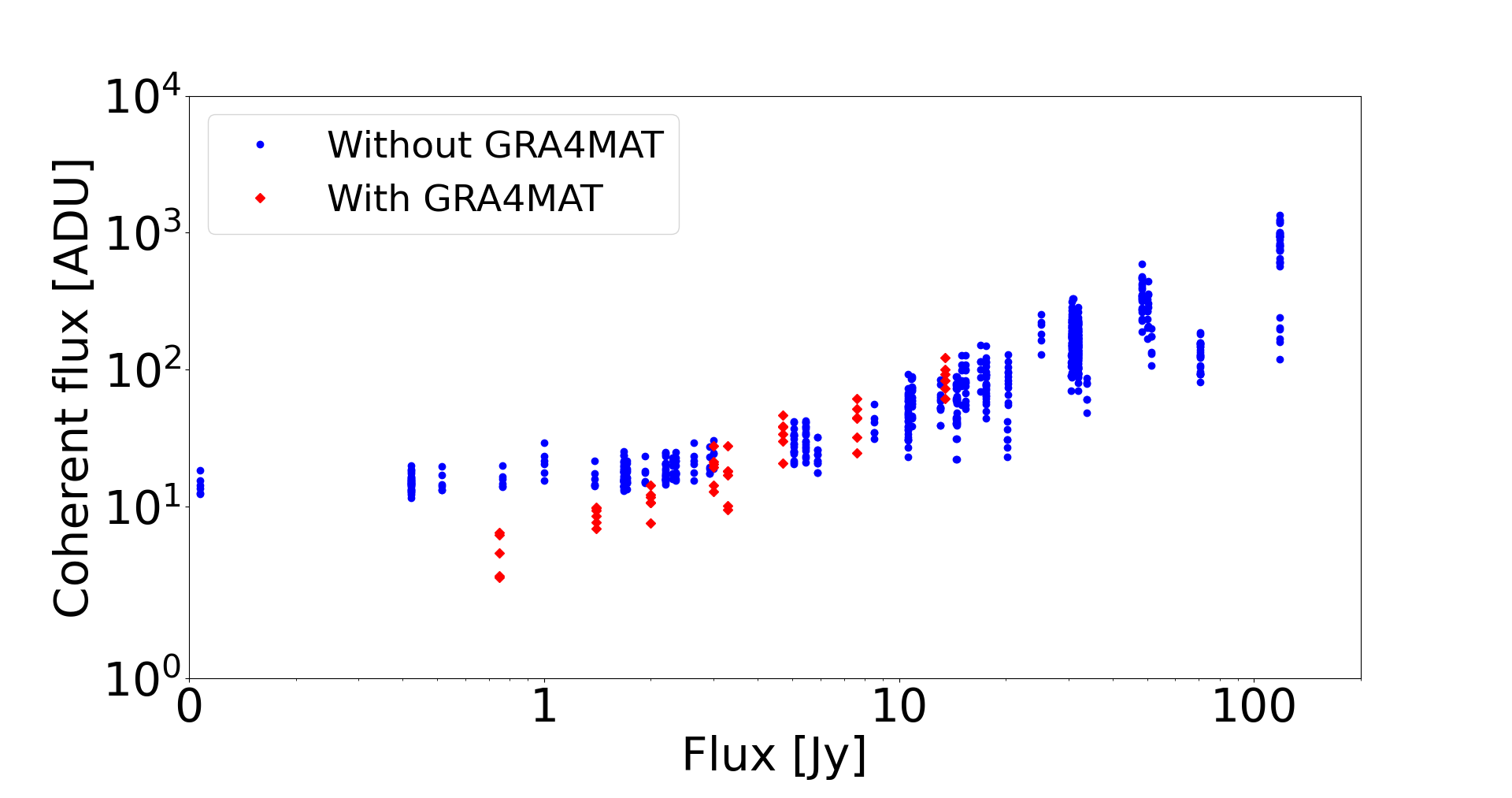}}
\caption{Measured coherent flux for various calibrators carried out in April 2019, as a function of the calibrator flux taken from the MDFC catalogue. Each dot (or diamond) corresponds to an average over a 1 min exposure and over a 1~$\mu$m spectral bandwidth (8~$\mu$m - 9~$\mu$m). The improvements provided by GRA4MAT are discussed in Section 6.} 
\label{fig:biasN}
\end{figure}

The following aspects of our analysis could improve the instrument performance:
\begin{itemize}
\item The differential phase is affected by the chromatic part of the atmospheric OPD. In our performance analysis, we estimated and removed the contribution from the chromatic OPD effect through a low-order polynomial fit. A future improvement may consist in using a physical model based on the refractivity of dry air and water vapour.
\item The differential phase estimation is done through a coherent integration of the frames over one exposure, which requires one to estimate the atmospheric OPD in each frame. We noticed that errors on the estimation of the atmospheric OPD, per frame, for faint sources yield a bias in the estimated coherent flux. This is illustrated in Fig.\ref{fig:biasN} for the N band. This bias limits to the level of $\sim$5~Jy at 8.5~$\mu$m (8~Jy at 11~$\mu$m) the possibility of obtaining reliable coherent fluxes and, therefore, directly impacts the sensitivity limits of the differential phase. More sophisticated methods (Berio et al., in preparation) based on the estimation of the atmospheric OPD in another spectral band can overcome this limitation and, therefore, improve the sensitivity limit of the coherent flux and differential phase.
\end{itemize}

\section{Illustrations of MATISSE capabilities and performance}
\label{secastro}
To illustrate MATISSE capabilities and performance, we present examples of measurements on objects of astrophysical interest, used as test cases during the MATISSE commissioning. 



\subsection{Image reconstruction of circumstellar disks}
\begin{figure*}[h!]
\centering
\includegraphics[width=0.6\textwidth]{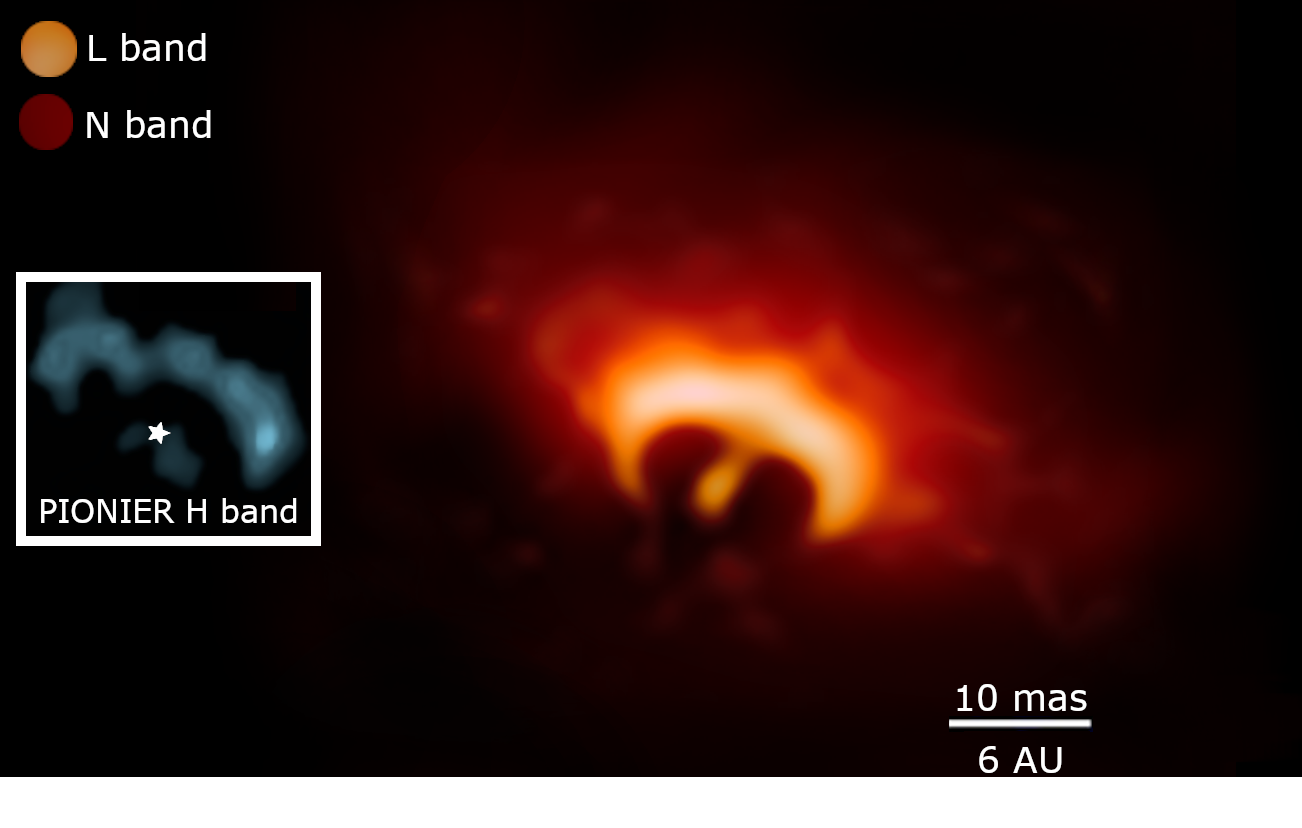}
\includegraphics[width=0.39\textwidth]{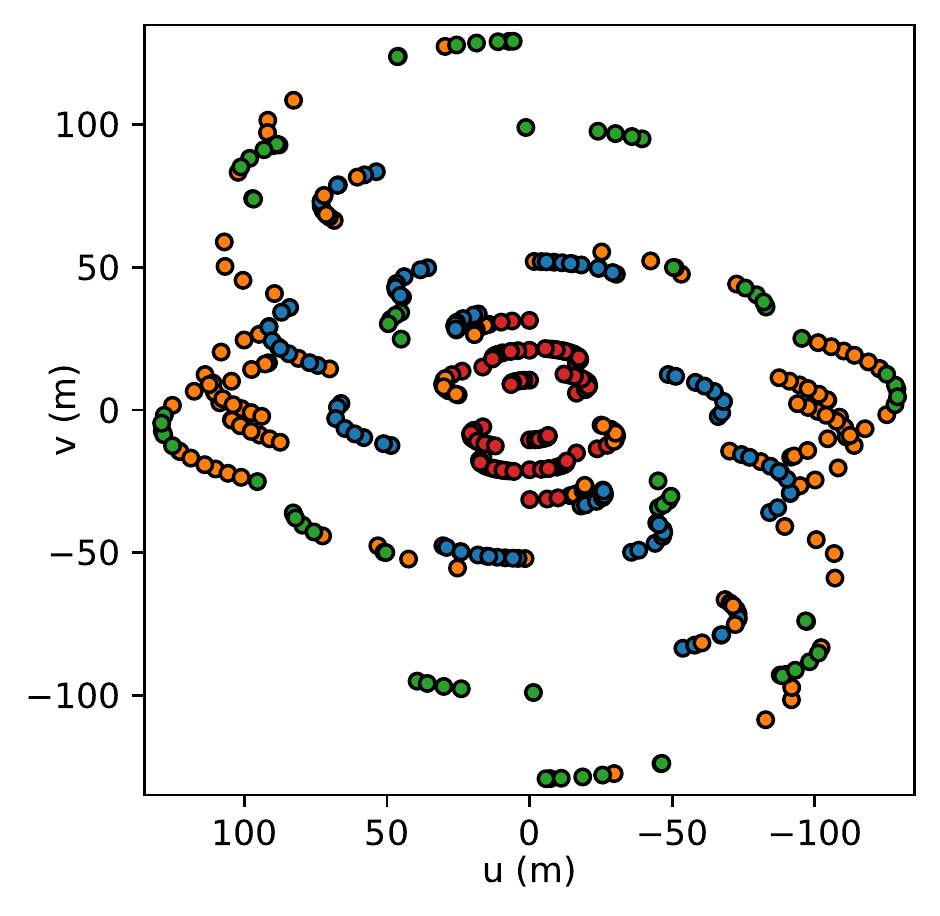}
      \caption{Image reconstruction of FS CMa. Left: FS CMa circumstellar disk image from MATISSE L- and N-band observations using MATISSE DRS image reconstruction software IRBis. The H band VLTI/PIONIER image from \cite{2020A&A...636A.116K} is shown as an inset at the same scale for comparison. Right: Corresponding uv-plane coverage with measurements obtained using the small (red), medium (blue), large (green), and three different intermediate AT configurations (orange).}
\label{Fig:FSCMa}
\end{figure*}
FS CMa is a hot star surrounded by a complex circumstellar environment. It exhibits the B[e] phenomenon, defined by a strong infrared excess due to dust located in a disk-like structure, emission lines of hydrogen produced in a dense gaseous environment, and forbidden emission lines of various elements including [FeII] and [OI] that are thought to be produced in a highly illuminated and very diluted environment \citep[][]{Lamers1998, Zickgraf1998}. Its evolutionary status remains highly debated. It is not yet clear if the disk is the remnant of a protostellar disk, in which case FS CMa would be a Herbig Be star, or if its material has been ejected from the stellar surface, either because of a combination of fast-rotation and radiative pressure \citep{Lamers1991}, or because of an interaction with an undiscovered companion as of yet.\\
FS CMa was selected to be one of the test objects to probe MATISSE imaging capability in both L and N bands. The star was observed during 12 commissioning nights in December 2018. Forty measurements representing a total of 40 x 6 uv points were obtained on this target. The data were reduced with the MATISSE standard DRS, and wide-band images in L and N bands were reconstructed separately using the MATISSE image reconstruction software IRBis \citep{2014A&A...565A..48H}\footnote{Image reconstructions were also attempted independently using MIRA \citep{Thiebaut2010}, exhibiting the same structures.}. \\
These first images are presented in the composite of Fig.~\ref{Fig:FSCMa} (yellow: L band, red: N band). In both images, the inner rim of the dusty disk is clearly resolved, and due to opacity effects and the intermediate inclination angle of the object, the rim appears skewed. As longer wavelengths probe colder material in the disk, the object appears much larger in the N band.\\
 In Fig.~\ref{Fig:FSCMa}, we inserted the H band VLTI/PIONIER image from \cite{2020A&A...636A.116K}, which shows a narrow inner rim as the H band is only sensitive to very hot dust close to the sublimation temperature. The PIONIER and MATISSE images are in good agreement in terms of the geometry of the inner rim (size, orientation, and skew). The ability of MATISSE to image objects from 3 to 13~$\mu$m, probing dust temperatures from 1000 to 200~K, is a very powerful tool to put constraints on the physical properties (density and temperature) of the dusty circumstellar environment. The full analysis of this dataset with chromatic image reconstruction and radiative transfer modelling is presented in \cite{Hofmann2021}.\\

\subsection{Low coherent flux observations and image reconstruction of AGN}
\begin{figure*}
\centering\includegraphics[width=0.2\textwidth]{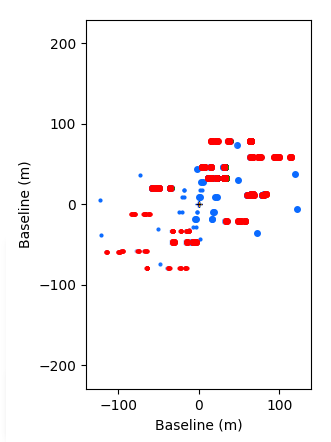}
\centering\includegraphics[width=0.79\textwidth]{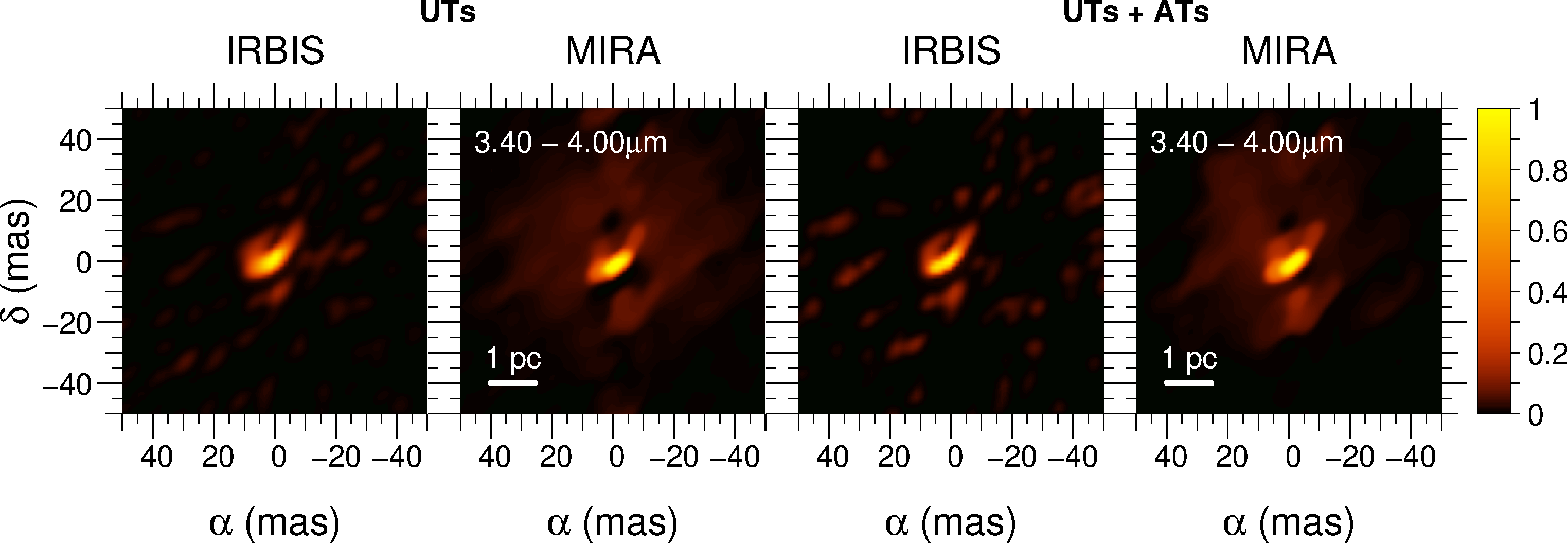}
\caption{{NGC1068 and the advantage for the image restoration of combining low frequency AT data with UT data. The left image shows the u-v coverage, with the AT baselines as blue dots, and the UT baselines are represented as red dots. The two left image reconstructions are with UT baselines only, while the two right ones combine the baselines of UTs and ATs. The IRBis images are grey reconstructions over the 3.4-4.0 $\mu m$ interval. The MIRA images are medians of all the 0.1 $\mu m$ obtained between 3.4 and 4.0 $\mu m$.}} %
\label{Fig:NGC_uv_UT_AT}
\end{figure*}

{One of the key programmes of MATISSE is the study of the inner dust environment of AGNs. These targets are challenging for the VLTI observations, involving the closing of the adaptive optics loops, since the AGNs are faint and extended in the visible. They are also faint and sometimes nebulous in the H and K bands, making them more complicated for image acquisition in the focal laboratory.  Finally, they are also faint in the MATISSE L and N bands. As a test of the MATISSE potential and limits for this class of targets, we observed the Seyfert 2 galaxy NGC~1068 with the ATs and the UTs during commissioning and during some of our first guaranteed time observations. ATs only appeared to be usable in the 'small configuration' that has a maximum baseline of 34 m. For longer baselines, the source is quite resolved, and the coherent flux is below the 300 mJy limit for MATISSE fringe tracking in the L band with the ATs. In total, we obtained 24 visibilities and 16 closure phases on UTs and six baselines and four closure phases on ATs.}
{In the north-eastern half of the L band image displayed in Fig.~\ref{Fig:NGC_uv_UT_AT}, a ring-like structure is observed.}

This L band image illustrates the possibility of obtaining complex high angular resolution images in the mid-infrared with a relatively small number of u-v points on rather faint objects. It also illustrates the importance of the short baselines in the image reconstruction. Fig.~\ref{Fig:NGC_uv_UT_AT} shows the strong improvement provided by the AT baselines that provide a better control of the diffuse structures that can contaminate the UT narrower field. This advantage comes with a complication due to the combination of fields of different sizes between ATs and UTs.
Detailed analysis of the L-,M-, and N-band data on NGC~1068 have provided new insights into the relation between its dusty strutures and its central supermassive black hole  \citep{Gamez2021}.

{A sample of other type 2 and type 1 AGN has been successfully observed, both during the commissioning and during the guaranteed time. The less than 60 mJy L band sensitivity of MATISSE gives access to a sample of AGN potentially larger than that observed by MIDI \citep{Burtscher2013}, particularly the type 1 AGN with hot, unobscured nuclei.  This increase in sensitivity will be partially limited by the current VLTI adaptive optics, but it nevertheless should allow mapping of AGN over a broader luminosity range than previously possible, and in turn opens the possibility of a wide analysis of the relative interactions of radiation and gravity on the dust accretion and polar wind launching mechanisms.}

\subsection{Probing the gas disks}
\begin{figure*}[h!]
\centering
\includegraphics[width=0.8\textwidth]{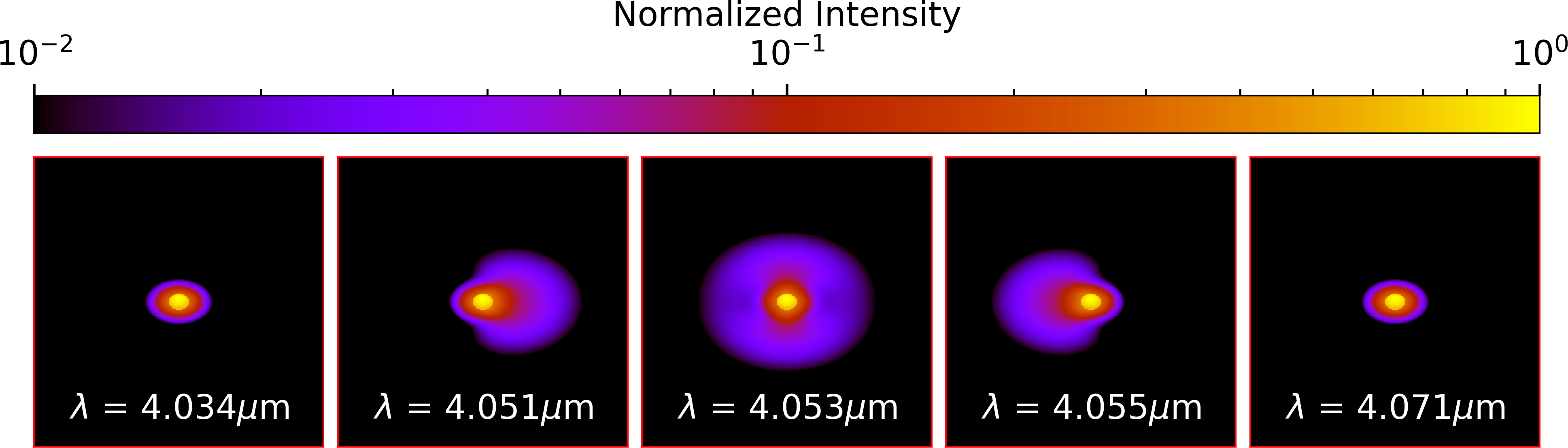}
\centering\includegraphics[width=0.8\textwidth]{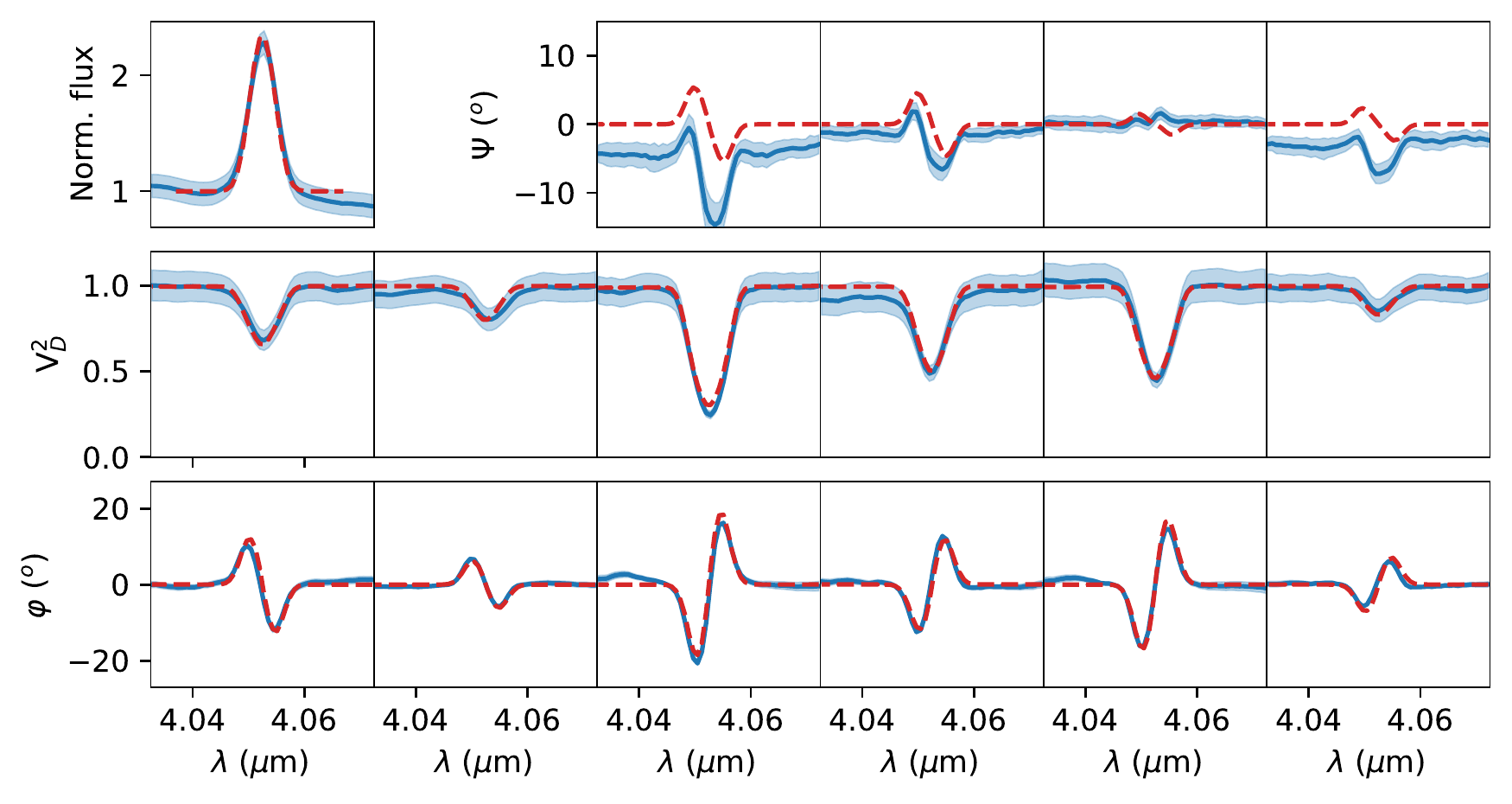}
\caption{MATISSE high spectral resolution (R$\sim$1000) observation of the Be star $\alpha$ Arae. Comparison of the observation in the 4.053$\mu$m Br$\alpha$ emission line (solid blue line) with the best-fit of a rotating disk toy model (red dashed line). Top: Images of the toy model at different wavelengths in the continuum and through the line. Bottom: All observable quantities from a single MATISSE observation, including the normalised spectrum, the four closure phases,  the six squared visibilities, and the six differential phases.}
\label{Fig:Aara}
\end{figure*}

Classical Be stars are fast-rotating stars showing an IR excess and emission lines (mainly of hydrogen) that stem from a gaseous equatorial disk. $\alpha$ Arae is one of the brightest stars of this type and, as a bright object with narrow spectral features in emission, it was a logical choice to test MATISSE's spectro-interferometric capability in its higher spectral modes. In 2005, it had been used to test a similar capability for the first generation instrument AMBER. These medium resolution data (R$\sim$1500) centred in the Br$\gamma$ line at 2.16~$\mu$m, published in \cite{Meilland2007}, helped to constrain the geometry, kinematics, and physics of the circumstellar disk, showing unambiguously for the first time that the disk was in Keplerian rotation with negligible expansion.\\
We observed $\alpha$ Arae several times during MATISSE commissioning in various instrumental configurations and both with the ATs and UTs. Here, we present an example of L band HIGH resolution (R$\sim$1000) data centred on the Br$\alpha$ emission line (4.053~$\mu$m). The observation was carried out on March 22, 2018 during MATISSE's first commissioning on the UTs. The normalised flux, closure phases, square visibility, and differential phases through the emission line are presented in Fig.~\ref{Fig:Aara}. \\
We modelled the data using a rotating disk described in detail in \cite{Meilland2012}. In order to check the capability of MATISSE to constrain the geometry and kinematics of the circumstellar environment in emission lines, we performed a fit on the following three free parameters: the object inclination angle, as well as the disk extension in the line ($\theta_{l}$) and position angle (PA). Values for other parameters were taken from \citet{Meilland2012}. To perform an automatic fit and estimate the parameter uncertainties, we used emcee\footnote{Available at https://emcee.readthedocs.io/en/stable/}, a python implementation of Goodman \& Weare’s Affine Invariant Markov chain Monte Carlo (MCMC) ensemble sampler.\\
The inclination angle and position angle determined from the fit, i=50.5$\pm$7.2$^o$ and PA=92.5$\pm$8.5$^o$, are fully compatible with the one found in \citet{Meilland2012} when analysing AMBER high resolution (R$\sim$12000) data. 
The size of the emission in Br$\alpha$, 8.1$\pm$0.8D$_\star$, is significantly larger than the one found in Br$\gamma$ 5.8$\pm$0.5D$_\star$, as expected from the population of corresponding energy levels in a hot circumstellar disk.\\
We note that the uncertainties on the parameters are only slightly larger than the one derived from the AMBER observation despite the 12 times lower spectral resolution of MATISSE. This clearly demonstrates that MATISSE HIGH resolution mode allows us to put significant constraints on the geometry and kinematics of a circumstellar gas with velocities of the order of a few hundreds of $km/s$. We note that the non-zero closure phase in the continuum as well as its asymmetric variation in the line, which is linked with asymmetries in the circumstellar disk, cannot be reproduced in our simple model. A dedicated modelling of all $\alpha$ Arae and other Be stars observed with MATISSE during the commissioning will be presented in a forthcoming paper.

One of the great advantages of MATISSE, in particular regarding the study of circumstellar environments, is its ability to observe simultaneously in the L, M, and N bands, probing different regions in the environment and putting additional constraints on its physical properties. Hence, during the high resolution observation of $\alpha$ Arae, N band data were also recorded and were compared, for instance, to the MIDI observations published in \cite{2009A&A...505..687M}. \\

\begin{figure*}[h!]
\centering
\centering\includegraphics[width=0.35\textwidth]{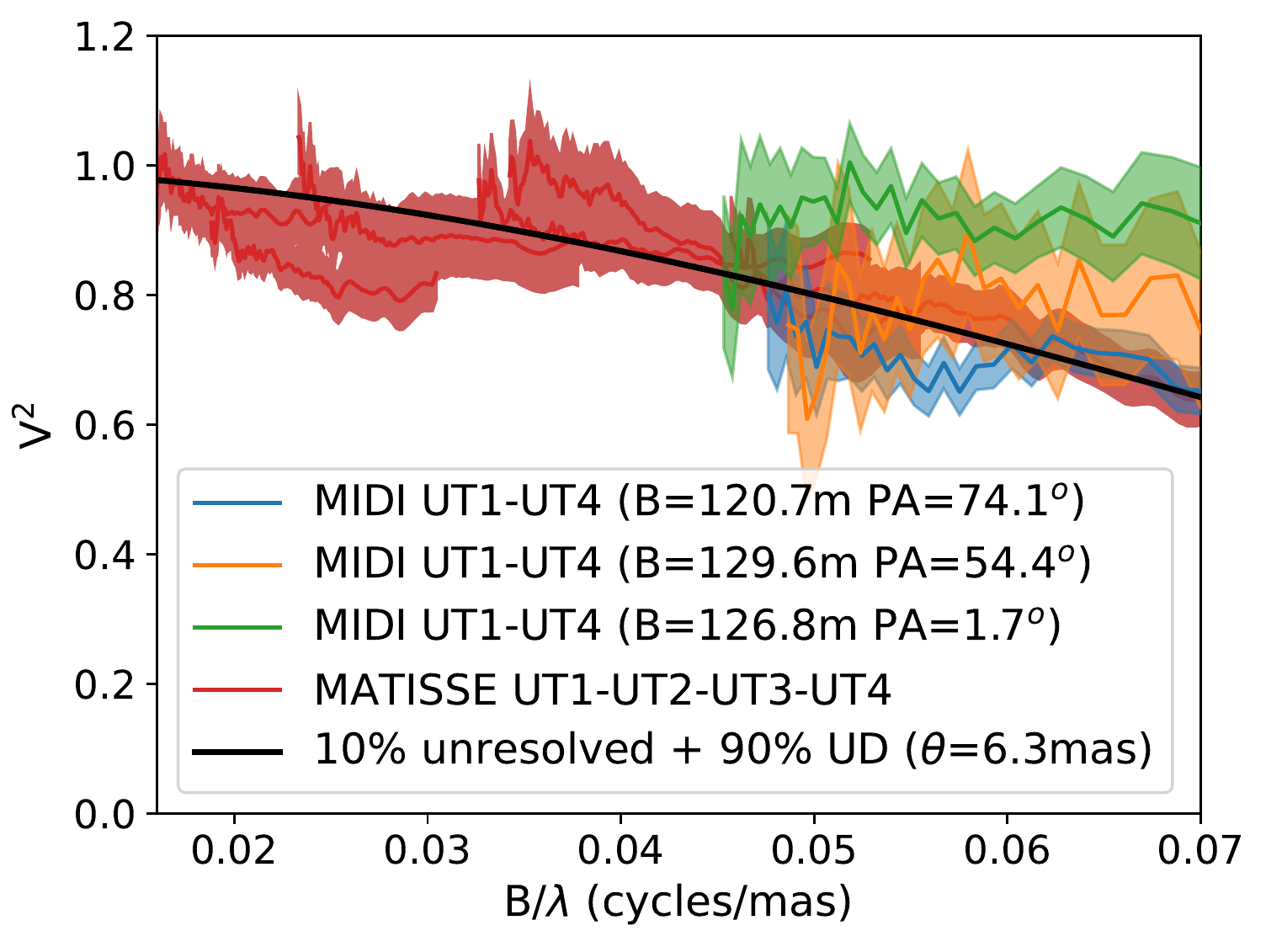}
\centering\includegraphics[width=0.285\textwidth]{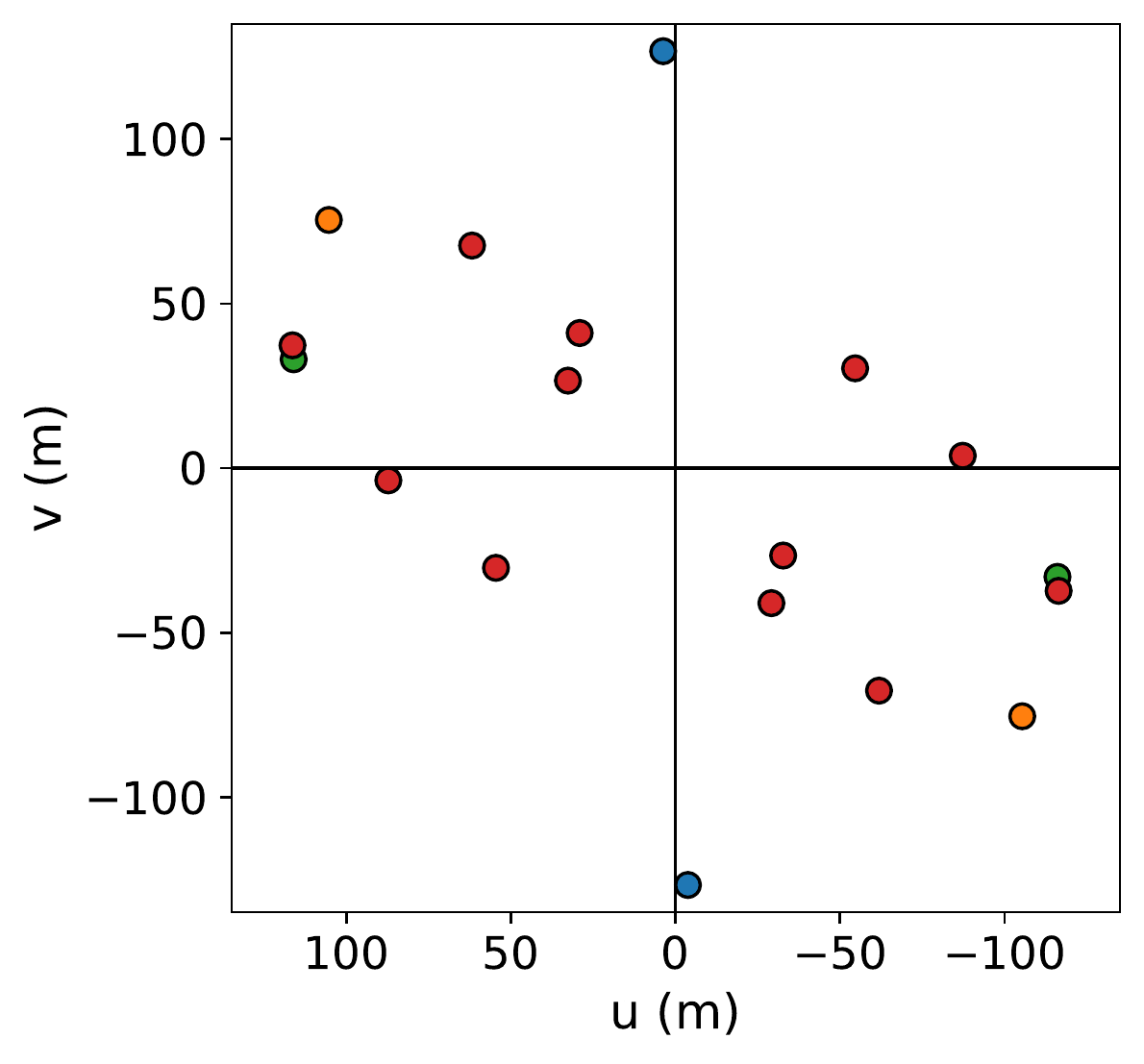}
\centering\includegraphics[width=0.345\textwidth]{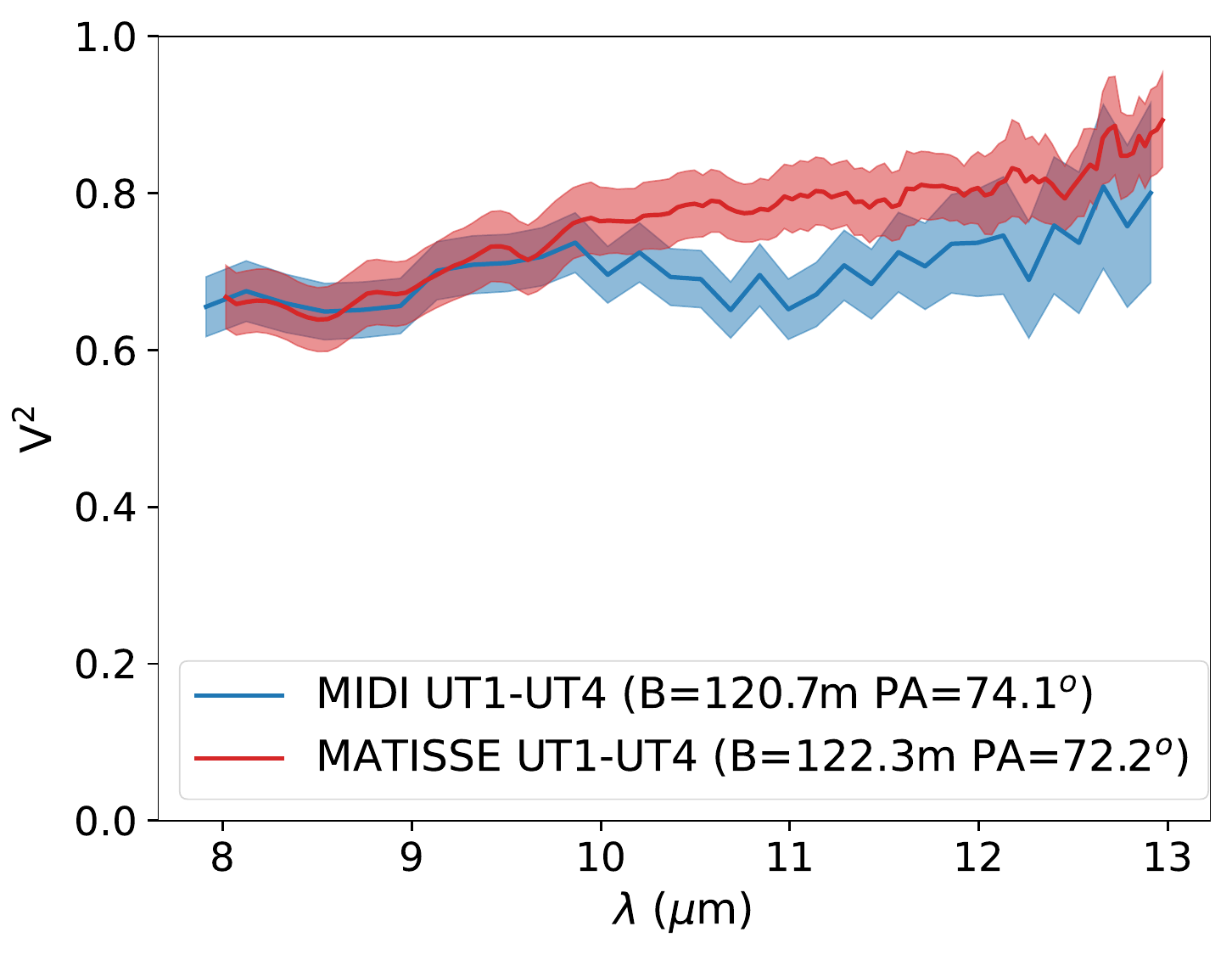}
\caption{Comparison between MATISSE N band and MIDI data obtained on the classical Be star $\alpha$ Arae using the UTs. Left: All data plotted as a function of the spatial frequency with a best unresolved + uniform disk model. Centre: Corresponding uv plane coverage with the same colours for MIDI and MATISSE observations. Right: Data plotted as a function of the wavelength for two similar baselines in terms of length and orientation.}
\label{Fig:aara_matisse_midi}
\end{figure*}

Fig.~\ref{Fig:aara_matisse_midi} summarises this comparison between our MATISSE observation with all UTs and the three MIDI observations with the UT1-UT4 obtained at various position angles (PA). MATISSE clearly resolved the object, and in modelling it as a simple unresolved star contributing to 10\% of the flux (the value derived from a simple fit of the SED with a Kurucz model) and a uniform disk component for the circumstellar disk, we derived a value of 6.23$\pm$0.08 mas (9.8$\pm$0.2 R$_\star$).\\
Looking closer at the data from two specific baselines, one from MATISSE and the other from MIDI,  with the similar projected length and orientation, we clearly see that they are fully compatible between 8 and 10~$\mu$m with a difference smaller than 1$\sigma$, and that they agree within 2$\sigma$ above this limit. We also note that the uncertainties on the measurements are also of the same order, that is between 5 to 10\% of the calibrated V$^2$. As already stated in the previous section, this data will be analysed in a forthcoming paper on the first MATISSE observation of classical Be stars.

\begin{figure}[h!]
\includegraphics[width=0.44\textwidth]{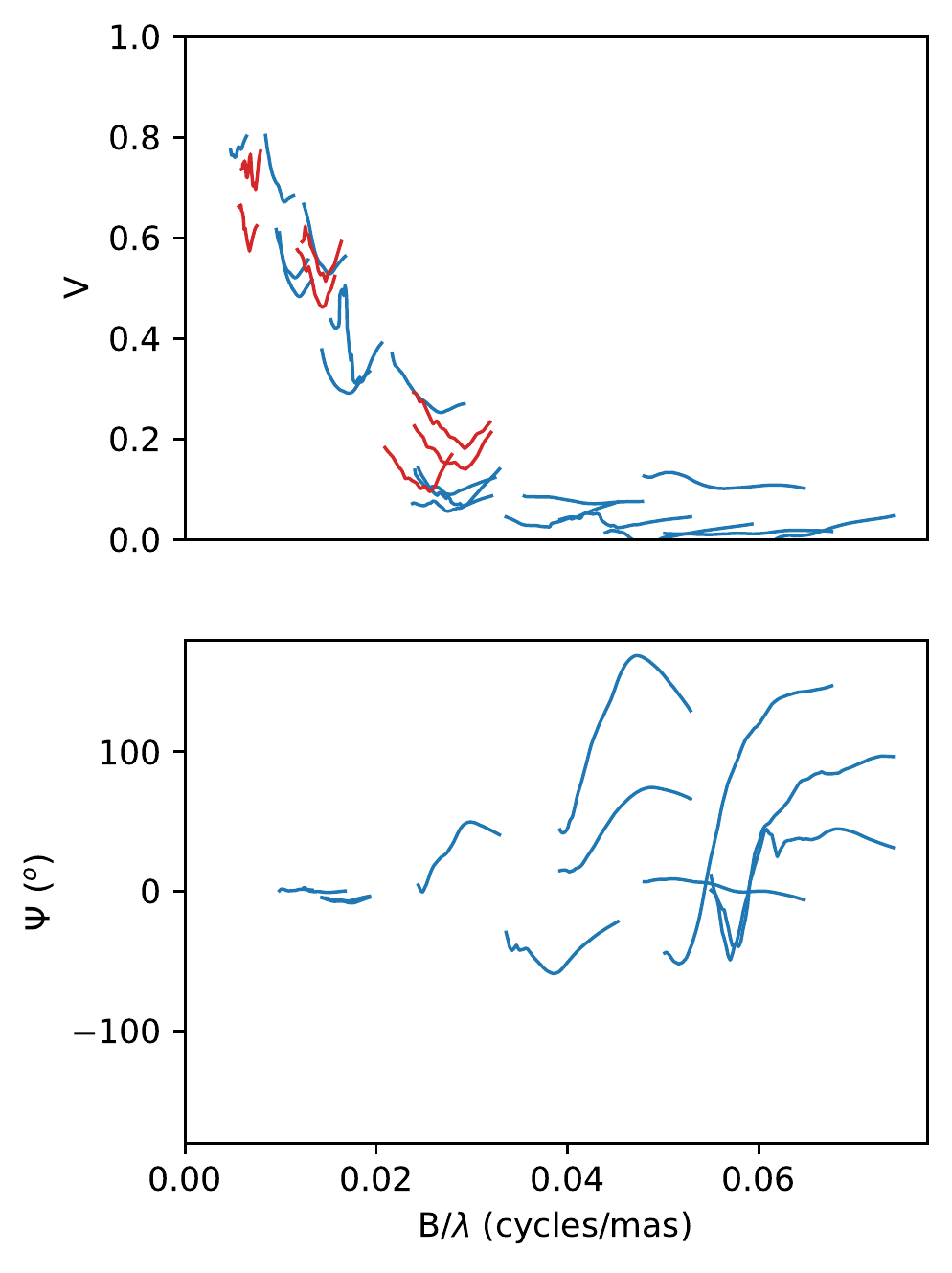}
\caption{Comparison between MATISSE N band (blue) and MIDI (red) data obtained on the A[e] supergiant l Pup. MATISSE data were taken during the GTO survey on B[e] stars and MIDI ones are from \cite{2010A&A...512A..73M}}
\label{Fig:lpup_matisse_midi}
\end{figure}

To illustrate the MATISSE and MIDI comparison in the N band further, in Fig.~\ref{Fig:lpup_matisse_midi}, we present visibility and closure phases from the A[e] supergiant star l Pup conjointly with MIDI visibility on the same object published in \cite{2010A&A...512A..73M}. MATISSE data were recorded in February 2020 during the MATISSE GTO survey on B[e] stars. Three measurements were obtained, one with each of the offered AT configurations (small, medium, and large). MIDI and MATISSE data are in full agreement. MIDI data already allowed us to put strong constraints on the circumstellar environment geometry, showing that the material was indeed, as suspected, in a circumstellar disk. But the lack of closure phases
prevented the authors from determining the disk's vertical stratification temperature law. 
As seen in Fig.~\ref{Fig:lpup_matisse_midi}, this limitation has been overcome.

\subsection{Spatially resolved spectroscopy of solid-state features}
\begin{figure}
\centering
\includegraphics[width=0.45\textwidth]{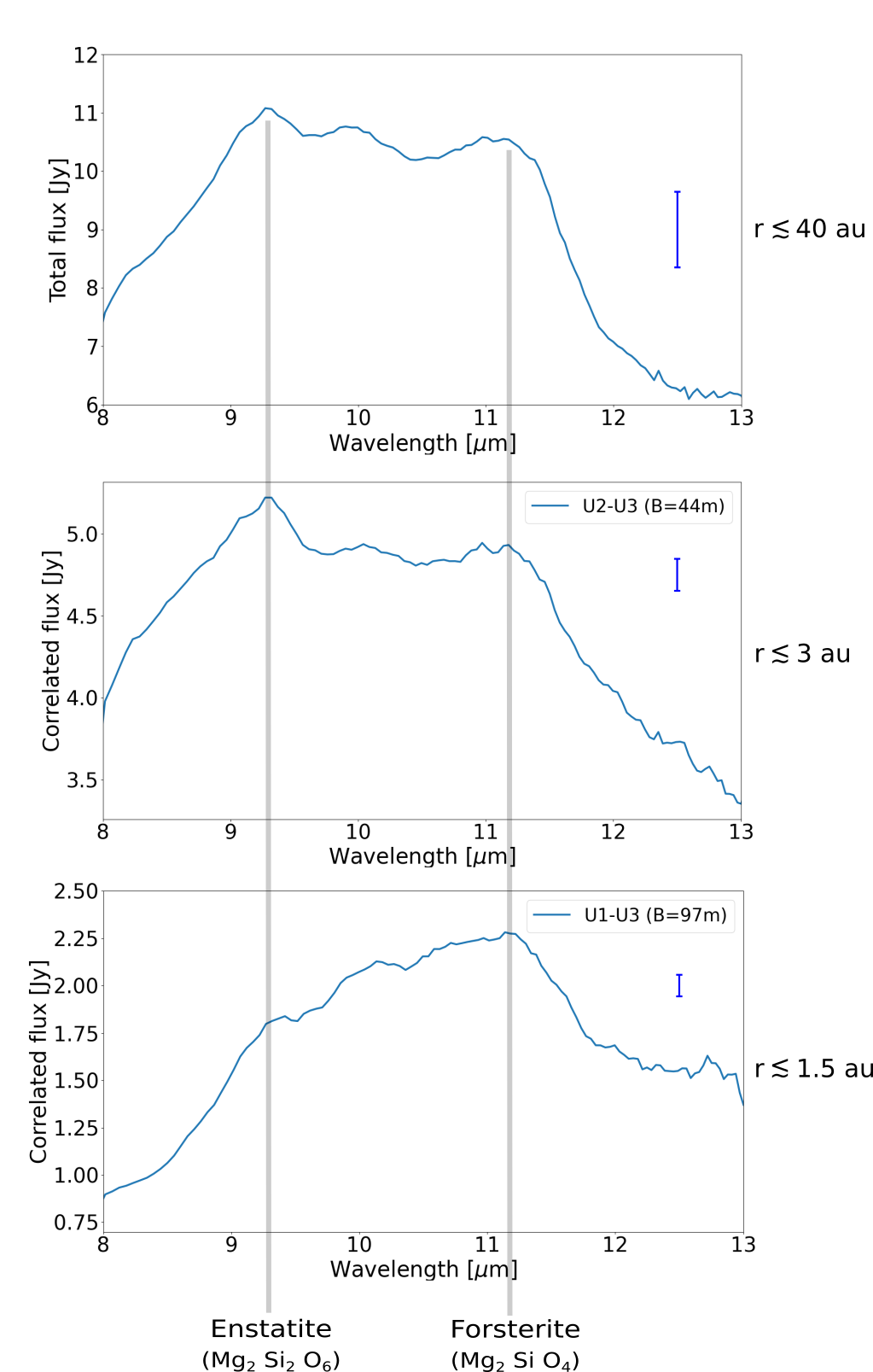}
      \caption{Left, from top to bottom: Low-resolution (R=30) N band total spectrum and correlated spectra of the Herbig Ae star HD~142527 obtained with MATISSE. The two correlated spectra shown here correspond to increasing baseline lengths (B=44m and B=97m). The disk region probed by the different spectra is indicated for each panel. The uncertainties affecting the measured spectra are indicated by the single error bars in the top right-hand corner of each plot.}
\label{Fig:HD142527}
\end{figure}

\begin{figure}
\centering
\includegraphics[width=0.45\textwidth]{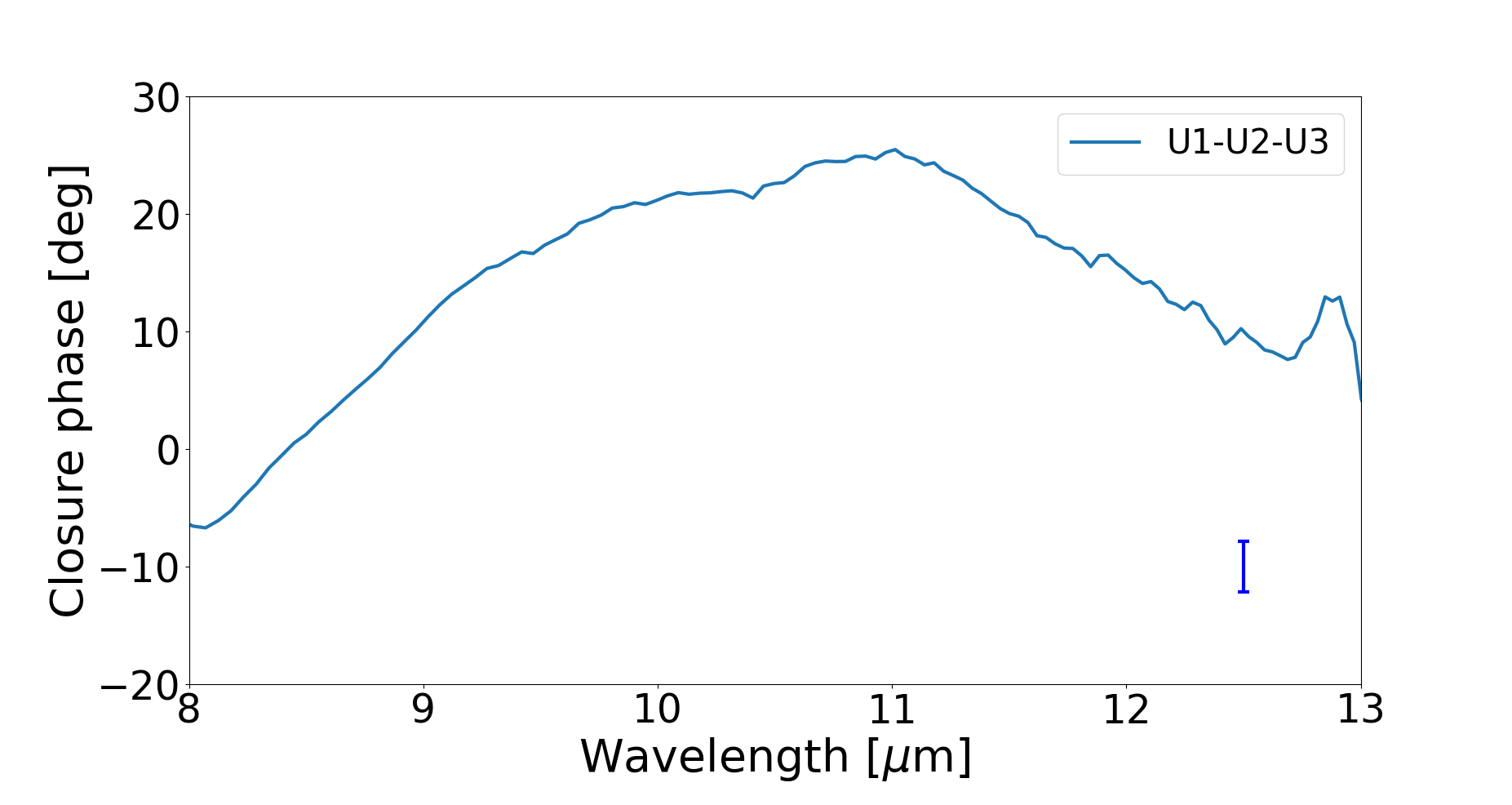}
      \caption{Example of low-resolution (R=30) N-band closure phase of HD142527 obtained with MATISSE. The bar in the bottom right part represents the average error level on the data.}
\label{Fig:HD142527_clo}
\end{figure}
Thanks to its spectroscopic and angular resolution capabilities, MATISSE can provide N band spectroscopic observations, down to the sub-au scale, of various solid-state features of hot and warm dust. This is especially important for studying the fine dust chemistry and mineralogy of protoplanetary disks in their inner region (0.1 to 10 au), which is the expected birthplace of telluric planets. Spatially resolving the chemical composition of the planet building blocks is thus a key aspect MATISSE can tackle in the context of understanding the formation of planetary systems and their diversity. For illustration purposes, in Fig.~\ref{Fig:HD142527}, we show the low spectral resolution (R=30) calibrated N band total spectrum and two of the six correlated spectra obtained on the Herbig Ae star HD142527 during the first MATISSE commissioning run on May 23, 2018 with the UTs. HD142527 is an intermediate-mass young star that is surrounded by a dusty disk made of a purported warped inner part, a large 100 au-wide gap, and an outer disk with large-scale structures \citep{2014ApJ...781...87A}. A low-mass companion was also found orbiting the central star at about 11 au, just outside the inner disk \citep{2018A&A...617A..37C,2019A&A...622A..96C}. Previous mid-infrared observations by the former VLTI instrument MIDI and the Spitzer space telescope suggested a high degree of crystallinity of the dust in both the inner and outer regions \citep{2004Natur.432..479V,2010ApJ...721..431J}. That is confirmed by our MATISSE data. As shown in Fig.~\ref{Fig:HD142527}, the three spectra clearly show prominent crystalline silicate features from enstatite (around 9.3~$\mu$m) and forsterite (at 11.3~$\mu$m). Moreover, our data suggest a radial variation in dust composition with a noticeable change in the relative amplitude of both features as the baseline length increases (i.e. as we get closer to the star). {In addition, Fig.~\ref{Fig:HD142527_clo} shows a significant closure phase signal in the silicate emission band, which suggests asymmetries in the silicate grains emission.}\\
The fine radial changes that MATISSE can thus probe in the dust composition represent direct constraints on the condensation sequence of solids in that disk and thus on its bulk composition \citep[see ][]{2020MNRAS.497.2540M}, which will then drive the composition of future planets. In that context, as shown in \citet[]{2020MNRAS.497.2540M}, the N band high spectral resolution mode (R=220) of MATISSE will be key to identify complex solid-state features accurately and to derive the detailed dust mineralogy and its distribution in the planet-forming region of disks.

\subsection{High accuracy stellar diameters}
\begin{figure}[htbp]
\begin{center}
\includegraphics[width=0.5\textwidth]{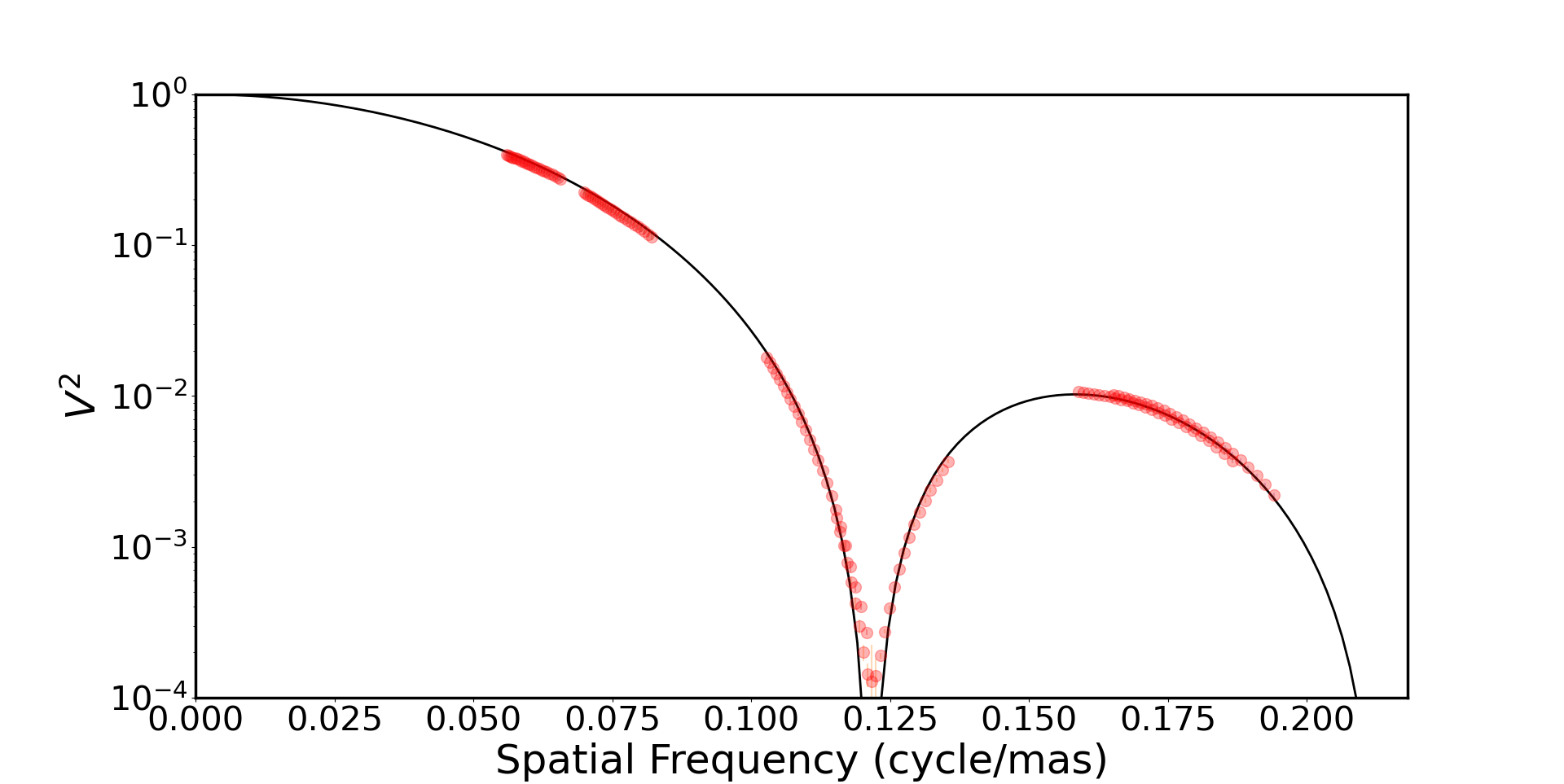}
\includegraphics[width=0.5\textwidth]{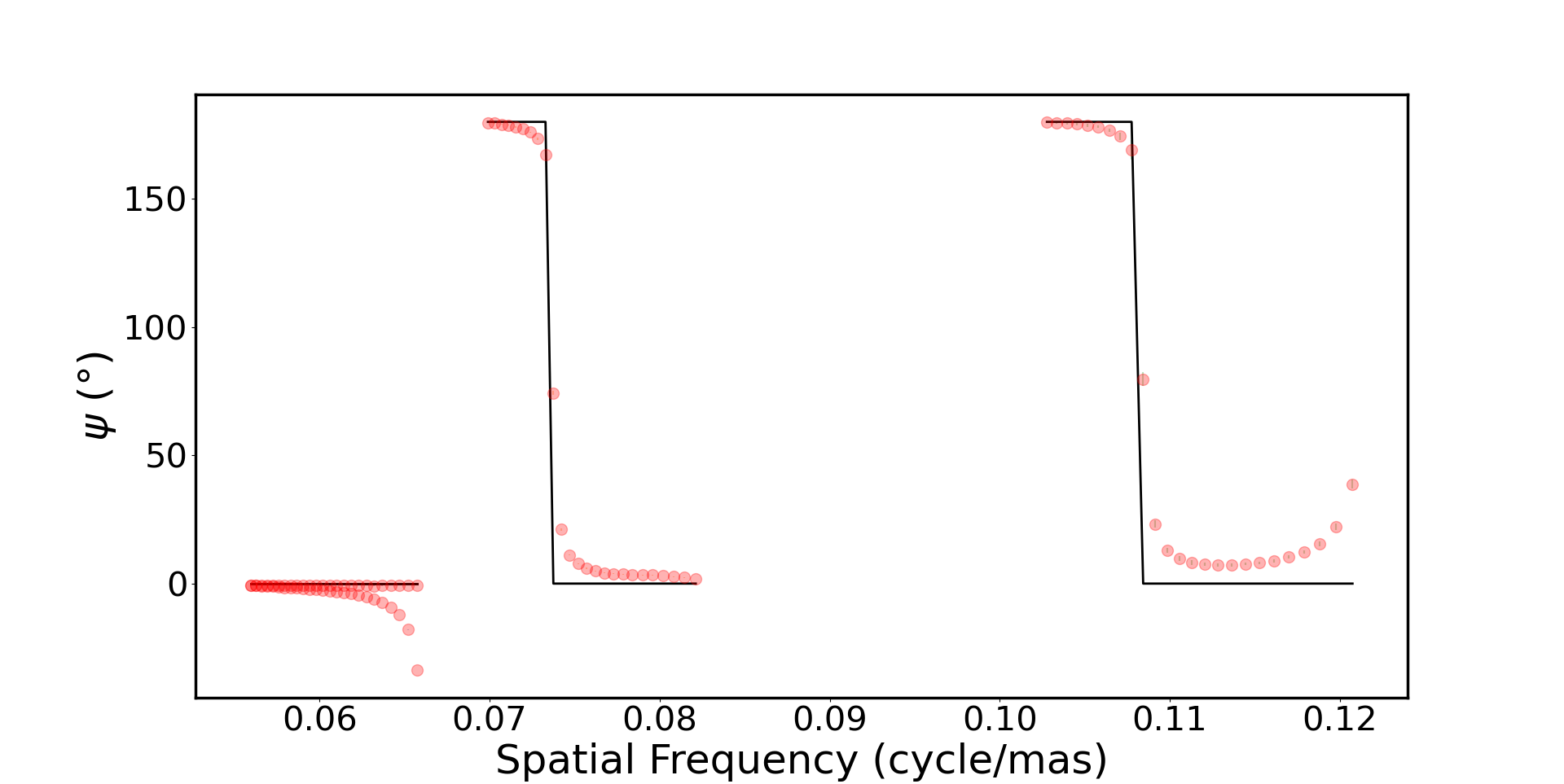}
\caption{Observation of $\delta Vir$ in the L band. Top: Measured squared visibilities (red) and limb-darkened model. Bottom: Measured closure phases (red) and limb-darkened model. For the closure phase, the spatial frequency corresponds to the longest baseline of the triplet}
\label{fig:delVir}
\end{center}
\end{figure}

In the night of May 20, 2018, we observed the star $\delta$ Vir, which is a red giant branch star of spectral type M3III. Fig.\ref{fig:delVir} presents the measured visibility and closure phase in the L band. These measurements were calibrated using the stars $\psi$ Vir and $\theta$ Cen, {for which the angular diameters were more accurately determined using the 'auto-calibration' method developed by  \citet{Robbe-Dubois}}. We adjusted a linear limb-darkened model to the MATISSE visibility and closure phase. We find a diameter of {10.565 $\pm$ 0.03~mas}, with a darkening coefficient of {0.33 $\pm$ 0.02}. The JSDC catalogue by \citet{Chelli2016} predicts a LDD diameter of $10.709$ mas, which is compatible with the MATISSE estimated diameter. This result illustrates the ability of MATISSE to measure accurate angular diameters and therefore to contribute to studies of fundamental importance for stellar physics.

\subsection{Binary star observations}
\label{binary}
\subsubsection{High-contrast binary systems}
\label{binary1}
Achernar is an intermittent emission-line B-type star (Be), which presents alternatively pure absorption Balmer lines, or weak emission, concomitant with the creation of a circumstellar disk \citep{2017A&A...601A.118D}. This star has a companion detected in the IR \citep{2007A&A...474L..49K}, and it is thought that its disk formation is related to periastron passages of that companion. The relative flux ratio between the two stars is measured to be roughly 30 in the near-infrared, and 56 at 11.25~$\mu$m \citep{2008A&A...484L..13K}.

\begin{figure}[htbp]
            {\includegraphics[width=0.49\textwidth]{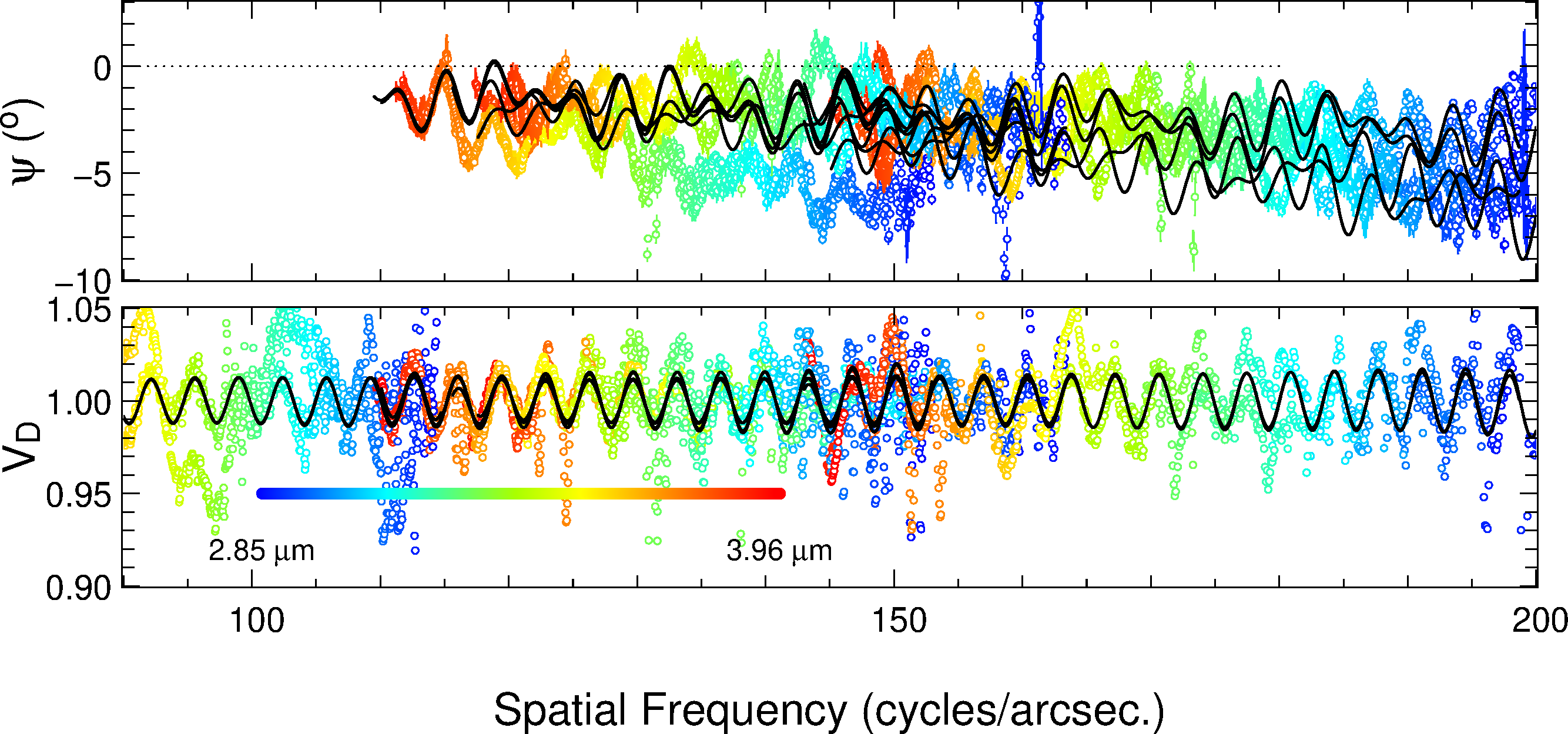}}
      \caption{Achernar differential visibilities and closure phases. The MATISSE data plot as a function of spatial frequencies, projected onto the direction of the binary star (-51$^\circ$), is in colour. The best-fit model is shown as a black line.}
      \label{puderef}
\end{figure}
We observed this star during the MATISSE commissioning (November 2018), as well as during the GRA4MAT fringe tracker commissioning (September 20, 21 and 25, 2019) in low, medium, and high spectral resolutions in the L band.
We focus here on the medium spectral resolution data from September 2019 that presents the best overall quality. Fig. \ref{puderef} shows, in colour, a subset of the MATISSE closure phases as well as the differential visibilities as a function of spatial frequency, projected onto the direction of the binary star (-51$^\circ$). These visibilities exhibit a  cosine modulation that is typical for a binary source. We modelled the system using an ad hoc model for the central star (an elongated uniform disk and an offset Gaussian disk, mainly affecting the overall closure phase offset of a few degrees) plus a point source representing the companion star. The model visibilities for a $\approx$1:100 flux ratio binary, with a separation of 293\,mas, are overplotted on top of the closure phases and differential visibilities as the black lines in Fig.~\ref{puderef}, showing the overall good match with the observations.




\subsubsection{Binary system parameters in the presence of a disk}
$\delta$ Sco, the famous eccentric binary Be star and its associated disk \citep{Suffak2020} was also observed in the course of the MATISSE commissioning during the night of February 20, 2018. The observations were carried out in low spectral resolution in the L band. We focus here on the binary star and only use the closure-phase observables since the visibilities mix information from both the binary and the disk.

\begin{figure*}[htbp]
            {\includegraphics[width=0.65\textwidth]{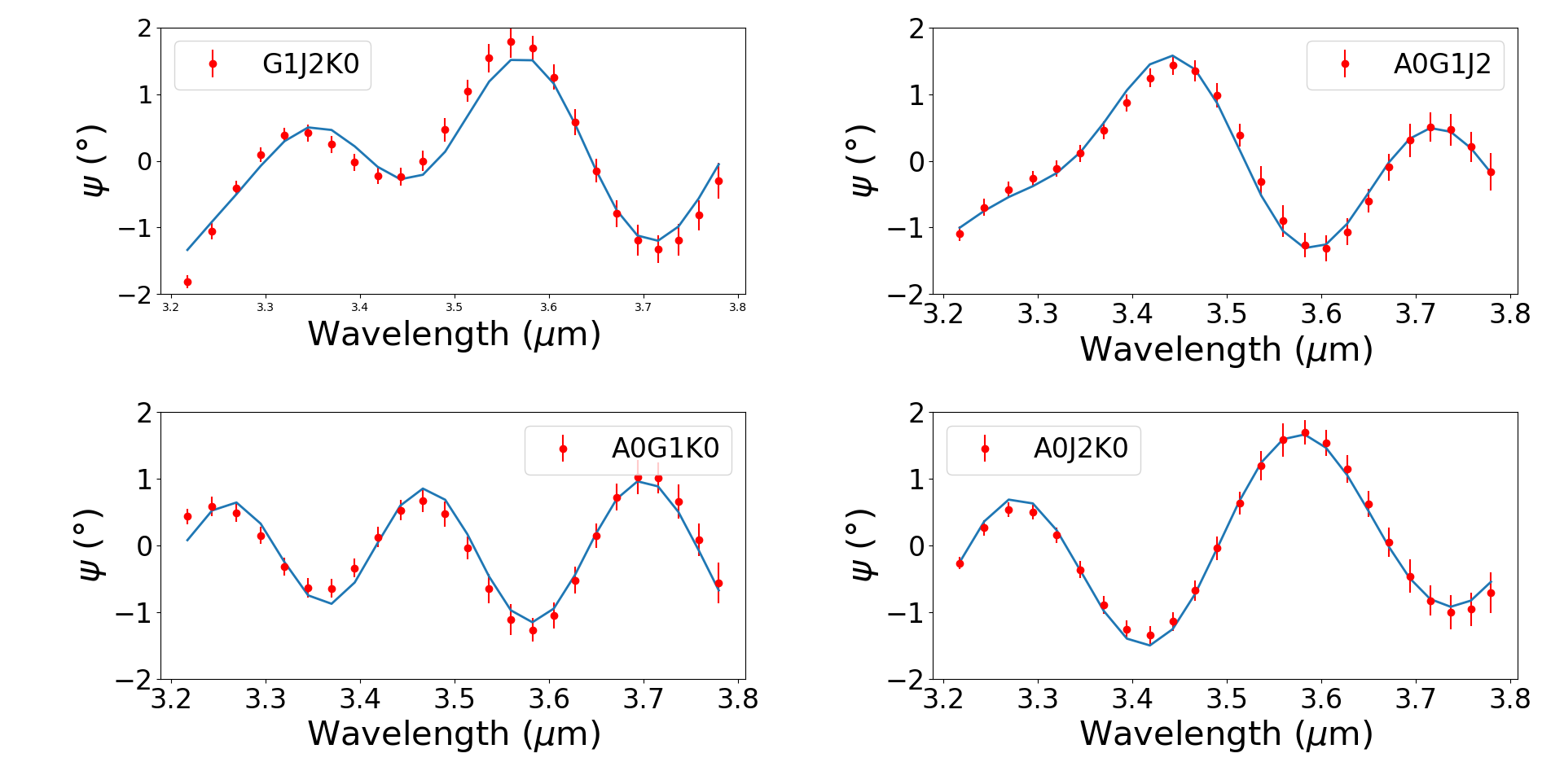}}
            {\includegraphics[width=0.35\textwidth]{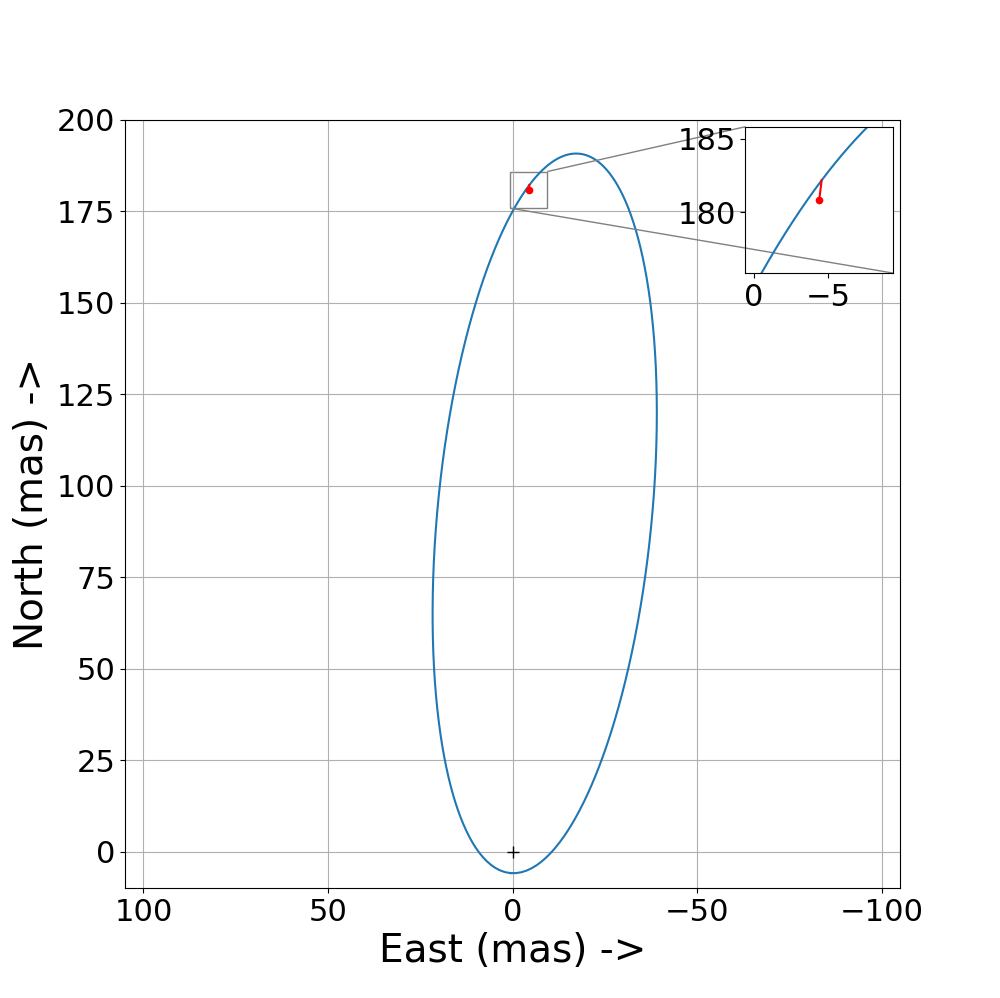}}
      \caption{Observation of $\delta$ Sco with MATISSE in L the band. Left panels: $\delta$ Sco closure phases (red ) and the best-fit binary star model (blue). Right panel: Comparison between the \cite{tycner2011} orbit (blue) and our astrometric point (red).}
         \label{fig:deltasco}
\end{figure*}

Using a two-point source model, we estimated the position (X,Y) of the secondary around the primary star and the flux ratio (R) between the two stars by adjusting this mode to the observed closure phases (Fig. \ref{fig:deltasco}). We found the following: $X=-4.41\pm 0.05$~mas, $Y=180.84\pm  0.04$~mas, and $R=58.79\pm 0.9$.

The reduced $\chi^2$ we obtain is 0.74 for the spectral interval 3.5-3.8~$\mu$m. For the spectral interval 3.2-3.8~$\mu$m, we obtain values for the position and flux ratio within the uncertainty, but the $\chi^2$ raises up to 1.07. This increase in the $\chi^2$ value could mean that the flux ratio of the binary is varying as a function of wavelength within the L band. We then compared that astrometric point to the best orbital elements of $\delta$ Sco that can be found in \cite{tycner2011}. That comparison is shown in the right panel of Fig.~\ref{fig:deltasco}, and it shows the excellent agreement between that MATISSE point and the previous orbital solution.

\section{Perspective of the GRA4MAT extension and of the availability of the VHR mode}
\label{secapersp}
\begin{table}[htbp]
\caption{K band coherent magnitude limit for GRA4MAT operation.}
\label{tab:gra4matLIMITS}
\begin{tabular}{ccccc}
\hline\hline
Conditions       & Good & Fair  & Poor \\ \hline
Seeing and $\tau_0$     & 0.6 $"$; 5.2 ms       & 1.0 $"$; 3.2 ms & 1.4 $"$; 1.6 ms  &  \\ \hline
K coherent limit        & 8.5   & 8     & 7      \\ \hline

\hline\hline
\end{tabular}
\tablefoot{The seeing bins in that table are the standard ESO ones (see https://www.eso.org/sci/facilities/paranal/instruments/matisse/inst.html).}
\end{table} 

\begin{table}[htbp]
\caption{Limits in Jansky for performance of MATISSE with ATs and GRA4MAT, per spectral channel after one minute of observation.}
\centering
\label{tab:gra4matPERF}
\begin{tabular}{ccccccc}
\hline\hline

\multirow{2}{*}{Resolution}     
 &
     \multicolumn{3}{c}{$\psi$} & \multicolumn{3}{c}{$\varphi$} \\
 
    &  L & M & N &L & M & N\\ 
 \hline\hline
 LOW     &0.25  &1.0  & <5.0 & 0.17 &   0.7 & <2.0 \\
 \hline
MED     & 1.5   & nc & -        & 1.0  &        nc & - \\

\hline
HIGH & 3 & - & nc & 2 & - & nc \\ 
\hline
VHR & 17 & 25 & - &     20 &    25 & - \\ 
\hline\hline
\end{tabular}
\tablefoot{In L and M bands, we used 1~$s$ frame times in 
LR and 10~$s$ in all higher resolutions. The symbol '-' means that the combination resolution-versus-band does not exist, and 'nc' means non-commissioned.}
\end{table}   

The {GRA}VITY {for} {MAT}ISSE (GRA4MAT) mode uses the GRAVITY Fringe Tracker (GFT) to freeze the fringes and allow longer frame times in MATISSE. The GFT operates in the K band \citep{2019Lacour} and this yields a specific problem for observations in L, M, and N bands: one lambda-fringe jump in the GFT, which has a marginal effect on observations in the tracking band (i.e. for science with GRAVITY), has a strong effect on the contrast of long exposure fringes at different wavelengths. The number of fringe jumps increases near the GFT limiting magnitudes, but it remains acceptable up to 0.5 to 1 magnitude below the GFT limits alone. This yields the limiting K magnitudes for GRA4MAT given in  Table \ref{tab:gra4matLIMITS}, which are the ones allowing 10 second frame times in the L and M bands with less than 10\% of frames being affected by fringe jumps. 

\begin{figure*}[htbp]
\centering
{\includegraphics[width=\textwidth]{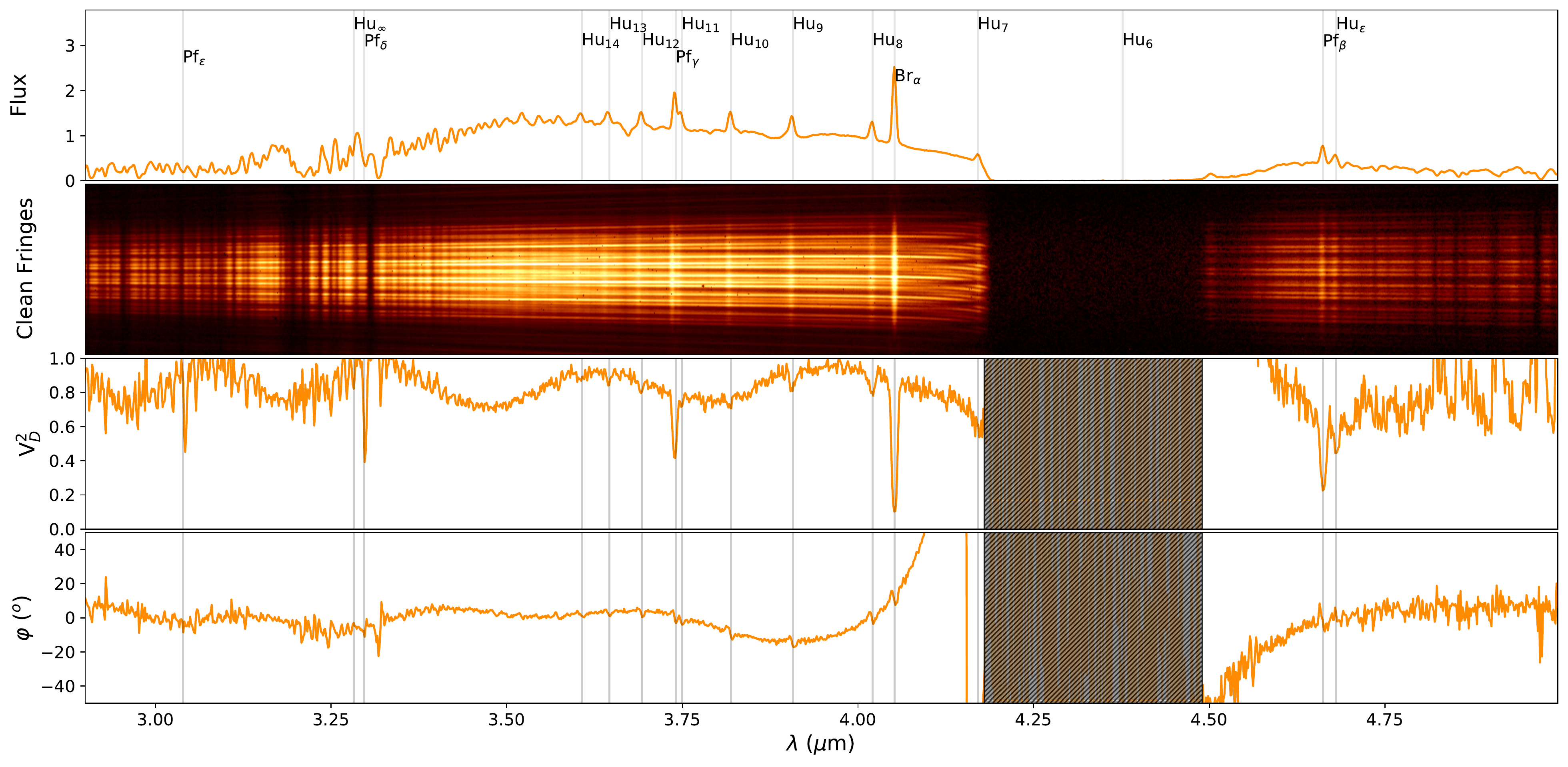}}
   \caption{L\&M bands medium resolution (R$\sim$500) observation of the Be binary $\delta$ Cen obtained in GRA4MAT mode. From top to bottom: Source spectrum, cleaned 10~s frame (i.e. with the mean sky removed) in the interferometric channel, differential squared visibility for one of the six baselines, and corresponding differential phase. The gap between the L and M band is shown in grey in the visibility and phase plots. The source spectrum shows a rich collection of hydrogen emission lines in Br$\alpha$ and the Humphrey and Pfund series.
   The interferometric channel of MATISSE composed of straight and stable high contrast fringes from 2.9 to 4.9 $\mu m$ shows a collection of bright source emission lines and dark telluric absorption lines.}
\label{fig:gra4mat}
\end{figure*}

GRA4MAT extends several of the MATISSE capabilities discussed hereafter, including the use of the very high-resolution (VHR) mode (R$\sim$3400 in L and M bands).
One immediate advantage of GRA4MAT is the gain in sensitivity for the MATISSE coherent flux, closure phase, and differential phase measurements. The improved sensitivity performance presented in Tab. \ref{tab:gra4matPERF} has to be compared to MATISSE stand-alone sensitivities displayed in Tab.
\ref{tab:fairandgood}. In L and M bands, observations are performed with a DIT of 1~s in LR and 10~s in MR, HR, and VHR, respectively.  In the N band, the DIT (20~ms in LR) remains constrained by the detector saturation due to the background level. However, in the N band, GRA4MAT allows the coherent integration analysis to be efficiently performed for a one minute duration, while the coherent integration with MATISSE stand-alone was limited to the duration of one modulation cycle equal to 200~ms. 
\\
Through DIT in L\&M bands that are long enough -- compared to the short DIT otherwise set by the coherence time of the atmosphere and limited to the reading of a few spectral channels -- the reading of the full detector width is now enabled for the interferometric and the photometric beams. 
This hence permits one to access all the spectral channels illuminating the detector. Fig. \ref{fig:gra4mat} illustrates this advantage of the Be binary star $\delta$ Cen, which shows the very rich information that can be extracted from a 10 s integration frame at medium spectral resolution. We can simultaneously observe many emission lines, each showing differential visibility and differential phase signatures that constrain the size, the kinematics, and the asymmetry of the gas envelope,  respectively. The small periodic oscillations of the visibility in the continuum reveal the faint companion of $\delta$ Cen. We can also observe the many telluric lines that can be seen between 3 and 3.5~$\mu$m and between 4.7 and 5~$\mu$m.


\begin{figure*}[htbp]
\centering\includegraphics[width=0.8\textwidth]{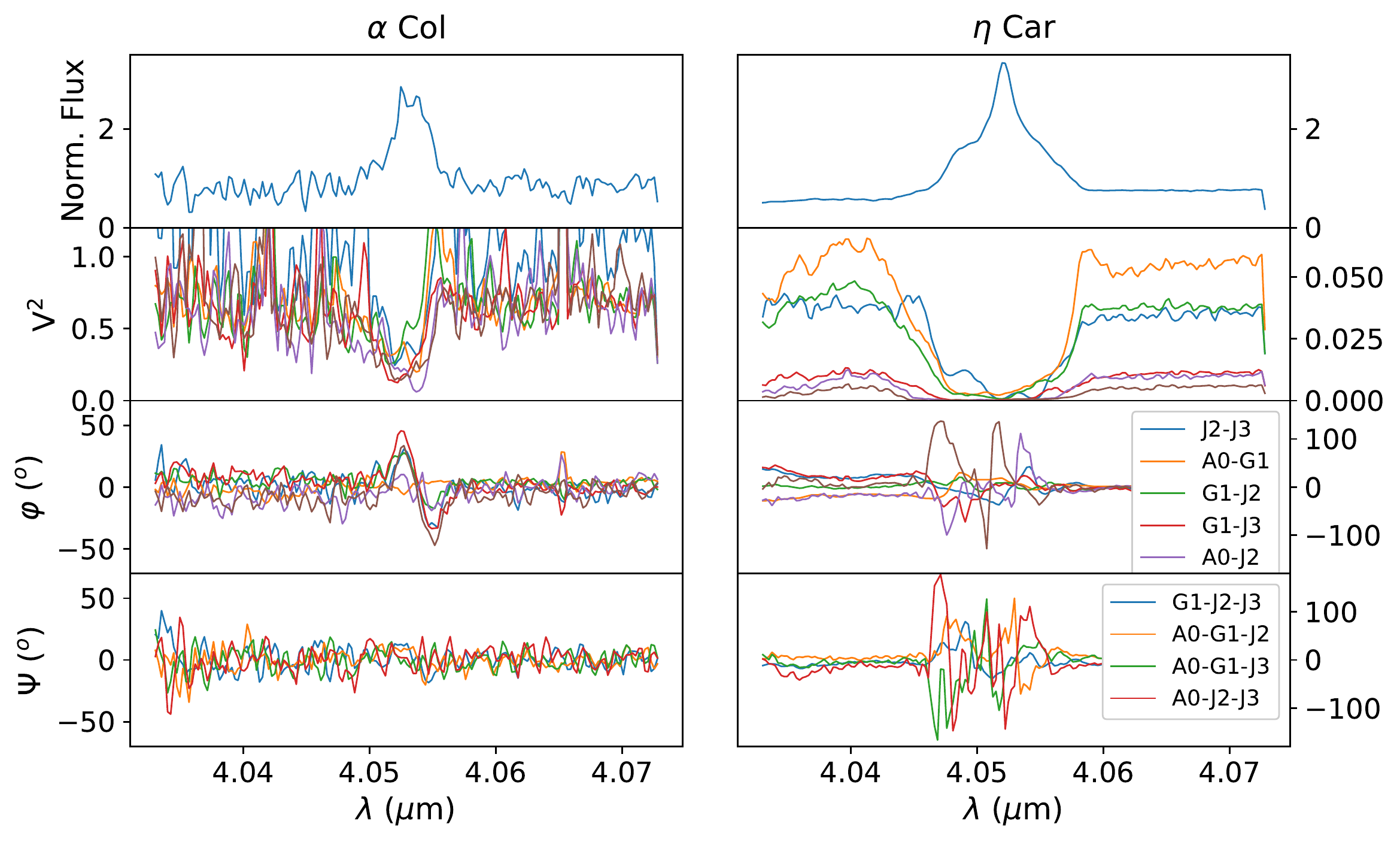}
\caption{MATISSE VHR (R=3370) observation of the classical Be star $\alpha$ Col (left) and LBV $\eta$ Car (right). From top to bottom: Normalised flux in the Br$\alpha$ line, square visibilities, differential phases and closure phases.}
\label{Fig:VHIGH}
\end{figure*}

MATISSE can also operate in L and M bands in a VHR mode with R$\sim$3400. With MATISSE used as stand-alone, the standard DIT, limited by the coherence time of an atmosphere, allows only the observations of a few very bright objects (brighter than 40 Jy in L and 55 Jy in M with ATs) and with a very short spectral coverage  (the L band window centred on the Br$\alpha$ line at 4.05~$\mu$m would be observable with a coverage of about 2.5~nm).
 The range of application is now broadened thanks to GRA4MAT to 20 Jy sources in the L band and 25 Jy in the M band on ATs. 
 As an illustration of this mode's capability, we present in Fig.~\ref{Fig:VHIGH} VHR data from two emission line stars: one very bright, the LBV $\eta$ Car, (F$_L\sim$1500Jy) and the other close to the VHR flux limit with ATs, the classical Be star $\alpha$ Col (F$_L\sim$20Jy).
The $\alpha$ Col data show the classic interferometric signature of classical Be stars, including the following: a drop in the visibility in the line, `S' shaped differential variation, and close-to zero closure phases.\\
$\eta$ Car is an interacting binary star composed of a LBV and a O or a WR star. Its complex and highly variable environment is the result of the collision of the strong radiatively driven winds that produced these components. The Br$\alpha$ data presented in Fig.~\ref{Fig:VHIGH} show a very broad line with a complex structure. The visibility, differential phase, and closure phase variations through the line are similar to those of Br$\gamma$ presented in \cite{Weigelt2016}.



The third advantage of observations with GRA4MAT, after the sensitivity and the spectral coverage, is that it stabilises the instrumental+atmosphere transfer function by reducing the piston jitter hence reducing the broadband calibration errors produced by the source and calibrator process.
Moreover, the jitter stabilisation in the N band mitigates -- or even cancels -- the bias effect in the coherent flux. As displayed in Fig.\ref{fig:biasN}, in the 8-9 $\mu$m range, a sensitivity approaching 1 Jy can be reached after a 1 min exposure from a coherent integration of N band frames (reference: Berio et al, in preparation).



\section{Conclusion}
 MATISSE has unique worldwide characteristics and performance presented in this article. The instrument capabilities offer two key breakthroughs. The first is the opening of the L and M bands (2.8–4.2 and 4.5–5.0 $\mu$m bands, respectively) for long-baseline mid-infrared interferometry, in addition to the N band which was available at the VLTI. The second breakthrough results from the four beam recombination and the access to a very efficient u-v coverage and mid-infrared imaging as illustrated for the FS CMa and NGC1068 examples, which can be performed with four UTs or ATs.

The performance of MATISSE surpasses what has been achieved in the past by MIDI. Not only is the N band sensitivity maintained, but also the four telescope recombination and access to the L and M band are made possible. 
As demonstrated in Table \ref{tab:fairandgood}, the MATISSE sensitivity limits, which are defined as the lowest flux in Jansky to achieve a visibility accuracy of 0.1, a closure phase accuracy of 5$^\circ$, a differential phase precision of 4$^\circ$, and a coherent flux S/N of 10, are below 0.1 $Jy$ in the L band and reach 0.3 $Jy$ in the N band for the differential phase and coherent flux in low spectral resolution with UTs. 
 
Indeed, the instrument capabilities go well beyond the sensitivity aspect. Each observation in the L and M band will give a totally new view of the astrophysical objects. Moreover, the four telescope observations provide access to phase closure observables and therefore to the study of asymmetric structures. 

The performances of MATISSE will be further improved in the near future by completing the commissioning of the GRA4MAT on UTs. This will boost the sensitivity and precision of MATISSE, in particular in the higher spectral resolution mode of the L and M bands on UTs, as it already did on ATs. 
The improvement of the UT adaptive optics for interferometry that is being considered by ESO as part of the GRAVITY+ project \citep{Eisenhauer2019} will multiply, by more than ten, the number of AGN accessible to MATISSE and allow us to study them in a much broader luminosity range which would bring the decisive unification of dust torus models. 
Improved access to the sky coverage through the use of reference stars \citep{Boskri2021} would boost the possibility of observing numerous Seyfert 2 AGN and protoplanetary disks very substantially.

This article shows how incomparable the instrument is to tackle the faint object science and the general astrophysical cases that require spectroscopic capabilities and/or accurate observable quantity measurements. Several examples are provided here to illustrate how those performances allow one to access the following: the stellar diameters; the binary star flux ratio and orbit point measurements; and the spectral line observations of the gas phase and the one of spectral bands providing access to solid materials presenting signatures of their mineralogy. Foreground results can be expected from this new instrument, which is opening a new observing window for long baseline interferometry, showing the most inner and hidden parts of the emitting sources and thus providing a field of potential discoveries. {MATISSE can be seen as a pathfinder for the future of mid-infrared interferometry. Several prospective projects such as the ground based Planet Finder Interferometer \citep[PFI,][]{Monnier2018} or future space missions similar to Darwin/TPF \citep[Terrestrial Planet Finder,][]{Cockell2009} could benefit from the current instrumental progress and achievement brought to the field of mid-infrared interferometry by MATISSE.}

\begin{acknowledgements}
MATISSE is defined, funded and built in close collaboration with ESO, by a consortium composed of French (INSU-CNRS in Paris and OCA in Nice), German (MPIA, MPIfR and University of Kiel), Dutch (NOVA and University of Leiden), and Austrian (University of Vienna) institutes. The Conseil Départemental des Alpes-Maritimes in France, the Konkoly Observatory and Cologne University have also provided resources for the manufacture of the instrument. A thought goes to our two nice colleagues, Olivier Chesneau and Michel Dugué, at the origin of the MATISSE project with several of us, and with whom we shared plenty of beautiful moments. We thank our anonymous referee for his useful questions and suggestions.
\end{acknowledgements}

\bibliographystyle{aa}
\bibliography{biblio}

\begin{appendix} \label{appendix}
\section{MATISSE noise model}\label{sec:noise_model}
\subsection{Fundamental noises' error}
To predict the variance $\sigma_{F}$ of the fundamental noises' error, we used a noise model based on the fact that all of MATISSE measurements are derived from the complex coherent flux estimates $C_{ij} (\lambda)$. The error on the coherent flux is set by the fundamental noises, including: source photon noise, thermal background photon noise, and detector readout noise $\sigma_{RON}$. 
The corresponding variance of the coherent flux per spectral channel and per frame is thus given by the following:
    \begin{equation}
         \sigma^2_{Cij}=N_T(n_{*}+n_{B})+N_{pix}\sigma^2_{RON}~,
    \end{equation}
where $n_{*}$ and $n_{B}$ are the average numbers of photons per telescope from the source and the thermal background, respectively, $N_{pix}$ is the number of pixels encoding the signal, and $N_T$ is the number of telescopes.
The coherent flux S/N per channel and per frame is given by
\begin{equation}
         S/N_{Cij1}=\frac{C_{ij}}{\sigma_{Cij}}=\frac{\sqrt{n_{i*}n_{j*}}V_{ij}}{\sqrt{N_Tn_{*}+N_Tn_{B}+N_{pix}\sigma^2_{RON}}}~,
    \end{equation}
where $n_{i*}$ is the source photon number for telescope $i$. When $N_{F}$ frames are added coherently and averaged over $N_{\lambda}$ spectral channels, the coherent flux S/N becomes
\begin{equation}
         S/N_{Cij}=S/N_{Cij1}\sqrt{N_{\lambda}N_{F}}~.
    \end{equation}
An incoherent integration of frames, that is an addition of interferogram power spectra ($C^2_{ij}$), averaged over the different spectral channels would yield the following:    
\begin{equation}
\label{eq:SNRC}
         S/N_{C^2_{ij}}=\frac{(S/N_{Cij1})^2}{\sqrt{(1+2(S/N_{Cij1})^2)}}\sqrt{N_{\lambda}N_{F}}~.
    \end{equation}
    
{We note that this S/N on the coherent flux obtained from incoherent integration decreases very rapidly when the S/N per frame $S/N_{Cij1}<1$. In practice, this nearly limits incoherent integration to S/N per frame larger than 1.}
\\
When the integrated $S/N_{Cij}>1$, the precision (standard deviation) on the phase of the coherent flux can be approximated by
 \begin{equation}
 \label{eq:precision}
         \sigma_{\varphi_{ ij}} \simeq \frac{1}{\sqrt{2}~S/N_{Cij}}~. 
    \end{equation}   

This is a good approximation of the error on the differential phase, as the error introduced by the much broader reference channel can be neglected. In the same conditions, the closure-phase precision can then be given by
 \begin{equation}
         \sigma^2_{\psi_{ijk}}\simeq\sigma^2_{\varphi_{ij}}+\sigma^2_{\varphi_{jk}}+\sigma^2_{\varphi_{ki}}\simeq 3 \sigma^2_{\varphi}~.
    \end{equation}    
    
When $\sigma_{\varphi_{ij}}>1$~rad, the closure phase error has a more complex expression. Nevertheless, we found that a two terms expansion fitted our precision estimates well on the closure phase
 \begin{equation}
         \sigma_{\psi_{ijk}}\simeq\sigma_{\varphi} \sqrt{3}+a\sigma^3_{\varphi}~, 
    \end{equation}  
   where $a$ is a constant associated with a given MATISSE set-up and which is usually set from the measured dispersion of the estimators. 
\\
Finally, as the absolute squared visibility is deduced from the squared modulus of the coherent flux divided by the photometry of the two involved beams (see Eq. \ref{eq:V2}), its fundamental precision also depends on the beam photometries $n_{i*}$ and $n_{j*}$:

\begin{equation}
         {\sigma_{V^2_{ij}}}=V^2_{ij}
        \left( {\frac{\sigma^2_{C^2_{ij}}}{C^2_{ij}}+\frac{\sigma^2_{n_{i}}}{n^2_{i*}}+\frac{\sigma^2_{n_{j}}}{n^2_{j*}}}
         \right)^{1/2}
         \label{eq:s2vvsv4}~.
    \end{equation}
Here, $\sigma_{F}(\lambda)$ is the predicted standard deviation of the error generated by the fundamental noises for each observable. The important input values such as the number of photons from the source, $n_*$, or the detector readout noise, $\sigma^2_{RON}$, were evaluated on the basis of the instrument characteristics measured in laboratory or updated using observations on sky (e.g. instrumental visibility, warm optics, cold optics and VLTI transmissions). 

\subsection{Broadband error}
{The calibration errors on the visibility and closure phase due to seeing and instrument variations are given in Table \ref{tab:broadbanderrors}. The absolute visibility is also affected by the broadband photometric error (denoted as $\epsilon_{p_{i}}$ hereafter), which is quadratically added to the fundamental noise, 
described in Eq. \ref{eq:s2vvsv4}, following}

\begin{equation}
         {\sigma_{V^2_{ij}}}=V^2_{ij}
        \left( {\frac{\sigma^2_{C^2_{ij}}}{C^2_{ij}}+\frac{\sigma^2_{n_{i}}}{n^2_{i*}}+\frac{\sigma^2_{n_{j}}}{n^2_{j*}}+\frac{\epsilon^2_{p_{i}}}{F^2_{i}}+\frac{\epsilon^2_{p_{j}}}{F^2_{j}}}
         \right)^{1/2}
         \label{eq:s2vvsv4new}~, 
    \end{equation}      
where $F_{i}$ is the source flux collected from telescope i.

\end{appendix}

\end{document}